\documentclass[11pt]{article}
\usepackage{jhep}
\usepackage[T1]{fontenc}
\usepackage{bm}
\usepackage{graphicx}
\usepackage{subfigure}
\usepackage{caption}
\usepackage[normalem]{ulem}
\usepackage{gensymb}
\usepackage{pifont}
\usepackage{bbold} % for \mathbold{1}
\usepackage{comment}
\usepackage{slashed}
\usepackage{wasysym}
\usepackage{fixmath}
\usepackage{snapshot}

\graphicspath{{./}{./figures/}}

\definecolor{GWcolour}{rgb}{1,0,0}
\definecolor{DM}{rgb}{1,0,1}
\definecolor{OGcolour}{rgb}{0,0.5,0}
\definecolor{azure}{rgb}{0.0, 0.5, 1.0}
\definecolor{PScolour}{rgb}{0.48, 0.07, 0.07}

\newcommand{\beq}{\begin{equation}}
\newcommand{\eeq}{\end{equation}}
\newcommand{\bea}{\begin{eqnarray}}
\newcommand{\eea}{\end{eqnarray}}
\newcommand{\bi}{\begin{itemize}}
\newcommand{\ei}{\end{itemize}}

\newcommand{\g}{\lambda_{\phi h}}

\newcommand{\nocontentsline}[3]{}
\newcommand{\tocless}[2]{\bgroup\let\addcontentsline=\nocontentsline#1{#2}\egroup}

\definecolor{mymagenta}{rgb}{1,0,1}

\newcommand\sumint[1]{\int\kern-1.5em\sum\nolimits_{#1}}% sum-integral symbol

\usepackage{axodraw2,pix}
\newcommand{\rmii}[1]{{\mbox{\tiny\rm{#1}}}}

\newcommand{\T}{\rmii{$T$}}

\newcommand{\Tint}[1]{{\hbox{$\sum$}\!\!\!\!\!\!\!\int\,}_{\!\!\!\!\raise-0.9ex\hbox{$\scriptstyle{#1}$}}}
\newcommand{\Tinti}[1]{{{\Sigma}\!\!\!\!\raise0.3ex\hbox{$\int$}_\rmii{${#1}$}}}
\newcommand{\Tintip}[1]{{{\Sigma'}\!\!\!\!\!\raise0.3ex\hbox{$\int$}_\rmii{${#1}$}}}

\title{The Electro-Weak Phase Transition at Colliders: Confronting Theoretical Uncertainties and Complementary Channels.}
\author[a,b]{Andreas Papaefstathiou,}

\author[c,d]{and Graham White}

\affiliation[a]{Higgs Centre for Theoretical Physics, University of Edinburgh, Peter Guthrie Tait Road, Edinburgh EH9 3FD, UK.}

\affiliation[b]{Department of Physics, Kennesaw State University, Kennesaw, GA 30144, USA.}

\affiliation[c]{TRIUMF, 4004 Wesbrook Mall, Vancouver, BC V6T 2A3, Canada.}

\affiliation[d]{Kavli IPMU (WPI), UTIAS, The University of Tokyo, Kashiwa, Chiba 277-8583, Japan.}

\emailAdd{apapaefs@cern.ch}
\emailAdd{graham.white@ipmu.jp}
\date{\today}
\abstract{We explore and contrast the capabilities of future colliders to probe the nature of the electro-weak phase transition. We focus on the real singlet scalar field extension of the Standard Model, representing the most minimal, yet most elusive, framework that can enable a strong first-order electro-weak phase transition. By taking into account the theoretical uncertainties and employing the powerful complementarity between gauge and Higgs boson pair channels in the searches for new scalar particles, we find that a 100~TeV proton collider has the potential to confirm or falsify a strong first-order transition. Our results hint towards this occurring relatively early in its lifetime. Furthermore, by extrapolating down to 27~TeV, we find that a lower-energy collider may also probe a large fraction of the parameter space, if not all. Such early discoveries would allow for precise measurements of the new phenomena to be obtained at future colliders and would pave the way to definitively verify whether  this is indeed the physical remnant of a scalar field that catalyses a strong first-order transition.}

\begin{document}
\maketitle

\section{Introduction}

One of the biggest scientific questions that could be answered by the largest next-generation experiments is ``What was the nature of the cosmological electro-weak symmetry breaking?'' \cite{Ramsey-Musolf:2019lsf,Caprini:2019egz,Benedikt:2653674}. If the electroweak phase transition (EWPT) was strongly first order, a property that we shall discuss below in detail, signatures would potentially be detectable at space-based gravitational wave observatories such as LISA~\cite{Caprini:2019egz} or Decigo~\cite{Sato_2017}, whose sensitivities peak at the millihertz to the decihertz range, frequencies that would occur if a phase transition took place at the electro-weak scale~\cite{Weir:2017wfa,Mazumdar:2018dfl,Zhou:2019uzq}.\footnote{Some models can predict signals at higher or lower frequencies, relevant to pulsar timing arrays \cite{Kobakhidze:2017mru,vonHarling:2017yew} and ground-based experiments like the Einstein Telescope, respectively~\cite{Beniwal:2018hyi,Punturo:2010zz}.} 

The Standard Model (SM) of particle physics by itself predicts the electro-weak transition to be a smooth crossover~\cite{Kajantie:1996mn}. Therefore, a strong first-order electro-weak phase transition (SFO-EWPT) dictates physics beyond the SM. Although the nature of the EWPT is an interesting inquiry in itself, the requirement of  new physics that enable it can also provide the right conditions to explain the baryon asymmetry of the Universe~\cite{Morrissey:2012db,White:2016nbo}. Next-generation collider experiments are expected to dramatically increase sensitivity to states that interact with the electro-weak sector, making such new phenomena a prime target. 

To design a collider that can answer both quantitative and qualitative questions about the nature of the EWPT, it is imperative to address the fact that there exists a multitude of models that can catalyse a SFO-EWPT~\cite{Pietroni:1992in,Cline:1996mga,Ham:2004nv,Funakubo:2005pu,Barger:2008jx,Chung:2010cd,Espinosa:2011ax,Chowdhury:2011ga,Gil:2012ya,Carena:2012np,No:2013wsa,Dorsch:2013wja,Curtin:2014jma,Huang:2014ifa,Profumo:2014opa,Kozaczuk:2014kva,Jiang:2015cwa,Curtin:2016urg, Vaskonen:2016yiu,Dorsch:2016nrg,Huang:2016cjm,Chala:2016ykx,Basler:2016obg,Beniwal:2017eik,Bernon:2017jgv,Kurup:2017dzf,Andersen:2017ika,Chiang:2017nmu,Dorsch:2017nza,Beniwal:2018hyi,Bruggisser:2018mrt,Athron:2019teq,Kainulainen:2019kyp,Bian:2019kmg,Li:2019tfd,Chiang:2019oms,Xie:2020bkl,Bell:2020gug}. Keeping to models that are reasonably minimal, the real singlet scalar field extension of the SM is an ideal test case, as it has no direct gauge interactions, making detection at a collider significantly challenging. Therefore, a collider powerful enough to falsify or confirm the relevant parameter space of this model should also be powerful enough to make a qualitative statement about the cosmological EWPT~\cite{Kotwal:2016tex,Huang:2017jws,Chen:2017qcz,Alves:2018oct,Ramsey-Musolf:2019lsf,Alves:2019igs}. 

To investigate the nature of the EWPT, one has to confront the breakdown of conventional analysis techniques of the EWPT~\cite{Croon:2020cgk}. In particular, the scale dependence and gauge dependence of thermal parameters, including the actual strength of a phase transition, can potentially result in theoretical uncertainties large enough to qualitatively alter conclusions derived using such techniques. Here, we strive to include the theoretical uncertainties when approaching the question of what collider specifications are needed to uncover the nature of the EWPT. The dramatic nature of these uncertainties was recently discussed in detail in context of the SM effective field theory~\cite{Croon:2020cgk}. The problems discussed in ref.~\cite{Croon:2020cgk} are even further accentuated if the theory catalysing the EWPT involves large couplings and scale hierarchies. 

The prospective high-energy upgrade of the Large Hadron Collider (HE-LHC) and the Future Circular Collider (FCC) at proton centre-of-mass energies of 27~TeV and 100~TeV, respectively, are the focus of our studies. To draw our conclusions, multiple detection channels of the new scalar resonance that appears in the real singlet model were considered. Specifically, parameter-space points that are difficult to detect in decays of the new scalar to two SM-like Higgs bosons, become easier to detect in its decays to vector bosons. Thus far in hadron collider analyses of the real singlet extension of the SM, the latter channels have been mostly neglected.\footnote{We note, however, that they have been considered in the context of the CLIC $e^+ e^-$ collider in ref.~\cite{No:2018fev}, discussed in qualitative terms in ref.~\cite{Alves:2020bpi}. They were also investigated in an extrapolation of 8~TeV results in ref.~\cite{Buttazzo:2015bka}.} We find that including decays to vector bosons dramatically improves the reach of a collider. Finally, we note that the theoretical uncertainty appears to grow with the mass of the new scalar particle. 

The paper is organised so as to summarise our methods and key results in its main portion, while deferring several technical aspects of interest to appendices. The structure is as follows: in section~\ref{sec:model} we outline the model that forms the focus of our investigations, the real singlet scalar field extension of the SM. As a complement, appendix~\ref{app:decaymodes} provides further details on the features of the model. In section~\ref{sec:phasetrans} we describe the method that we employ to calculate the order of the electro-weak phase transition, including the treatment of theoretical uncertainties that leads to our phase-space segmentation into categories which encapsulate varying degrees of theoretical uncertainties. Appendix~\ref{app:potential} provides details on the form of the one-loop potential. In section~\ref{sec:constraints} we provide a summary of the collider information that we employ in our phenomenological analysis, deferring a host of detailed information to appendices~\ref{app:current},~\ref{app:highlumi},~\ref{app:ee} and~\ref{app:futurepp}, where we discuss current collider constraints, high-luminosity LHC prospects, electron-positron collider constraints and future hadron collider analyses, respectively. We present the phenomenological results, obtained for a 100~TeV and extrapolated to a 27~TeV collider in section~\ref{sec:results}.  There, we also give a selection of benchmark points in the form of parameter sets with a selection of associated relevant observables. In addition, in appendix~\ref{app:altresults} we briefly discuss results originating from the alternative parameter-space parametrisation and in appendix~\ref{app:finetuning} we study the degree of fine tuning present in the model. We provide our conclusions and outlook in section~\ref{sec:conclusions}.

\section{Standard Model Augmented by a Real Singlet Scalar Field}\label{sec:model}
We focus on the SM extended by a real
singlet scalar field. Then, the most general form of the scalar potential that depends on the Higgs doublet field, $H$, and a gauge-singlet scalar field, $S$, is given by (see, e.g.~\cite{OConnell:2006rsp, Profumo:2007wc, Barger:2007im, Espinosa:2011ax, Pruna:2013bma, Chen:2014ask, Kotwal:2016tex, Robens:2016xkb, Englert:2020gcp, Adhikari:2020vqo}):
%\begin{eqnarray}\label{eq:xsm}
%V(H,S) = &-&\mu^2 (H^\dagger H) + \lambda (H^\dagger H)^2 + \frac{a_1}{2} (H^\dagger H) S \\ %\nonumber
%             &+& \frac{a_2}{2} (H^\dagger H) S^2 + \frac{b_2}{2} S^2 + \frac{b_3}{3} S^3 + \frac{b_4}{4} S^4 \;,
%\end{eqnarray}
\begin{eqnarray}\label{eq:xsm}
V(H,S) &=& \mu^2 (H^\dagger H) + \frac{1}{2} \lambda (H^\dagger H)^2 + K_1 (H^\dagger H) S \\ \nonumber
             &+& \frac{K_2}{2} (H^\dagger H) S^2 + \frac{M_S^2}{2} S^2 + \frac{\kappa}{3} S^3 + \frac{\lambda_S}{2} S^4 \;,
\end{eqnarray}
where the interactions proportional to $K_{1,2}$ constitute the Higgs ``portal'' that links the SM with the singlet scalar. Note that we do not impose a $\mathbold{Z}_2$ symmetry that would preclude terms of odd 
powers of $S$. Such terms are often key in catalysing a tree-level barrier between the electro-weak symmetric and broken phases, thus resulting in a stronger transition. 

After electro-weak symmetry breaking (EWSB) occurs, the Higgs doublet and the singlet scalar fields both attain vacuum expectation values (vevs) $v_0$ and $x_0$, respectively. To obtain the physical states, we expand about these: $H \rightarrow (v_0 + h) / \sqrt{2}$, with $v_0 \simeq 246$~GeV and $S \rightarrow x_0 + s$. Inevitably, the two states $h$ and $s$ mix through both the Higgs portal parameters $K_1$ and $K_2$ as well as the singlet vev and hence they do not represent mass eigenstates. Therefore, upon diagonalising the mass matrix one obtains two eigenstates,
\begin{eqnarray}
h_1 &=& h \cos \theta + s \sin \theta \;, \\ \nonumber 
h_2 &=& - h \sin \theta + s \cos \theta \;. 
\end{eqnarray}
where $\theta$ is a mixing angle that can be expressed in terms of the parameters of the model. For $\theta \sim 0$, $h_1 \sim h$ and $h_2 \sim s$. We will identify the eigenstate $h_1$ with the state observed at the LHC, and hence set $m_1 = 125.1$~GeV. We will only consider $m_2 > m_1$ here.\footnote{The case $m_1 > 2 m_2$ in the context of SFO-EWPT in the real singlet extension of the SM was investigated in ref.~\cite{Kozaczuk:2019pet}}

All the couplings of $h_{1,2}$ to the rest of the SM states are simply obtained by rescaling:
\begin{equation}\label{eq:rescaledcouplings}
g_{h_1 XX} = g_{hXX}^{\mathrm{SM}} \cos \theta\;,\;\; g_{h_2 XX} =  - g_{hXX}^{\mathrm{SM}}  \sin \theta\;,
\end{equation}
with $XX$ any SM final state, i.e.\ fermions or gauge bosons. These allow for constraints to be imposed on $\theta$ through the measurements of SM-like Higgs boson (i.e.\ $h_1$) signal strengths and for searches of $h_2$ decaying to SM particles. In addition, if $m_2 \geq 2 m_1$, then $h_2 \rightarrow h_1 h_1$ becomes kinematically viable and potentially significant, providing an additional search channel. 

\begin{figure}[htp]
  \centering
  \includegraphics[width=0.7\columnwidth]{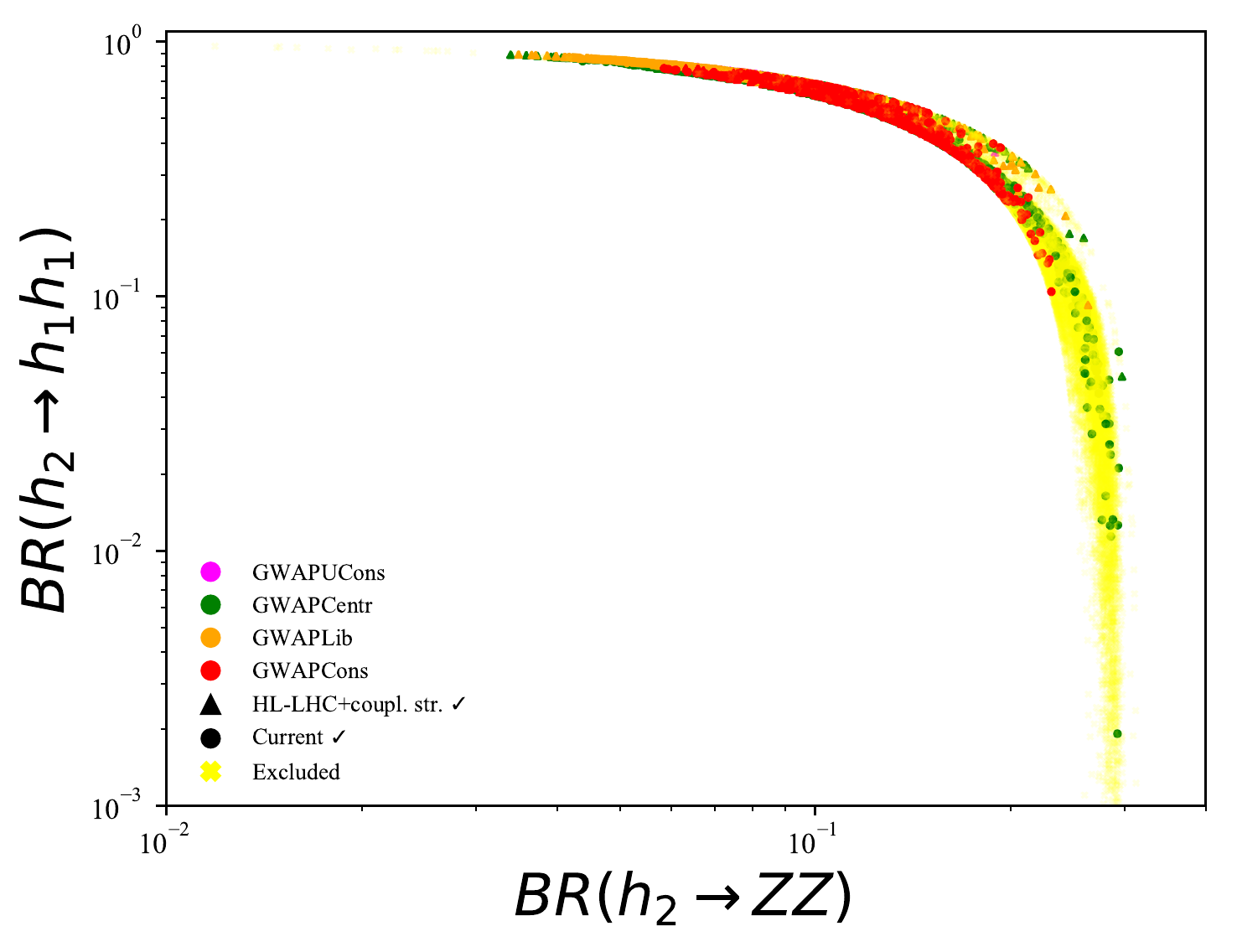}
\caption{The branching ratio BR$(h_2 \rightarrow h_1 h_1)$ plotted against BR$(h_2 \rightarrow ZZ)$, over the parameter-space points of the present study for the various categorisations (see section~\ref{sec:paramspace}). We show points that pass the current constraints ($\CIRCLE$) or the HL-LHC constraints plus future signal strength ($| \sin \theta | < 0.1$) constraints ($\blacktriangle$), as well as those that are currently excluded (faint $\color{yellow}\pmb{\times} \color{black}$).}
\label{fig:BRHH_ZZ}
\end{figure}

We discuss the decay modes of the $h_2$ boson within the context of our study in detail in appendix~\ref{app:decaymodes}. Here, we wish emphasise the complementarity between the decay modes $h_2 \rightarrow h_1 h_1$ and $h_2 \rightarrow VV$ ($V=Z,W$), exemplified by fig.~\ref{fig:BRHH_ZZ}, where the branching ratio of BR$(h_2 \rightarrow h_1 h_1)$ is plotted against BR$(h_2 \rightarrow ZZ)$, over the parameter-space points of the present study (see section~\ref{sec:paramspace} for their definitions). From fig~\ref{fig:BRHH_ZZ}, it also becomes clear that since the $h_2 \rightarrow ZZ$ decay (and hence $h_2 \rightarrow W^+W^-$) never becomes less frequent than $\mathcal{O}(\mathrm{few}~\%)$, it is has the potential to be overall a more powerful probe of the parameter space than $h_2 \rightarrow h_1 h_1$, which can  attain small BR values, down to $\mathcal{O}(10^{-3})$. We also note that there are exist viable SFO-EWPT parameter-space points with $m_2 < 2 m_1$, where $h_1 h_1$ production is not kinematically allowed. In appendix~\ref{app:decaymodes} we show an alternative view of the complementarity between the $h_1 h_1$ and $ZZ$ final states, through the ratio of BR$(h_2 \rightarrow h_1 h_1)$/BR$(h_2 \rightarrow ZZ)$ versus the mass of $h_2$, in the bottom-right panel of fig.~\ref{fig:BRs}.

\section{Calculating the order of the phase transition}\label{sec:phasetrans}
Perturbative methods of treating the effective potential at finite temperature suffer from both gauge \cite{Patel:2011th} and a break down of perturbation theory~\cite{Linde:1980ts}. The breakdown of perturbation theory arises due to Linde's infamous infrared problem --  during the phase transition as the expansion parameter is not the actual coupling but the coupling multiplied by the mode occupation instead: $g n_{B} \sim g T/m $ (where $T$ is the temperature, $n_B$ is the mode occupation, $m$ is a mass and $g$ is some coupling), which clearly diverges for $m\to 0$. A probe of how well perturbation theory is working is to measure the scale dependence of the theory. If perturbation theory is converging quickly, we can define our potential at one loop order and the scale dependence from running should approximately cancel the scale dependence at one loop, any remaining scale dependence is formally higher order. A large sensitivity therefore shows that the theory is particularly sensitive to higher-loop corrections which points to the failure of perturbation theory to converge quickly. \par 

The scale dependence is the dominant theoretical uncertainty if one keeps within the perturbative region of the theory~\cite{Croon:2020cgk}. A partial solution to the scale dependence is to perform a resummation of the masses. Doing so makes the uncertainty in the critical temperature of the SM effective theory non-trivial but manageable, in contrast to the uncertainty in the gravitational wave amplitude generated from a first-order transition which is at the multi-order of magnitude level~\cite{Croon:2020cgk}. In the present paper we consider bosons that couple to the SM with potentially large couplings - in fact past papers that studied the EWPT, considered portal couplings of $K_2 \sim \mathcal{O} (10)$ \cite{Kotwal:2016tex}. For that reason we expect the theoretical uncertainties even in the strength of the phase transition to be very large. \par 

The gauge dependence can be avoided via an expansion in the Planck constant, $\hbar$, instead of the usual loop expansion, but unless one expands to second order in $\hbar$, the scale dependence is unmanageable \cite{Chiang:2018gsn}.\footnote{It should be noted that the running of the couplings was not included in ref.~\cite{Chiang:2018gsn} and the true scale dependence in this prescription is not yet known, though it is suspected to be worse than the ``Arnold-Espinosa'' method~\cite{Arnold:1992rz} due to the lack of resummation at the first order.} Going to two loops is cumbersome and we will not pursue it here as a solution. Another solution is to integrate out the heavy ``Matsubara'' modes, which results in a theory that is effectively three-dimensional~\cite{Appelquist:1981vg,Nadkarni:1982kb,Laine:1995np,Kajantie:1995dw,Karjalainen:1996rk,Losada:1996ju,Cline:1997bm,Andersen:1998br,Laine:2000kv,Vepsalainen:2007ji,Brauner:2016fla,Gorda:2018hvi,Niemi:2018asa}. The dimensionally-reduced theory is manifestly gauge-invariant and includes all-orders of resummation, resulting in a substantial reduction in the uncertainties~\cite{Croon:2020cgk}. For the SM extended by a real singlet scalar field, there can be two light dynamical modes, rendering the analysis a theoretical challenge that we leave to future work.

Here we instead advocate to analyse the electro-weak phase diagram within a given model via an approach that balances convenience, while limiting theoretical uncertainties. Given that we are interested in assessing the order of the phase transition and not gravitational wave phenomenology, the uncertainties are often large but nonetheless manageable\footnote{This statement appears to break down for larger $h_2$ masses, see fig.~\ref{fig:deltaM1}.} in the usual four-dimensional perturbative calculation, where the gauge and scale dependence are treated as theoretical uncertainties. 

To assess the question whether a collider rules out a SFO-EWPT, we consider the {\it maximum} value of the order parameter, requiring
\begin{equation}
    {\rm Max} \left( \frac{\Delta h}{T _C} \right) >  1  \;,
\end{equation}
for a given parameter point with physical quantities matched at the $Z$ pole and with the renormalisation scale and gauge parameters varied. In the above, $T_C$ is the critical temperature at which the theory has a degenerate ground state and $\Delta h$ is the difference in the Higgs field between the two vacua at the critical temperature. By contrast, to assess the discovery prospects we instead look at the {\it minimum} value, requiring
\begin{equation}
    {\rm Min} \left( \frac{\Delta h}{T _C} \right) >  1  \;,
\end{equation}
when varying over the same range of the scale and gauge parameters. It is therefore useful to classify which points might be strongly first order, as they sometimes are when vary scale and gauge parameters, and when they are robust to theoretical uncertainties. The specific implementation of this and how we categorise our points we discuss in section~\ref{sec:paramspace}.

\subsection{Gauge and Scale Dependence}

To capture the theoretical uncertainties coming from the gauge dependence, we construct the effective potential in the general covariant gauge (also known as Fermi gauges) rather than the usual $R_\xi$ gauge~\cite{Andreassen:2013hpa}. The $R_\xi$ gauge uses a different gauge for every value of the scalar field(s). It is not a priori clear that this is possible~\cite{Arnold:1992fb,Laine:1994bf}. To construct the potential we follow the techniques of ref.~\cite{Andreassen:2013hpa}.

The potential at one loop in the covariant gauge is given by
\begin{equation}
    V_{\rm 1-loop} = V_{\rm tr}(h,s,Q ) + V_{\rm CW} (h,s,Q,\xi_W ,\xi _B) +V_T(h,s,Q ,\xi _W , \xi _B)\;,
\end{equation}
where $V_{\rm tr}$ is the tree-level potential, $V_{\rm CW}$ the Coleman-Weinberg (CW) term, $V_T$ the thermal term, $Q$ is the renormalisation scale and the $\xi _i$ ($i=W,B$) are the gauge parameters. The remaining details of the effective potential at finite temperature in the covariant gauge are given in appendix~\ref{app:potential}.

To minimise scale dependence, we use ``Arnold-Espinosa'' resummation~\cite{Arnold:1992rz} of the masses and allow all parameters in the effective potential to run. We use the program \texttt{SARAH}~\cite{Staub:2008uz} to derive the one-loop renormalisation group equations (RGEs) and we vary the RGE scale, $Q$, that defines the coupling and the CW potential, by an order of magnitude: $m_Z/2<Q<m_Z\times 5$, where $m_Z$ is the $Z$ boson mass. 

\subsection{Numerical calculation of the phase transition}\label{sec:numphase}

\texttt{PhaseTracer} is a publicly-available \texttt{C++} package~\cite{Athron:2020sbe} that traces the thermal evolution of the effective potential using the algorithm developed in ref.~\cite{Wainwright:2011kj}. It was designed to be robust,\footnote{We find very few points that cause a time out. These points are rare and we simply be analyse them by re-running the program.} with an improved treatment of thermal functions~\cite{Fowlie:2018eiu} and discrete symmetries. \par
We have developed a module for \texttt{PhaseTracer} for the real scalar singlet extension of the SM in the covariant gauge with ``Arnold-Espinosa'' resummation of thermal masses. Since the choice of gauge only enters the effective potential through the field-dependent masses, the covariant gauge can be readily implemented in \texttt{PhaseTracer} by simply providing expressions for these. We can then treat the gauge parameters as inputs we vary. \par 

The requirement that we have a SFO-EWPT typically requires large portal couplings, which implies that the unphysical scale dependence at one loop can become quite large, even at zero temperature. This, in addition to the unphysical gauge dependence of the minimum calculated at one loop, motivates a generous tolerance for the zero-temperature vev value of the Higgs field. In particular, if we wish to claim that a certain collider can exclude the possibility of the cosmological phase transition being strongly first order, we choose to be liberal and not too aggressive in excluding points that could be potential candidates. We discuss our analysis of the phase structure in detail in the next sub-section.

\subsection{Parameter-space categorisation}\label{sec:paramspace}

\begin{table}[h]
    \centering
    \begin{tabular}{c|c}
        Parameter & Range  \\ \hline 
        $\lambda _s$ & $[10^{-4},3]$ \\
                $K_1$ & $[-2000,0]$ GeV \\
                $K_2$ & $[0,10]$ \\
            $M_S$ & $[-2000,2000]$ GeV \\
        $\kappa$ & $-[1800,1800]$ GeV \\
    \end{tabular}
    \caption{The range of parameters used in our scans of the parameter space of the real singlet scalar extension of the SM.}
    \label{tab:parameters}
\end{table}

\begin{figure}
  \centering
  \includegraphics[width=0.9\columnwidth]{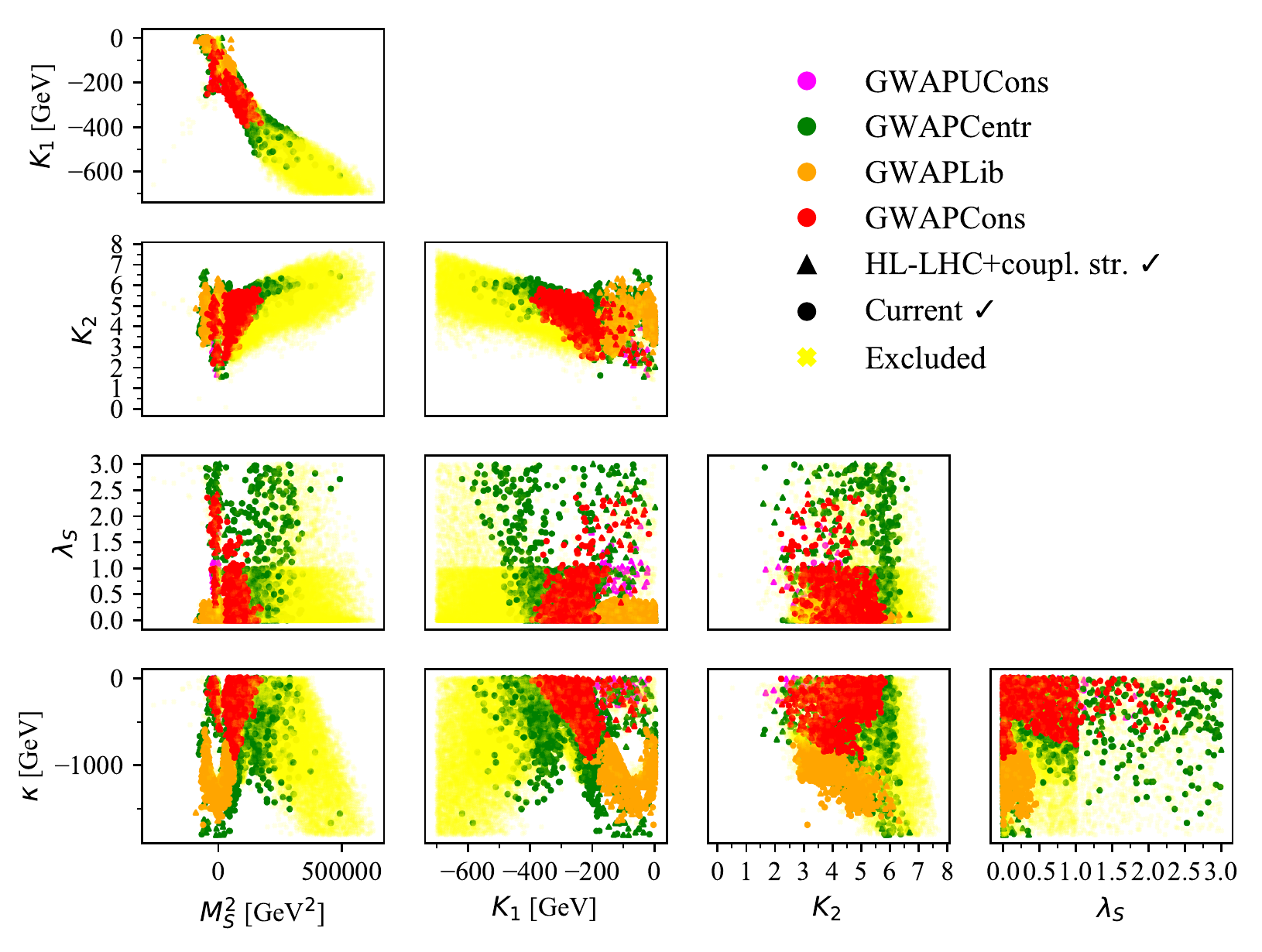}
\caption{Two-dimensional projections of the parameter space showing points obtained in our scan for the five free parameters that we scanned over. We show points that pass the current constraints ($\CIRCLE$) or the HL-LHC constraints plus future signal strength ($| \sin \theta | < 0.1$) constraints ($\blacktriangle$), as well as those that are currently excluded (faint $\color{yellow}\pmb{\times} \color{black}$).}
\label{fig:correlation}
\end{figure}

\begin{figure}
  \centering
  \includegraphics[width=0.8\columnwidth]{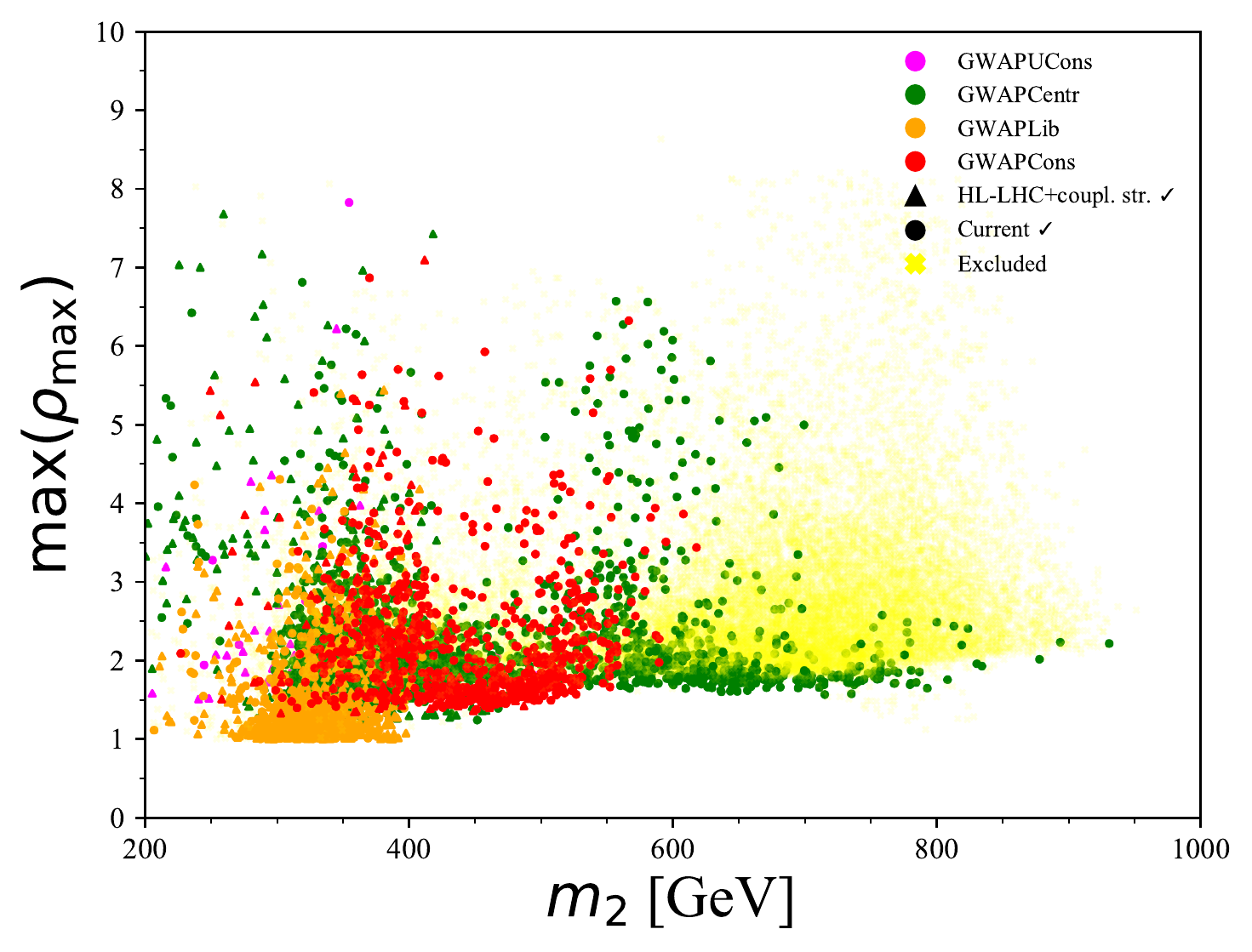}
\caption{The maximum value of $\rho_\mathrm{max} = \phi_C/T_C$ over the set of eight scale/gauge variations for each of the parameter-space points against the mass of $h_2$, $m_2$. We show points that pass the current constraints ($\CIRCLE$) or the HL-LHC constraints plus future signal strength ($| \sin \theta | < 0.1$) constraints ($\blacktriangle$), as well as those that are currently excluded (faint $\color{yellow}\pmb{\times} \color{black}$).}
\label{fig:rhomax}
\end{figure}

To derive parameter-space points that satisfy SFO-EWPT, we perform random scans over the five free parameters in the potential. We generate points with viable zero-temperature phenomenology, leaving a large leeway for theoretical uncertainties. Table~\ref{tab:parameters} presents our scan range, which was performed for $\mathcal{O}(10^6)$ points. We impose a cutoff on dimensionful parameters, at around 2~TeV, as well as on the dimensionless portal coupling $K_2$, in order to lie well below the perturbativity bound. This is because the theoretical uncertainties become unmanageable for larger couplings and large scale hierarchies. Note that our scan does not represent a uniform sampling of the parameter space: this implies that our results should not be interpreted so as to indicate densities of points in parameter space, but rather to map out its boundaries. We create a set of eight points for each parameter point with $\{ \mu , \xi _{W,Z} \}  \in \{ (m_Z/2, m_Z, 2m_Z, 5m_Z ), (0,3) \}$. We require the Higgs boson mass to equal $125.1$~GeV at $\mu=m_Z,~\xi_{W,Z}=0$ and we impose a restriction on the one-loop value of the sine of the mixing angle: $|\sin \theta| < 0.25$. Each of the eight points is individually passed to \texttt{PhaseTracer} for further processing. \texttt{PhaseTracer} calculates the phases and the transitions between them. The transitions are classified according to the value of $\rho = \phi_C / T_C$, where $\phi_C$ is the Higgs field vev and $T_C$ is the critical temperature. \par 

\texttt{PhaseTracer} has multiple advantages \cite{Athron:2020sbe} over \texttt{CosmoTransitions} \cite{Wainwright:2011kj}. However, \texttt{PhaseTracer} does not calculate the nucleation rate at present. We therefore consider approximate criteria in categorizing a strong first order phase transition. If there exist\footnote{This condition is to limit the cases where the tunneling rate is too slow for the phase transition to complete. We leave an analysis of the tunneling to future work. For now the reader is referred to refs.~\cite{Profumo:2014opa,Baum:2020vfl} for further discussion on this topic.} phase transitions with $T_c > 30$~GeV and $\rho > 1$, and no other transitions with $\rho \in [0.1, 1.0]$\footnote{The choice of a $0.1$ here rather than zero is because \texttt{PhaseTracer} can on occasion find small transitions that do not appear to be second order transitions, but appear to be spurious. \texttt{PhaseTracer} gives information on the minimum a lot lower than the critical temperature and such a minimum is expected to be large for a genuinely second order transition.} with higher $T_c$ than the $\rho > 1.0$ transition, then we label a point as satisfying the ``SFO condition''. Additionally, the maximum value of $\rho$, denoted as $\rho_\mathrm{max}$, is kept for each of the eight scale/gauge variations. Furthermore, if for an individual point there exists a Higgs field vev within $v_0 = 246 \pm 30$~GeV, that also corresponds to a transition to the absolute (i.e.\ deepest) minimum of the potential, we label it as satisfying the ``first type of vev condition''. The subset of points that satisfy the ``SFO condition'' are considered for additional post-processing. We also save the value of the Higgs field vev at the deepest minimum for each of these points. During the post-processing of the points, we also consider an additional property, pertaining to the group of eight scale/gauge variations: if at least one point has for the deepest minimum $v_0 < 246$ GeV and another has for the deepest minimum $v_0> 246$ GeV, we label it as satisfying the ``second type of vev condition''. We use these conditions to categorise the points. To best take into account the theoretical uncertainties, we consider two types of classification. The first type of classification separates points into:
\begin{itemize}
\item \textit{Ultra-Conservative} points: \textit{all eight} of them satisfy the SFO and \textit{all eight} also satisfy the first type of vev condition.
\item \textit{Conservative} points: \textit{all eight} of them have to satisfy SFO and \textit{at least one point} has to satisfy the first type of vev condition. Ultra-Conservative points are excluded from this category.
\item \textit{Centrist} points: \textit{at least one} point within the eight satisfying SFO and first type of vev conditions \textit{simultaneously}. (Ultra-)Conservative points are excluded from this category.
\item \textit{Liberal} points: \textit{at least one} point within the eight satisfies SFO and \textit{any other} satisfies the first type of vev conditions. Centrist and (Ultra-)Conservative points are excluded from this category. 
\end{itemize}
The second classification separates points into two mutually-exclusive categories:
\begin{itemize}
    \item \textit{Tight} points: \textit{all eight} of them have to satisfy SFO and the group has to satisfy the second type of vev condition.
\item \textit{Loose} points: \textit{at least one} point within the eight satisfying SFO and the group has to satisfy the second type of vev condition. 
\end{itemize}

In the main part of the article we show results for the first type of classification (GWAPUCons, GWAPCons, GWAPCentr and GWAPLib as defined above, respectively), with a selection of results for the second type (GWAPTight, GWAPLoose, respectively) shown in Appendix~\ref{app:altresults}. Our results show that the ``Centrist'' and ``Loose'' categories behave in a similar fashion, whereas the ``Liberal'' category of the first classification has no correspondence in the second and allows for larger theoretical uncertainties.  

We show two-dimensional projections of the parameter-space points obtained in our scan, for the five free parameters that we scanned over in fig.~\ref{fig:correlation}. Notice that the theoretical uncertainties get large when there is either a large scale hierarchy (including a large ratio of dimensionful parameters) or a large dimensionless coupling. Specifically if $\lambda _s \gtrsim 2$ or $K_2 \gtrsim 4$ then the theoretical uncertainties begin to get large. This is precisely what is to be expected if the uncertainties mostly arise due to the unphysical scale dependence coming from the breakdown of perturbativity. In fig.~\ref{fig:rhomax} we show the resulting maximum value of $\rho_\mathrm{max}$ over the eight scale and gauge variations for each of the parameter-space points. In both figs.~\ref{fig:correlation} and~\ref{fig:rhomax} we show points that pass the current constraints ($\CIRCLE$) or the HL-LHC constraints plus future signal strength ($| \sin \theta | < 0.1$) constraints ($\blacktriangle$), as well as those that are currently excluded (faint $\color{yellow}\pmb{\times} \color{black}$), see the next section and the relevant appendices for further clarification of these conditions. We emphasise that, in all of the phenomenological studies that follow, we employ the ``central'' parameter-space point in the group, i.e.\ with $\{ \mu , \xi _{W,Z} \}  = \{ m_Z, 0 \}$, irrespective of whether it satisfies the above conditions.

%Remarkably, a scalar field that is heavier that 2 TeV is part of the potentially-viable parameter space. A more advanced treatment of the phase transition is needed to determine how heavy a state can be and still catalyse a SFO-EWPT. 

\section{Summary of collider constraints}\label{sec:constraints}
There exist several categories of collider constraints that affect the parameter space of the real singlet extension of the SM. These come through direct searches for heavy scalar resonances (i.e.\ $h_2$), SM-like Higgs boson signal strength measurements (i.e.\ $h_1$) and electro-weak precision observables. To assess the potential of future colliders for exclusion or discovery of the real singlet model in the context of a SFO-EWPT, at the end of the lifetime of the high-luminosity run of the LHC (HL-LHC), we consider the most up-to-date collider constraints coming from these categories. 

To impose current constraints coming from heavy Higgs boson searches and Higgs boson measurements, we employ the \texttt{HiggsBounds}~\cite{Bechtle:2008jh,Bechtle:2011sb,Bechtle:2013gu,Bechtle:2015pma} and \texttt{HiggsSignals}~\cite{Bechtle:2013xfa, Stal:2013hwa, Bechtle:2014ewa} packages. In addition, we consider constraints coming from resonant Higgs boson pair production, $h_2 \rightarrow h_1 h_1$ and also incorporate the latest 13 TeV ATLAS and CMS SM-like Higgs boson global signal strengths, $\mu = \sigma^{\mathrm{measured}}/\sigma^{\mathrm{SM}}$, that are currently not included in \texttt{HiggsSignals}. Further details on these can be found in Appendix~\ref{app:current}. 

For the HL-LHC, we consider the prospects for the Higgs boson signal strength measurement, as well as various analyses assessing the heavy Higgs boson prospects of the HL-LHC in final states originating from $h_2 \rightarrow h_1 h_1$, $h_2 \rightarrow ZZ$ and $h_2 \rightarrow W^+W^-$. We combine these with extrapolations of results from 13 TeV where appropriate. For further details, see Appendix~\ref{app:highlumi}.

Electron-positron colliders may also provide relevant constraints on the model. In particular, electro-weak precision observable measurements (EWPO) coming from LEP and future lepton colliders, such as the International Linear Collider (ILC) can also probe the effects of new scalar particles. To consider these, we follow the treatment of refs.~\cite{Profumo:2014opa, Kotwal:2016tex, Huang:2017jws}. We found the resulting constraints to be generally weaker than either the direct heavy Higgs boson searches or the Higgs signal strength measurements and we defer the detailed description and results to Appendix~\ref{app:ee}. There, we also discuss phenomenological results obtained for the Compact Linear Collider (CLIC)~\cite{No:2018fev}, which we briefly contrast to the hadron collider results. 

To derive the prospects of future hadron colliders to either discover or rule out the real singlet extension of the SM in regards to SFO-EWPT, we have constructed detailed phenomenological analyses at the Monte Carlo level, considering the processes $h_2 \rightarrow h_1 h_1$, $h_2 \rightarrow ZZ$ and $h_2 \rightarrow W^+W^-$ at a 100~TeV proton collider in the mass range $m_2 \in [200, 1000]$~GeV. These were inspired by current LHC analyses, with appropriate modifications at 100~TeV. To obtain the HE-LHC constraints we have performed an extrapolation of the 100~TeV results down to 27~TeV. These phenomenological analyses and associated results are discussed in detail in Appendix~\ref{app:futurepp}.

To estimate the future colliders constraints on the mixing angle through signal strength measurements, we note the several signal strength and coupling measurement projections for various channels that are found in ref.~\cite{Abada:2019lih}. These point towards a $\mathcal{O}(1\%)$ uncertainty or better at electron-positron colliders and at a 100~TeV proton collider. In the absence of a full signal strength combination at these colliders, to remain conservative, we will assume the 95\% C.L. constraint $\sin^2 \theta < 0.01$. We  apply this constraint in addition to those coming from direct searches for heavy Higgs bosons at future colliders. 

Throughout this article we will only show parameter-space points that pass the current constraints described in detail in Appendix~\ref{app:current} (denoted by a $\CIRCLE$), or the HL-LHC constraints described in detail in Appendix~\ref{app:highlumi} plus future signal strength ($| \sin \theta | < 0.1$) constraints (denoted by a $\blacktriangle$). For completeness, we also show those that are currently excluded (faint $\color{yellow}\pmb{\times} \color{black}$) without indicating the category they fall into.

\section{Phenomenological results}\label{sec:results}

\subsection{Theoretical uncertainties}

\begin{figure}[htp]
  \centering
  \includegraphics[width=0.8\columnwidth]{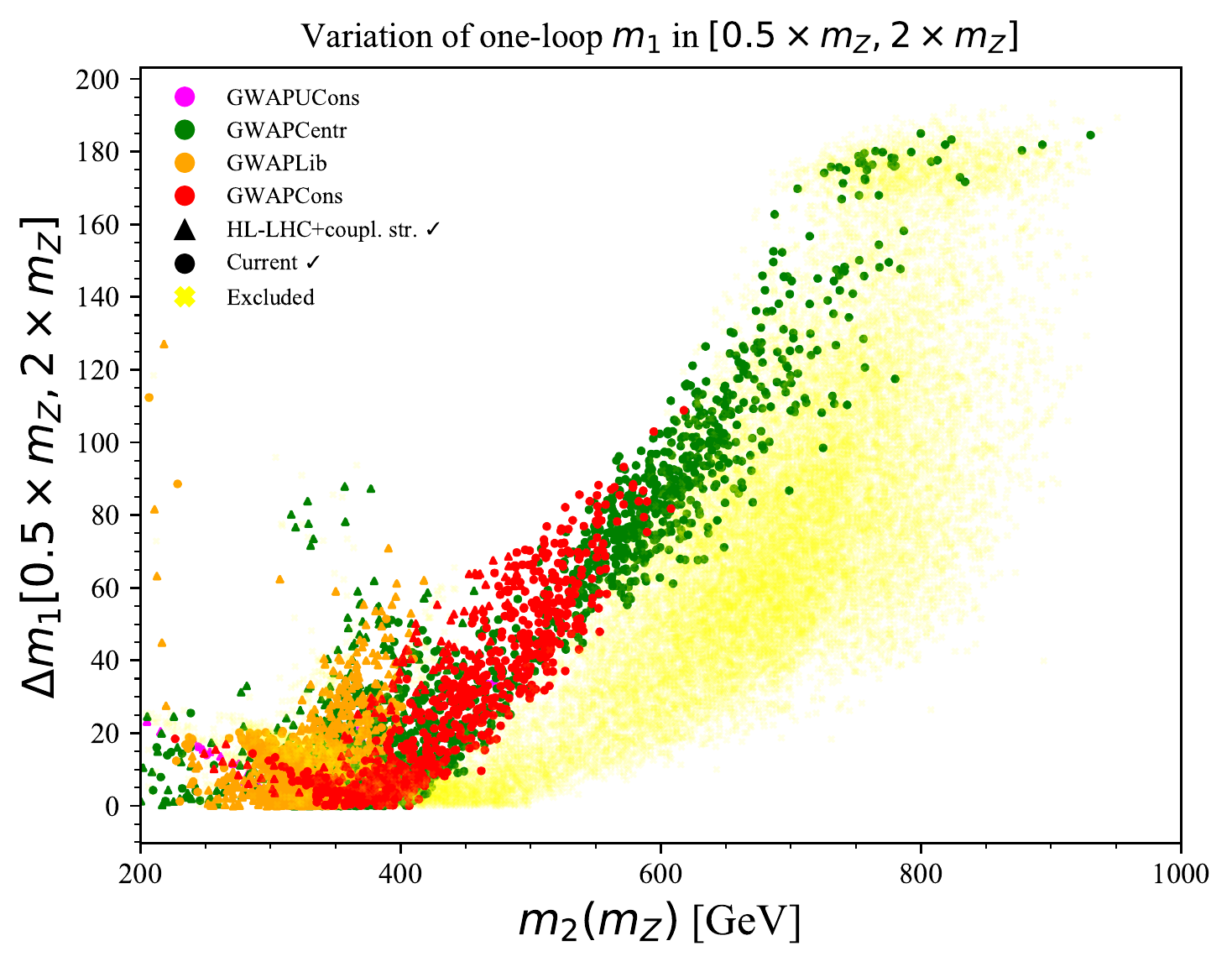}
\caption{The change of the one-loop SM-like Higgs boson ($h_1$) mass, $m_1$, with scale taken in $[0.5 m_Z, 2 m_Z]$. We show points that pass the current constraints ($\CIRCLE$) or the HL-LHC constraints plus future signal strength ($| \sin \theta | < 0.1$) constraints ($\blacktriangle$), as well as those that are currently excluded (faint $\color{yellow}\pmb{\times} \color{black}$).}
\label{fig:deltaM1}
\end{figure}

To assess the severity of the theoretical uncertainties at the (zero-temperature) phenomenological level, we have calculated the variation of the one-loop mass of the SM-like Higgs boson with scale, $\Delta m_1$, taken in $[0.5 m_Z, 2 m_Z]$.\footnote{Note that in doing so, we are calculating the mass using the one-loop effective potential rather than calculating the pole mass from the full self energy. We do not expect a qualitative change in the results from doing this.} The resulting values are shown in fig.~\ref{fig:deltaM1}. One can observe that the ``Conservative'' points generally possess a lower variation, $\Delta m_1 \lesssim 80$~GeV, whereas the ``Centrist'' plus the ``Liberal'' points possess variations as large as 100~GeV at $\sim 600$~GeV, which then reach 150--200~GeV around $m_2 \sim 900$~GeV. It is also evident that the current constraints ``push'' the parameter-space points towards higher theoretical uncertainties, removing a large proportion of the higher-$m_2$ points with a relatively lower value of the variation. 

\subsection{Future collider prospects}\label{sec:futureresults}

\subsubsection{Proton colliders at 100 TeV}\label{sec:100tevsignificance}

\begin{figure}[htp]
  \centering
  \includegraphics[width=0.48\columnwidth]{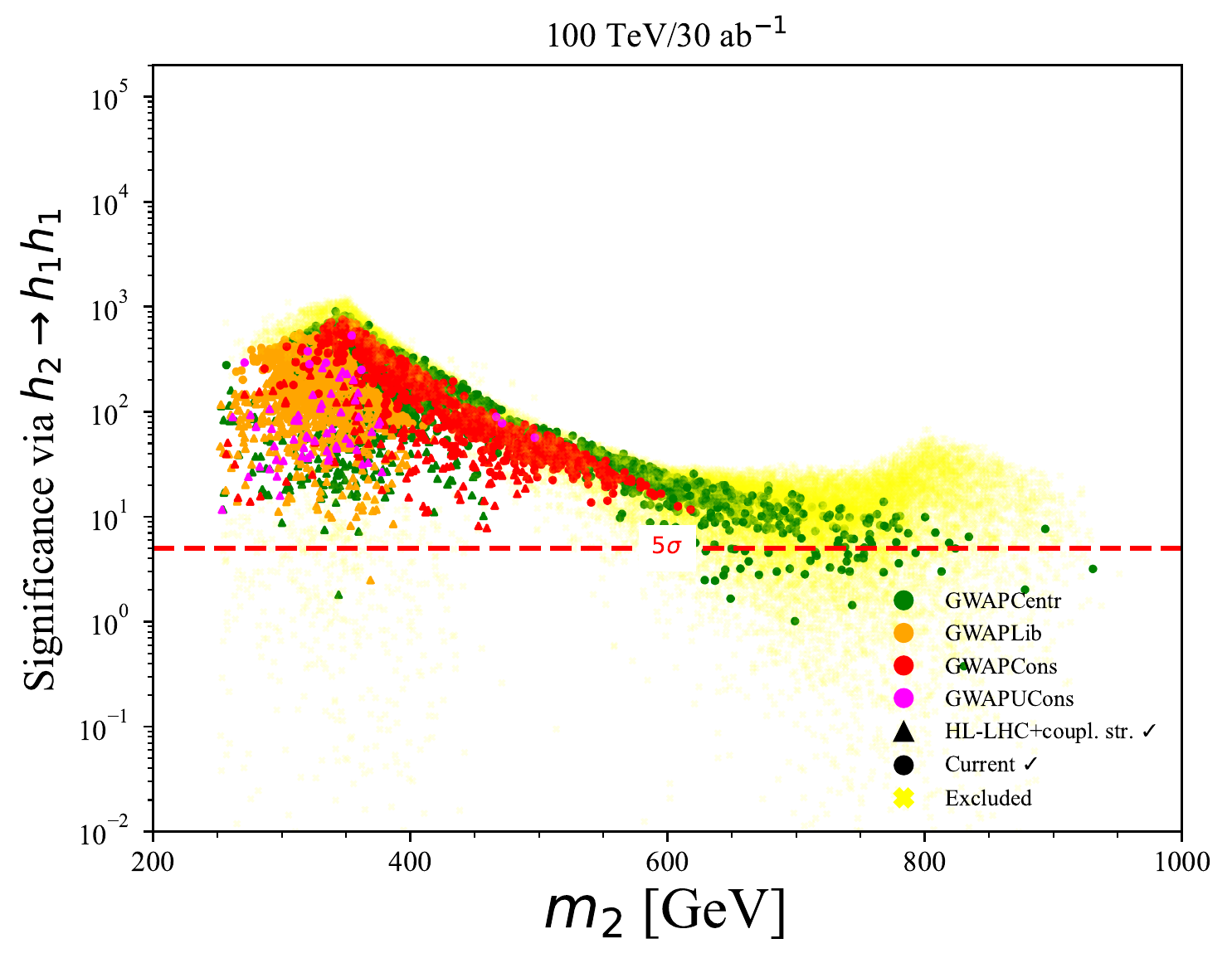}
  \includegraphics[width=0.48\columnwidth]{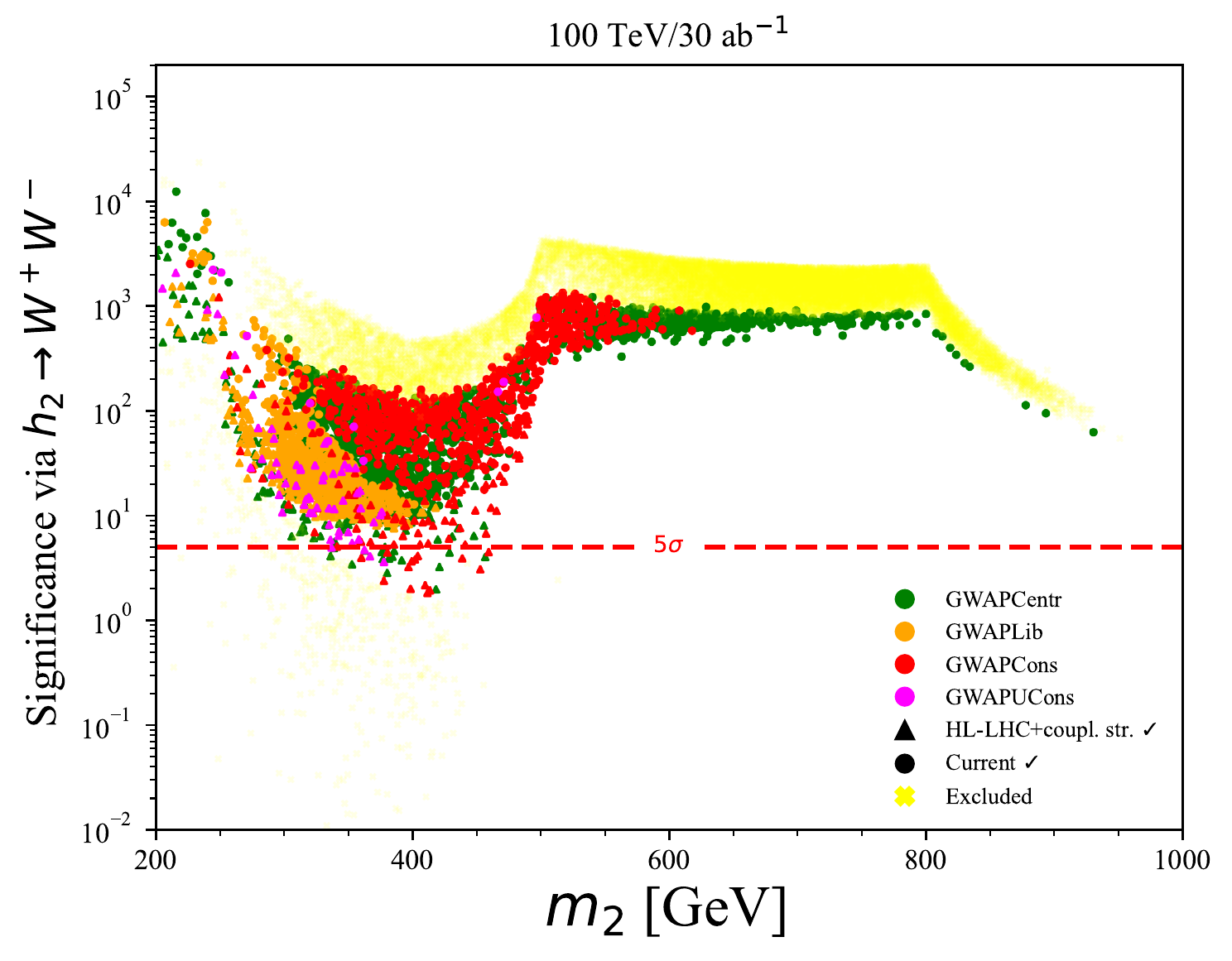}
  \includegraphics[width=0.48\columnwidth]{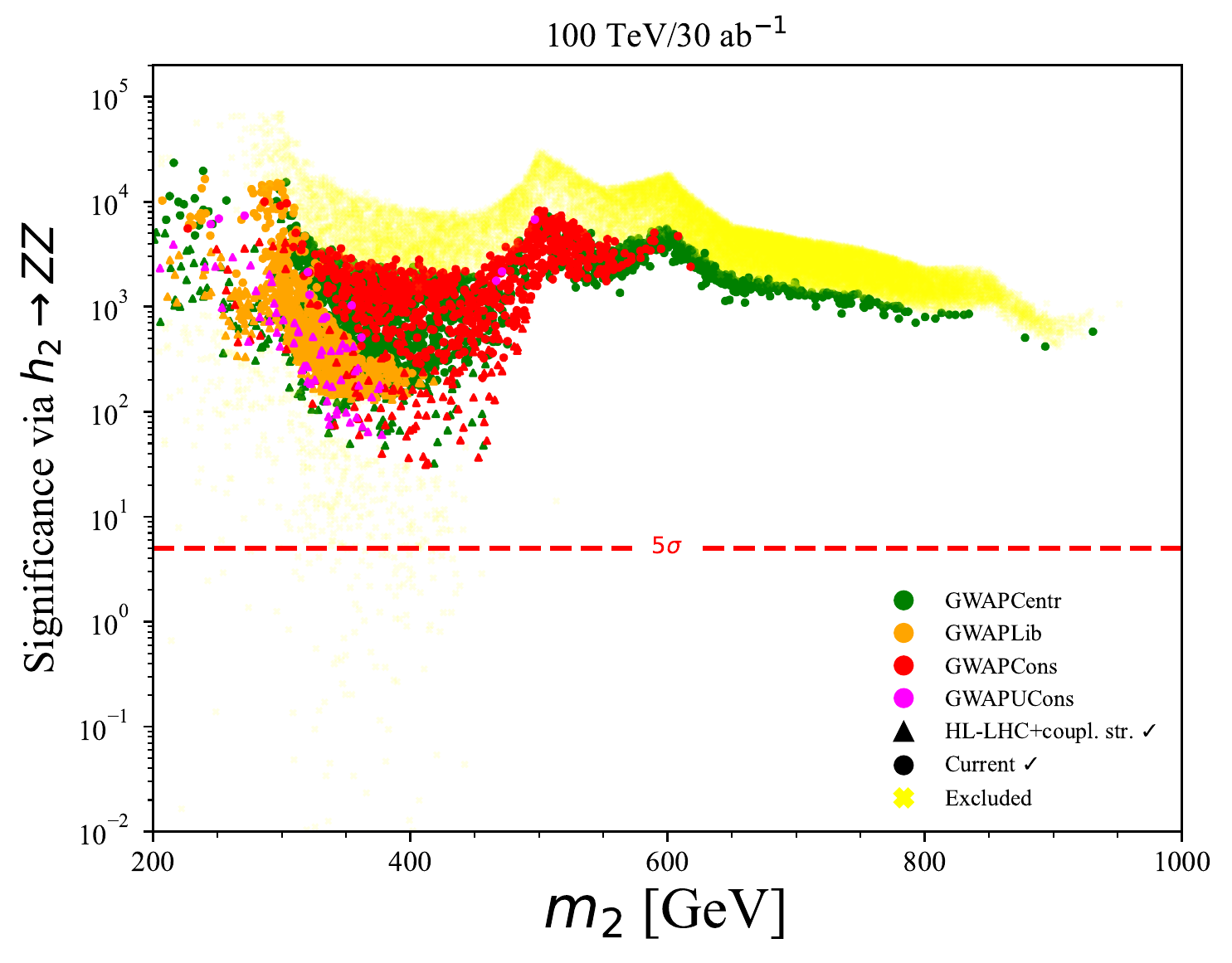}
\caption{The statistical significances for a heavy Higgs boson ($h_2$) signal at a 100 TeV collider with an integrated luminosity corresponding to $30$~ab$^{-1}$, for each of individual analyses: $pp \rightarrow h_2 \rightarrow h_1 h_1$, $pp \rightarrow h_2 \rightarrow W^+W^-$ and $pp \rightarrow h_2 \rightarrow ZZ$. We indicate the $5\sigma$ (discovery) boundaries by the red dashed lines.  We show points that pass the current constraints ($\CIRCLE$) or the HL-LHC constraints plus future signal strength ($| \sin \theta | < 0.1$) constraints ($\blacktriangle$), as well as those that are currently excluded (faint $\color{yellow}\pmb{\times} \color{black}$).}
\label{fig:signif100channels}
\end{figure}

Following the Monte Carlo-level phenomenological analyses for $h_2$ resonant searches we have developed, presented in Appendix~\ref{app:futurepp} in detail, we have derived the expected statistical significance at a 100 TeV proton-proton collider, for each of the parameter-space points for a lifetime integrated luminosity of 30~ab$^{-1}$. The significances for each of the individual analyses: $pp \rightarrow h_2 \rightarrow h_1 h_1$, $pp \rightarrow h_2 \rightarrow W^+W^-$ and $pp \rightarrow h_2 \rightarrow ZZ$ are shown in fig.~\ref{fig:signif100channels}. It is evident that  the $pp \rightarrow h_2 \rightarrow h_1 h_1$ process can probe a large part of the parameter space, at an integrated luminosity of 30~ab$^{-1}$ on its own, apart from a limited number of points, mostly at higher values of $m_2$. The $pp \rightarrow h_2 \rightarrow ZZ$ and $pp \rightarrow h_2 \rightarrow W^+ W^-$ channels yield high significances even for points with a large branching ratio $h_2 \rightarrow h_1 h_1$. This is due to the fact that the parameter-space branching ratios for $h_2 \rightarrow ZZ$ and $h_2 \rightarrow W^+ W^-$ remain relatively significant, at least $\mathcal{O}(\mathrm{few}~\%)$, even for large BR$(h_2 \rightarrow h_1 h_1)$, and due to the power of the gauge boson analyses themselves. In turn, the $pp \rightarrow h_2 \rightarrow ZZ$ analysis performs better than the $pp \rightarrow h_2 \rightarrow W^+ W^-$ analysis that we have constructed, which fails to yield high enough significances for discovery in the intermediate mass range. The sensitivity of the $ZZ$ analyses is driven by the $ZZ \rightarrow (2\ell)(2\nu)$ channel, which yields the most stringent constraints on the $pp \rightarrow h_2 \rightarrow Z Z$ cross section. The $ZZ$ analysis itself represents by far the highest significance for any parameter-space point.

\subsubsection{Proton colliders at 27 TeV}

\begin{figure}[htp]
  \centering
  \includegraphics[width=0.48\columnwidth]{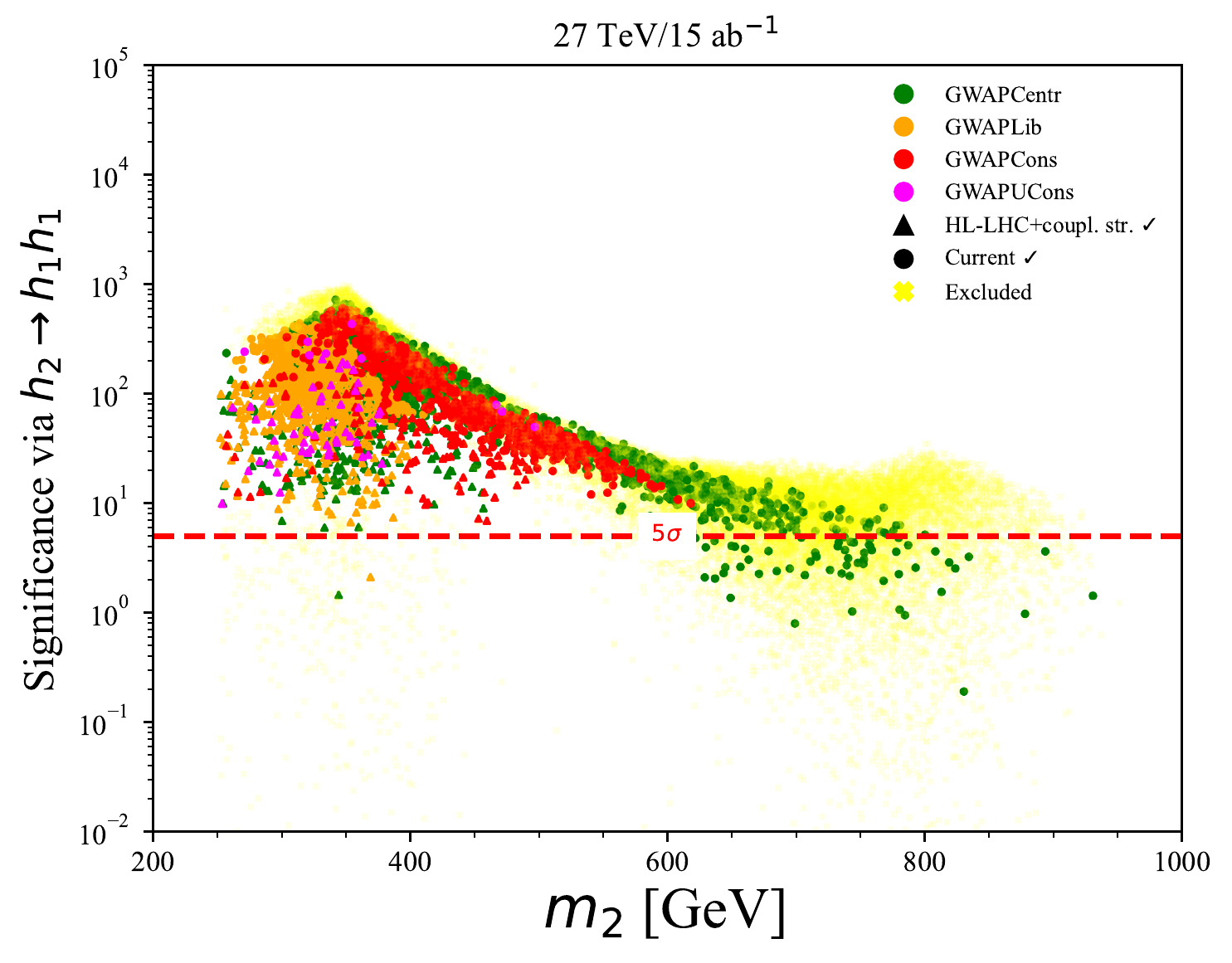}
  \includegraphics[width=0.48\columnwidth]{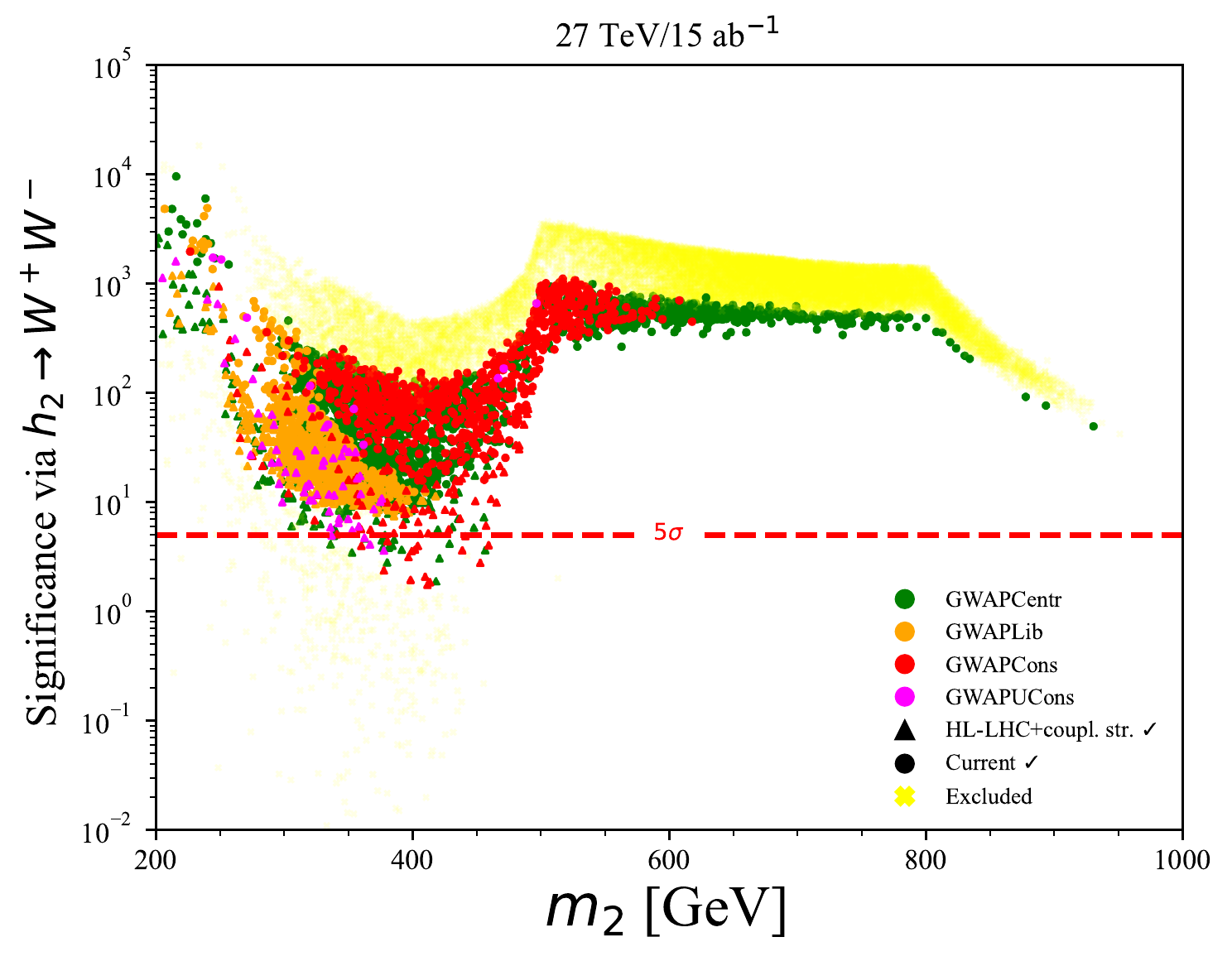}
  \includegraphics[width=0.48\columnwidth]{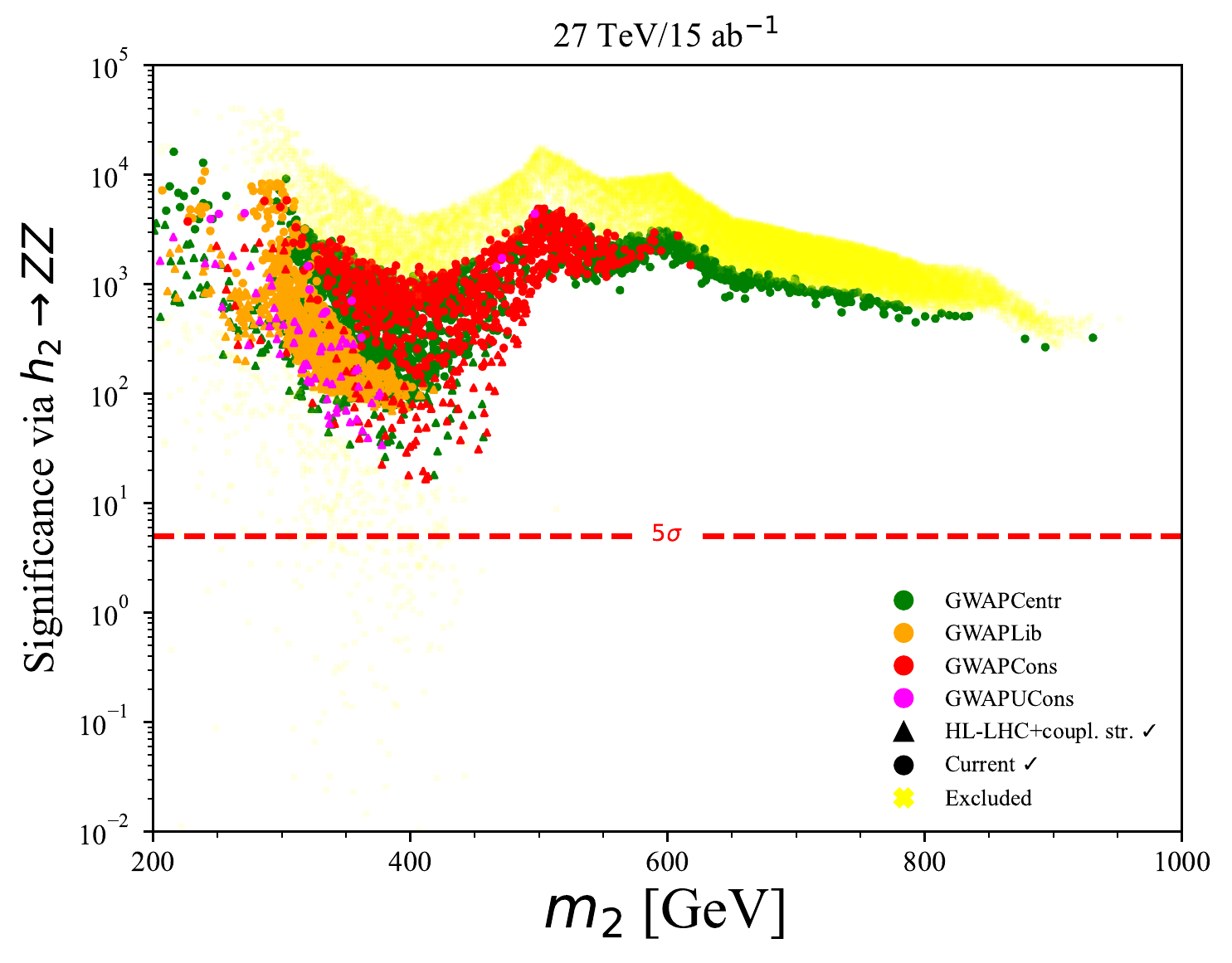}
\caption{The statistical significances for a heavy Higgs boson ($h_2$) signal at 27 TeV collider with an integrated luminosity corresponding to $15$~ab$^{-1}$, for each of individual analyses: $pp \rightarrow h_2 \rightarrow h_1 h_1$, $pp \rightarrow h_2 \rightarrow W^+W^-$ and $pp \rightarrow h_2 \rightarrow ZZ$. We indicate the $5\sigma$ (discovery) boundaries by the red dashed lines.  We show points that pass the current constraints ($\CIRCLE$) or the HL-LHC constraints plus future signal strength ($| \sin \theta | < 0.1$) constraints ($\blacktriangle$), as well as those that are currently excluded (faint $\color{yellow}\pmb{\times} \color{black}$).}
\label{fig:signif27channels}
\end{figure}

Following the extrapolation described in appendix~\ref{app:27tev}, we show equivalent plots for the significances obtained in the $h_1 h_1$ and gauge boson channels in fig.~\ref{fig:signif27channels}, for the case of the HE-LHC 27 TeV proton-proton collider with an integrated luminosity of 15~ab$^{-1}$. It is evident that a 27~TeV collider could be powerful enough to exclude or discover a large fraction of the SFO-EWPT parameter space, with reductions in significance of $\mathcal{O}(\mathrm{a~few})$. We emphasise, however, that this result should be validated by a full phenomenological analysis, taking into account the signal and background kinematics at 27 TeV. 

\subsubsection{Summary}

It is evident that both a 27~TeV collider with an integrated luminosity of 15~ab$^{-1}$ and a 100~TeV collider with 30~ab$^{-1}$ may be able to exclude or discover the whole of the parameter-space points. We note that since the 27~TeV result originates in an extrapolation down from 100~TeV/30~ab$^{-1}$, it is essential that future studies verify this directly through detailed phenomenological analyses that take into account the changes in kinematics at 27~TeV directly. Nevertheless, it is clear that a 100~TeV machine possesses the power to potentially quite efficiently discover the real singlet scalar extension of the SM. The very high statistical significances found all over the parameter space, particularly in the $ZZ$ final state, point to an early discovery. This would allow for further precise measurements of the properties of the heavy scalar to be obtained during the lifetime of a 100~TeV machine, so as to definitively verify whether indeed this is the physical remnant of a real singlet scalar field that catalyses SFO-EWPT.  

\subsection{Selected benchmark points}

\begin{table}[h]
    \centering
    \begin{tabular}{|l|c|c|c|c|c|c|}
    \hline
    \multicolumn{7}{|c|}{(Ultra-)Conservative benchmarks} \\ \hline
    \bf Name & UCons1 & UCons2 & Cons1 & Cons2 & Cons3 & Cons4 \\ \hline 
    $\mathbold{x_0}$ \textbf{[GeV]}& 49.2 & 48.8 & 36.3 & 46.2 & 40.1 & 35.0 \\\hline 
    $\mathbold{\mu^2}$ \textbf{[GeV}$\mathbold{^{2}}$\textbf{]}& -2204.7 & -1466.7 & -3328.2 & -2146.1 & -3407.8 & -7955.2 \\\hline 
    $\mathbold{M_S^2}$ \textbf{[GeV}$\mathbold{^{2}}$\textbf{]}& -8129.0 & 77814.0 & 53473.6 & 98974.7 &  100232.2 & 160581.1  \\\hline 
    $\mathbold{K_1}$ \textbf{[GeV]}& -204.3 & -299.5 & -228.1 & -357.1 & -336.7 & -358.9 \\\hline 
    $\mathbold{K_2}$ & 4.33 & 4.53 & 5.35 & 4.61 & 5.58  & 5.75 \\\hline 
    $\mathbold{\kappa}$ \textbf{[GeV]}& -61.0 & -219.0 & -229.1 & -23.47 & -17.6 & -37.1 \\\hline 
    $\mathbold{\lambda_S}$ & 0.52 & -0.0002 & 0.25 & 0.45 & 0.08 & 0.26\\\hline 
    $\mathbold{\lambda}$ & 0.15 & 0.16 & 0.03 & 0.17 & 0.02 & 0.01\\\hline \hline 
    \textbf{max}$\mathbold{(\rho_{\mathrm{max}})}$ & 1.58 & 2.05 & 1.49 & 2.24 & 1.88 & 3.43 \\\hline 
    \textbf{sin}$\mathbold{\theta}$ & -0.034 & 0.138 & 0.068 & 0.152 & 0.118 & 0.129\\\hline 
    \bf $\mathbold{m_2}$ [GeV] & \textbf{377.5} & \textbf{466.29} & \textbf{480.3} & \textbf{527.0} & \textbf{550.8} & \textbf{617.5}\\\hline 
    \bf $\mathbold{\Gamma_2}$ [GeV] & 0.098 & 1.686 & 0.604 & 2.32 & 1.89 & 2.69 \\ \hline
    \bf BR($\mathbold{ZZ}$) & 0.074 & 0.147 & 0.109 & 0.197 & 0.167 & 0.207 \\\hline 
    \bf BR($\mathbold{W^+W^-}$) & 0.161 & 0.309 & 0.228 & 0.410 & 0.347 & 0.427 \\\hline 
    \bf BR($\mathbold{h_1 h_1}$) & 0.736 & 0.430 & 0.578 & 0.246 & 0.364 & 0.234 \\\hline 
    \bf BR($\mathbold{t\bar{t}}$) & 0.028 & 0.11 & 0.085 & 0.148 & 0.121 & 0.131 \\\hline 
    \bf $\mathbold{\sigma_{13}}$ [pb] & 0.012 & 0.113 & 0.024 & 0.085 & 0.041 & 0.030  \\\hline 
    \bf $\mathbold{\sigma_{27}}$ [pb] & 0.060 & 0.535 & 0.115 & 0.399 & 0.195 & 0.143 \\\hline 
    \bf $\mathbold{\sigma_{100}}$ [pb] & 0.497 & 4.873 & 1.062 & 3.880 & 1.948 & 1.532 \\ \hline 
    \end{tabular}
    \caption{The ``(Ultra-)Conservative'' benchmark points (UCons1--Ucons2) Cons1--Cons4 in increasing mass of the $h_2$ from left to right. The top panel shows the real singlet-extended SM potential parameters and the bottom panel shows some useful derived quantities, including the significant branching ratios and the gluon-fusion cross sections at 13, 27 and 100~TeV. Note that conservative points have tend to have reasonably large $K_2$ and no large hierarchies of scale either in $M_S/|\mu |$ or $K_1/|\mu |$.}
    \label{tab:consbenchmarks}
\end{table}

\begin{table}[h]
    \centering
    \begin{tabular}{|l|c|c|c|c|c|c|}
    \hline
    \multicolumn{7}{|c|}{Centrist benchmarks} \\ \hline
    \bf Name & Centr1 & Centr2 & Centr3 & Centr4 & Centr5 & Centr6\\ \hline 
    $\mathbold{x_0}$ \textbf{[GeV]}& 3.84 & 26.65 & 17.9 & 43.1 & 38.9 & 39.7 \\\hline 
    $\mathbold{\mu^2}$ \textbf{[GeV}$\mathbold{^{2}}$\textbf{]}& -6665.1 & -4711.8 & -5997.3 & -1724.8 & -3701.0 & -13203.1 \\\hline 
    $\mathbold{M_S^2}$ \textbf{[GeV}$\mathbold{^{2}}$\textbf{]}& -46929.5 & -7926.1 & 9804.3 & 96682.3 & 117494.1 & 291088.2 \\\hline 
    $\mathbold{K_1}$ \textbf{[GeV]}& 0.06 & -76.3 & -89.1 &  -289.4 & -374.1 & -432.9 \\\hline 
    $\mathbold{K_2}$ & 4.06 & 5.22 & 4.85 & 5.60 & 5.55 & 6.07 \\\hline 
    $\mathbold{\kappa}$ \textbf{[GeV]}& -603.2 & -1488.3 & -205.7 & -787.6 & -122.7 & -650.5\\\hline 
    $\mathbold{\lambda_S}$ & 0.004 & -0.0003 & 1.75 & 1.485 & 1.487 & 0.897  \\\hline 
    $\mathbold{\lambda}$ & 0.196 & 0.125 & 0.088 &  0.022 & 0.029 & 0.022 \\\hline \hline 
     \textbf{max}$\mathbold{(\rho_{\mathrm{max}})}$ & 2.73 & 1.52 & 1.33 & 3.03 & 3.40 & 1.73 \\\hline 
    \textbf{sin}$\mathbold{\theta}$ & -0.050 & -0.055 & 0.006 & 0.098 & 0.115 & 0.146 \\\hline 
    \bf $\mathbold{m_2}$ [GeV] & \textbf{299.6} & \textbf{364.7} & \textbf{434.8} & \textbf{528.1} & \textbf{605.8} & \textbf{732.9} \\\hline 
    \bf $\mathbold{\Gamma_2}$ [GeV] & 0.133 & 0.185 & 0.004 & 1.479 & 1.925 & 5.002 \\ \hline
    \bf BR($\mathbold{ZZ}$) & 0.046 & 0.087 & 0.202 & 0.130 & 0.215 & 0.247 \\\hline 
    \bf BR($\mathbold{WW}$) & 0.104 & 0.188 & 0.404 & 0.271 & 0.444 & 0.506 \\\hline 
    \bf BR($\mathbold{h_1 h_1}$) & 0.849 & 0.704 & 0.349 & 0.501 & 0.201 & 0.127 \\\hline 
    \bf BR($\mathbold{t\bar{t}}$) & 0.0 & 0.020 & 0.043 & 0.098 & 0.140 & 0.120 \\\hline 
    \bf $\mathbold{\sigma_{13}}$ [pb]& 0.024 & 0.032 & 0.0003 & 0.0355 & 0.0255 & 0.1572 \\\hline 
    \bf $\mathbold{\sigma_{27}}$ [pb] & 0.114 & 0.159 & 0.001 & 0.166 & 0.123 & 0.080 \\\hline 
    \bf $\mathbold{\sigma_{100}}$ [pb] & 0.893 & 1.291 & 0.012 & 1.616 & 1.299 & 0.956 \\ \hline 
    \end{tabular}
    \caption{The ``Centrist'' benchmark points Centr1--Centr6 in increasing mass of the $h_2$ from left to right. The top panel shows the real singlet-extended SM potential parameters and the bottom panel shows some useful derived quantities, including the significant branching ratios and the gluon-fusion cross sections at 13, 27 and 100~TeV. The extra theoretical uncertainty is mostly driven by the large hierarchies in scale - either a large $|M_S|/|\mu|$ or a large $K_1/|\mu |$.}
    \label{tab:centrbenchmarks}
\end{table}

We present a selection of benchmark points within the ``(Ultra-)Conservative'' and ``Centrist'' categories that we have defined in section~\ref{sec:paramspace}.\footnote{We note the study of~\cite{Lewis:2017dme} considered a series of benchmarks focusing on double Higgs boson production within the real singlet scalar extension of the SM.} These represent the (very) ``low'' and ``medium'' theoretical uncertainty categories of points. Tables~\ref{tab:consbenchmarks} and~\ref{tab:centrbenchmarks} show these points in increasing mass of the $h_2$ from left to right. The top panel shows the real singlet-extended potential parameters and the bottom panel shows some useful derived quantities: the maximum value of $\rho_\mathrm{max}$ over the variations of scale and gauge parameters, $\sin \theta$, the width of $h_2$, the significant branching ratios of $h_2$ and the gluon-fusion production cross sections of $h_2$ at 13, 27 and 100~TeV. The cross sections are given at NNLO+NNLL according to the CERN ``Yellow Report 4''~\cite{deFlorian:2016spz} at 13~TeV and calculated at N$^3$LO+NNLL as described in Appendix~\ref{sec:100tev} via the \texttt{ihixs} program (v2.0)~\cite{Dulat:2018rbf} at 27~TeV and 100~TeV. We also show the branching ratio to top quarks, which was not part of our analysis. It is nevertheless of interest, as it could provide the only \textit{direct} way to measure the coupling of the $h_2$ to fermions. 

\section{Conclusions}\label{sec:conclusions}

We have performed a multi-channel analysis of the electro-weak phase transition at future colliders utilising the real scalar extension of the SM as the test model. Considering decays into vector bosons dramatically improves the reach of colliders to the point that a substantially ``weaker'' collider, particularly in terms of the integrated luminosity, may be needed to probe the nature of the electro-weak transition than previously thought. This is true even when theoretical uncertainties are taken into account, dramatically expanding the potentially relevant parameter space. 

However, it is not clear that the entire relevant parameter space can be probed, as theoretical uncertainties become unmanageable for large scale hierarchies and/or large portal couplings. Improved techniques, such as a higher-loop expansion, an improved resummation or dimensional reduction, will be needed to verify if the real singlet extension of the SM can indeed catalyse the eletro-weak transition within its viable parameter space. Furthermore, an improved analysis could probe larger portal couplings than we were able to consider here. Nevertheless, we emphasise that the present treatment of theoretical uncertainties is generous, and we expect our conclusions to be robust.

Finally, if a new heavy scalar particle is discovered early on at any future collider experiment, as our studies indicate, the challenge would then be to comprehend whether it is indeed the remnant of a new scalar field related to the electro-weak phase transition. This endeavour, part of the so-called ``inverse'' problem, should be pursued in future investigations. 

\section*{Acknowledgements}

We would like to thank Michael Ramsey-Musolf, Djuna Croon, Oli Gould, Phillip Schicho, Tuomas Tenkanen, Tania Robens and Jos\'e Zurita for helpful discussions. We would also like to thank GridPP's Scotgrid for the continuous use of the computing resources during the execution of the project. AP is supported by the UK's Royal Society. This work was supported by World Premier International Research Center Initiative (WPI), MEXT, Japan.

\appendix

\section{Heavy Higgs boson decay modes}\label{app:decaymodes}
Since we concentrate on the scenario $m_2 > m_1$, no new decay modes appear for the $h_1$. For $m_2 \geq 2 m_1$, the $h_2 \rightarrow h_1 h_1$ decay mode opens up, with width:
\begin{equation}
\Gamma_{h_2 \rightarrow h_1 h_1} = \frac{ \lambda^2_{112} \sqrt { 1 - 4 m_1^2/ m_2^2 } } { 8 \pi m_2 } \;,
\end{equation}
where $\lambda_{112}$ is the $h_1- h_1 -h_2$ coupling contained in the scalar potential after electroweak symmetry breaking, $V(h_1, h_2) \supset \lambda_{112} h_1 h_1 h_2$, and $m_1$, $m_2$ are the masses of the $h_1$ and $h_2$ particles, respectively.  For completeness, we give the full list of the tree-level triple couplings between the scalars $h_1$ and $h_2$, representing terms of the form $V(h_1, h_2) \supset \lambda_{ijk} h_i h_j h_k$, $i, j, k = \{1, 2\}$ (see, e.g.~\cite{Papaefstathiou:2019ofh}):
\begin{eqnarray}\label{eq:xsmtriple}
    \lambda_{111} &=& \lambda v_0 c_\theta^3 + \frac{1}{2} (K_1 + K_2 x_0)  c_\theta^2 s_\theta\;,\\ \nonumber
                  &+& \frac{1}{2} K_2 v_0 s_\theta^2 c_\theta + \left(\frac{\kappa}{3} + 2 \lambda_S  x_0\right) s_\theta^3 \;, \\ \nonumber
  \lambda_{112} &=& v_0 ( K_2 - 6 \lambda )  c_\theta^2 s_\theta - \frac{1}{2} K_2 v_0 s_\theta^3 \\ \nonumber
 &+& ( -K_1 - K_2 x_0 + \kappa + 6 \lambda_S x_0 ) c_\theta s_\theta^2 + \frac{1}{2}  (K_1 + K_2 x_0 ) c_\theta^3 \;, \\ \nonumber
    \lambda_{122} &=&  v_0  ( 6 \lambda - K_2 )  s_\theta^2  c_\theta + \frac{1}{2}  K_2 v_0 c_\theta^3 \\ \nonumber
&+& (\kappa + 6 \lambda_S x_0 - K_1 - K_2 x_0 ) s_\theta c_\theta^2 + \frac{1}{2} ( K_1 + K_2 x_0 )  s_\theta^3 \;, \\ \nonumber
    \lambda_{222} &=& \frac{1}{12}\left[4 (\kappa + 6 \lambda_S x_0)  c_\theta^3 - 6  K_2 v_0 c_\theta^2 s_\theta \right.\\ \nonumber
                  &+& \left.6 (K_1 + K_2 x_0) c_\theta s_\theta^2 - 24 \lambda v_0 s_\theta^3 \right] \;,
\end{eqnarray}
where we have defined $c_\theta \equiv \cos\theta$ and $s_\theta \equiv \sin \theta$.

The total width of the $h_2$ scalar is given by:
\begin{equation}
\Gamma_{h_2} = \sin^2 \theta ~ \Gamma^{\mathrm{SM}} (m_2) + \Gamma_{h_2 \rightarrow h_1 h_1} \;,
\end{equation}
where $\Gamma^{\mathrm{SM}} (m_2)$ corresponds to the width of a scalar boson of mass $m_2$ possessing the same decay modes as the SM Higgs. In what follows, we have used the $\Gamma^{\mathrm{SM}} (m_2)$ corresponding to the CERN ``Yellow Report 4'' (YR4)~\cite{deFlorian:2016spz}, interpolated for intermediate values up to $m_2 = 1000$~GeV. The branching ratios corresponding to $h_2 \rightarrow xx$, for $x \neq h_1$, are then given by:
\begin{equation}
\mathrm{BR}(h_2 \rightarrow x x) = \sin^2 \theta \frac{\Gamma_{xx}^{\mathrm{SM}} (m_2)}{\Gamma_{h_2}} \;,
\end{equation}
where $\Gamma_{xx}^{\mathrm{SM}} (m_2)$ corresponds to the SM-like width of a scalar boson of mass $m_2$ into the final state $xx$, i.e.\ in the limit $\lambda_{112} \rightarrow 0$. The values $\Gamma_{xx}^{\mathrm{SM}} (m_2)$ were likewise obtained from the CERN YR4 and interpolated. We note that owing to the current constraints on $\sin \theta$, the width of $h_2$ for viable parameter-space points is small compared to its mass. Therefore, throughout this paper, we have assumed that $\Gamma_{h_2} \ll m_2$. 

\begin{figure}
  \centering
  \includegraphics[width=0.8\columnwidth]{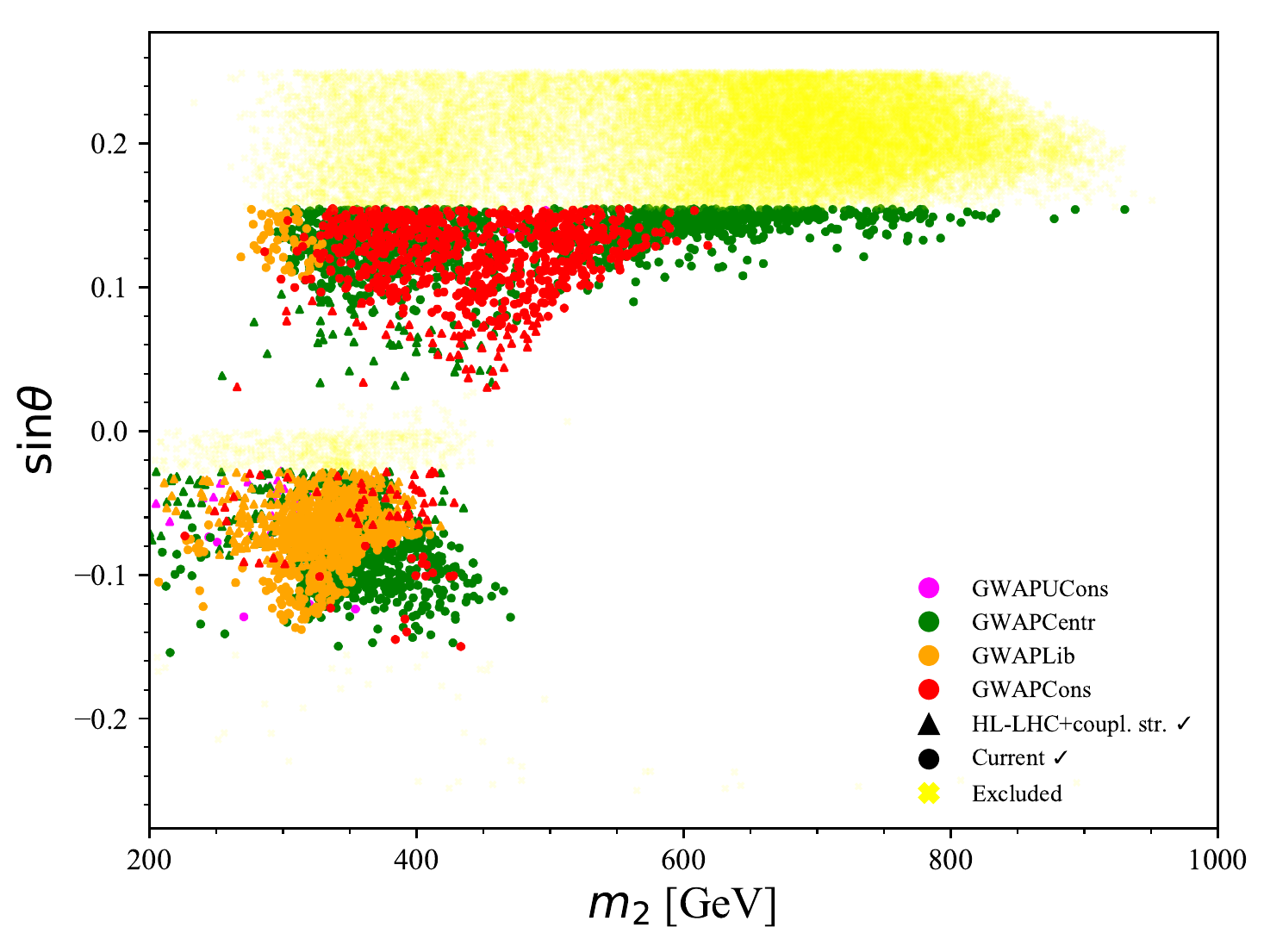}
\caption{The value of $\sin \theta$ for each of the parameter-space points against the mass of $h_2$, $m_2$. We show points that pass the current constraints ($\CIRCLE$) or the HL-LHC constraints plus future signal strength ($| \sin \theta | < 0.1$) constraints ($\blacktriangle$), as well as those that are currently excluded (faint $\color{yellow}\pmb{\times} \color{black}$).}
\label{fig:sintheta}
\end{figure}

In fig.~\ref{fig:sintheta} we show the one-loop value of $\sin \theta$, the most crucial ingredient when calculating the collider production cross sections of $h_2$. We note that this can also attain negative values, which dominate the lower-$m_2$ region of parameter space. Although the sign of $\sin \theta$ does not affect single scalar production, it has an impact on multi-scalar production processes. The portion of parameter space above $|\sin \theta| \sim 0.16$ is excluded by current experimental results. 

\begin{figure}[htp]
  \centering
  \includegraphics[width=0.48\columnwidth]{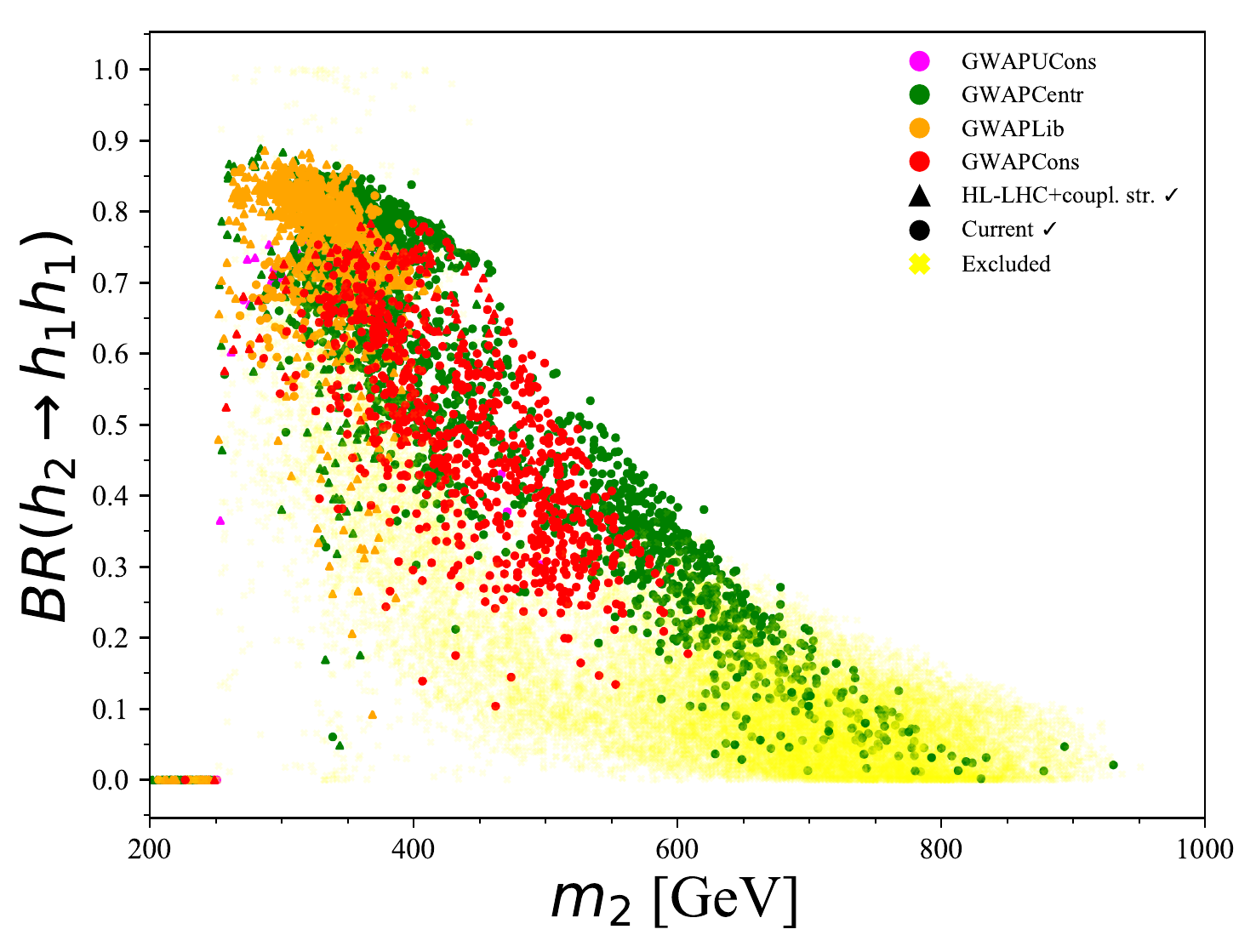}
  \includegraphics[width=0.48\columnwidth]{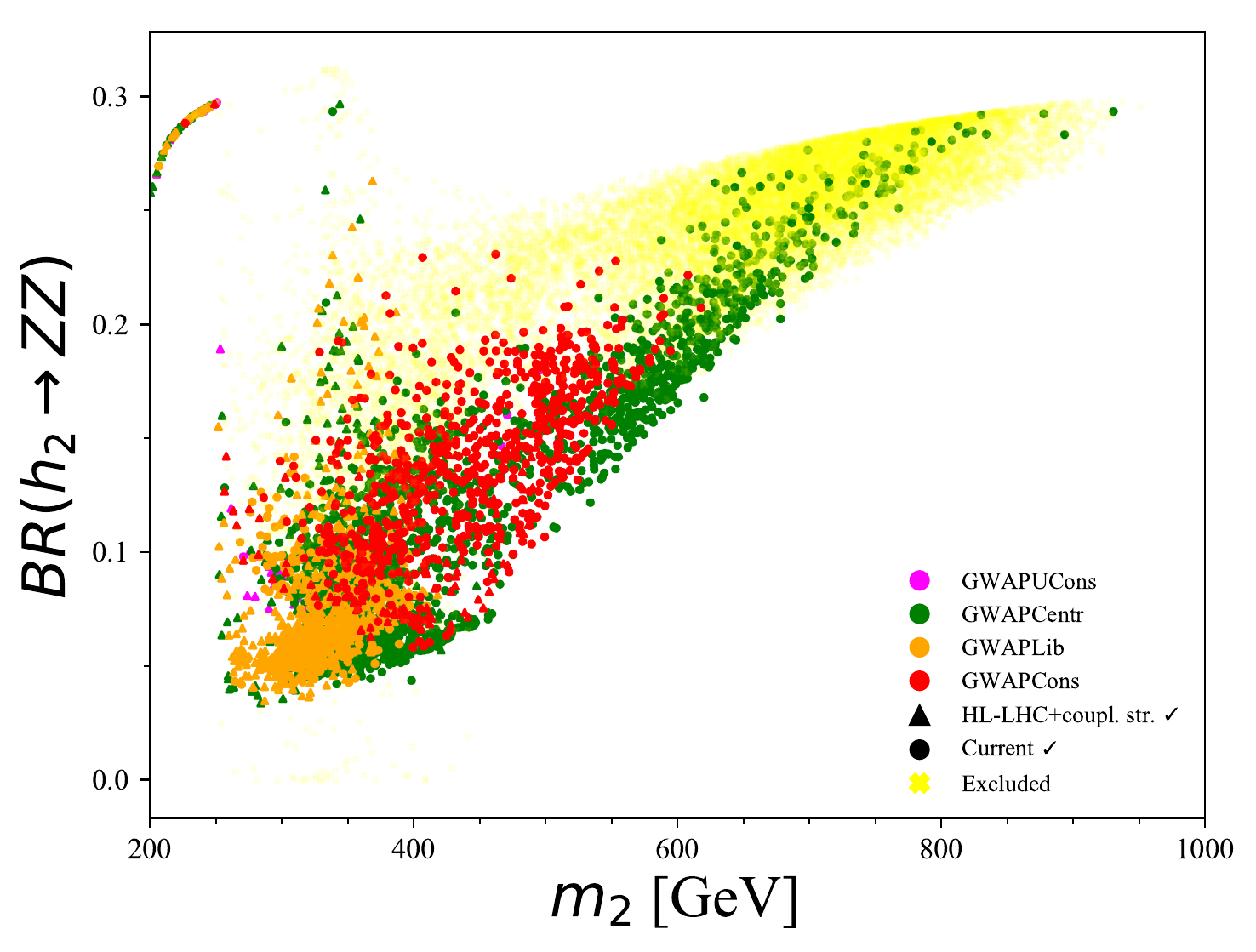}
  \includegraphics[width=0.48\columnwidth]{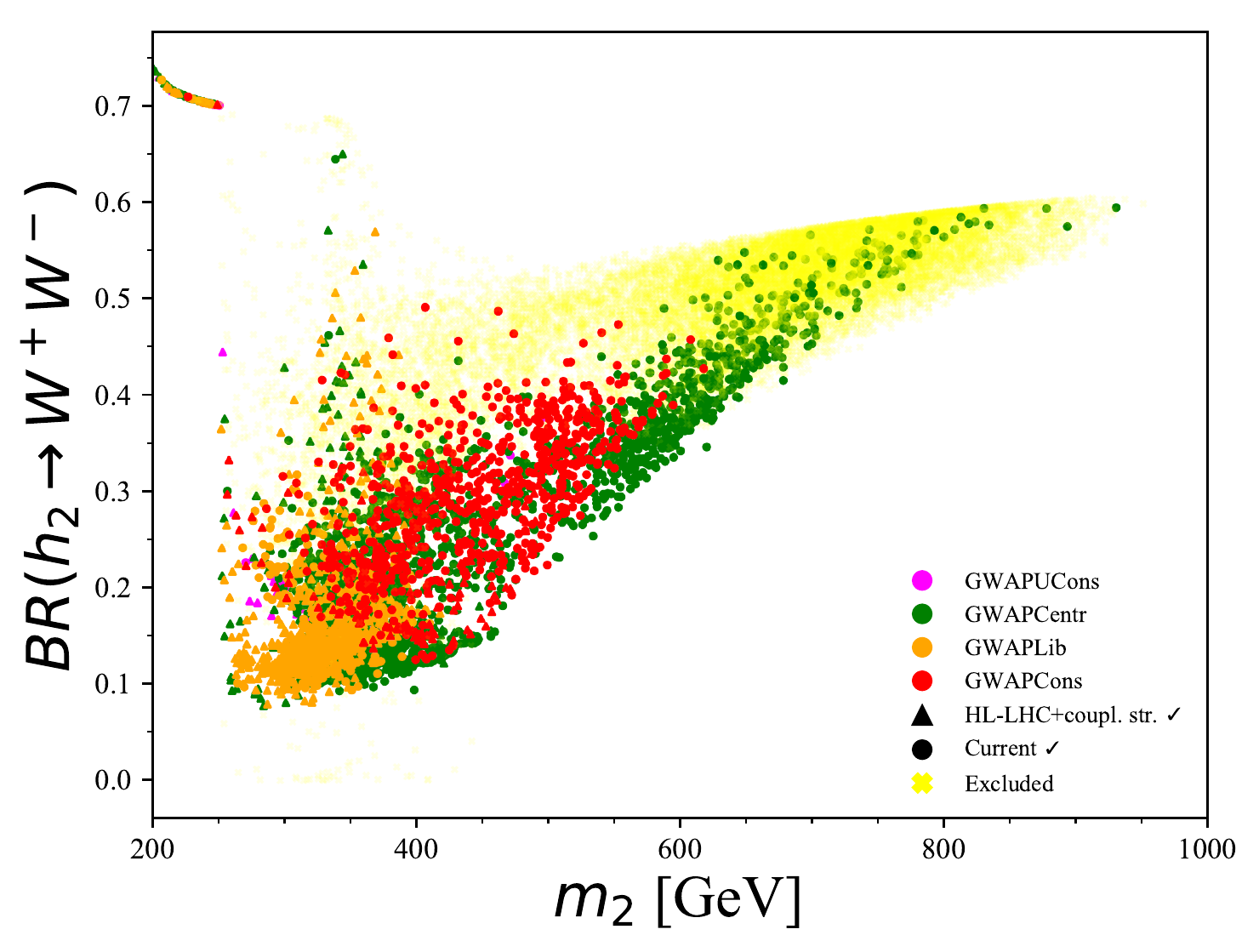}
  \includegraphics[width=0.48\columnwidth]{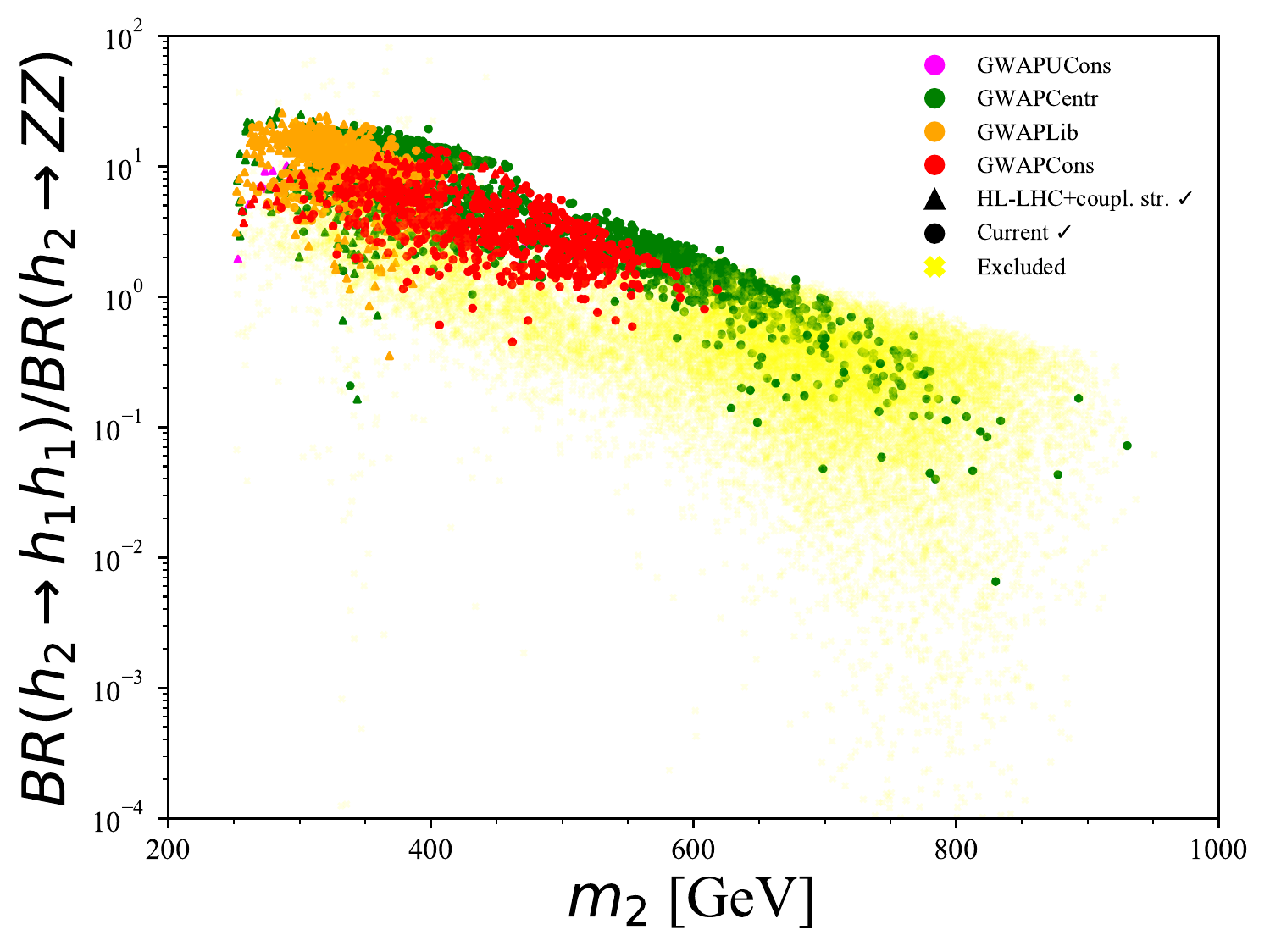}
\caption{The branching ratios for the heavy Higgs boson, $h_2 \rightarrow h_1 h_1$ (top right), $h_2 \rightarrow ZZ$ (top left) and $h_2 \rightarrow W^+ W^-$ (bottom left) for the parameter-space points in the ``Conservative'' (red), ``Liberal'' (orange) and ``Centrist'' (green) categories, plotted against its mass, $m_2$. The bottom right plot shows the ratio of the $h_2 \rightarrow h_1 h_1$ branching ratio to that of $h_2 \rightarrow ZZ$. We show points that pass the current constraints ($\CIRCLE$) or the HL-LHC constraints plus future signal strength ($| \sin \theta | < 0.1$) constraints ($\blacktriangle$), as well as those that are currently excluded (faint $\color{yellow}\pmb{\times} \color{black}$).}
\label{fig:BRs}
\end{figure}

In fig.~\ref{fig:BRs}, we show the branching ratios of the $h_2$ for the parameter-space points that we have generated. The ``(Ultra-)Conservative'' points exhibit a relatively high branching ratio to $h_2 \rightarrow h_1 h_1$, between $\sim 0.2 - 0.75$, whereas the ``Centrist'' plus ``Liberal'' points span values from $\mathcal{O}(10^{-2})$ up to $\sim 0.9$. The lower-right plot of fig.~\ref{fig:BRs} shows the ratio of the rate to $h_2 \rightarrow h_1 h_1$ over that to $h_2 \rightarrow ZZ$.\footnote{Since $h_2 \rightarrow ZZ$ and $h_2 \rightarrow W^+W^-$ are related by custodial symmetry, we show only the former here.} It is clear that the gauge boson final states can play a crucial in discovering or excluding the existence of $h_2$, particularly in the higher-mass regions ($m_2 > 600$~GeV), where the decay to $h_1 h_1$ is sub-dominant.

\section{One-loop corrections to the effective potential in the covariant gauge}\label{app:potential}
In this appendix we give details about how to derive the effective potential at zero and finite temperature in the covariant gauge (also known as Fermi gauges). We follow the prescription given in ref. \cite{Andreassen:2013hpa}. We begin by writing the Higgs doublet components as
\begin{equation}
    H = \left( \begin{array}{c}
         \phi _1 +i \psi _1  \\
         \phi _2 + i \psi _2
    \end{array}
    \right) \ .
\end{equation}
It then becomes useful to write a vector of the five real scalars, including the singlet component which we denote as $s$,
\begin{equation}
    \Psi _a = \left( \begin{array}{c}
         \phi _1  \\
          \phi _2 \\
          \psi _1 \\
          \psi _2 \\
          s
    \end{array} \right) \ .
\end{equation}
Similarly, we write a vector of gauge bosons
\begin{equation}
    V_\mu = \left( \begin{array}{c}
         W^1 _\mu  \\
          W^2 _\mu \\
          W^3 _\mu \\
          B_\mu 
    \end{array} \right) \ .
\end{equation}
The one-loop Lagrangian can be written as:
\begin{equation}
    {\cal L} _{\rm 1-loop} = V(H,s) - \frac{1}{2} \Phi ^\dagger \left[ \begin{array}{cc}
    D^{ab}     &  M^a_\mu  \\
    M_\mu ^a{}^\dagger      & \bar{\Delta } _{\mu \nu}  
    \end{array} \right] \Phi   +{\cal L} _{\rm fermion} \supset \frac{1}{2} \Phi ^\dagger \Sigma \Phi \;,
\end{equation}
where $\Phi = (\Psi _a , V_\mu ) ^{\rm T}$ and
\begin{eqnarray}
D^{ab} &=& \left( \begin{array}{ccccc}
    -p^2 +d_{11} & 0  & 0  & 0  & 0  \\
    0 & -p^2+d_{22} & 0 & 0 & d_{25} \\
    0 & 0 & 0  -p^2 +d_{33} & 0& 0 \\
    0 & 0 & 0  &  -p^2 +d_{44} & 0 \\ 
    0 & d_{52} & 0 & 0 & -p^2 +d_{55} 
\end{array} \right) \;, \\ 
M ^a _\mu  &=& \left( \begin{array}{cccc}
    0 & \frac{i}{2} g_2 \phi _2 p_\mu & 0 & 0   \\
    0 & 0 & 0 & 0 \\
    \frac{i}{2} g_2 \phi _2 p_\mu & 0 & 0 & 0 \\
    0 & 0 & - \frac{i}{2} g_2 \phi _2 p_\mu & \frac{i}{2} g_1 \phi _2 p_\mu \\ 0 & 0 & 0 & 0
\end{array} \right) \;,\\
\end{eqnarray}
\begin{eqnarray}
    \bar{\Delta } _{\mu \nu} &=& \left( \begin{array}{cccc}
    \Delta ^{11}_{\mu \nu}  & 0 & 0 & 0  \\
     0 & \Delta ^{11} _{\mu \nu } & 0 & 0 \\
     0 & 0 & \Delta ^{33} _{\mu \nu } & \Delta ^{34} _{\mu \nu} \\
     0 & 0 &\Delta ^{43} _{\mu \nu } & \Delta ^{44} _{\mu \nu}  
\end{array} \right) \;, \\
d_{11} &=& d_{22} = d_{33}=d_{44}-\mu^2 + \frac{\lambda}{2} \phi _2 ^2 + K_1 s+ \frac{K_2}{2} x^2 \;,\\
d_{25}&=& d_{52} = K_1 \phi _ 2 + \phi _2 s K_2 \;, \\
d_{55} &=& \frac{K_2}{2} \phi _2 ^2 + M_S^2 +2 s \kappa +6 \lambda _s s^2 \;, \\
\Delta ^{11} _{\mu \nu } &=& \Delta ^{22}_{\mu \nu } =\Delta ^{33}_{\mu \nu } g_{\mu \nu } \left( p^2 -\frac{1}{4} g_2^2 \phi _2^2 \right)- \left( 1 - \frac{1}{\xi _W} p_\mu p_\nu \right) \;, \\
\Delta ^{34}_{\mu \nu } &=& \Delta ^{43} _{\mu \nu} = \frac{1}{4} g_1 g_2 \phi _2 ^2 g_{\mu \nu} \;, \\
\Delta ^{44} _{\mu \nu} &=& g_{\mu \nu } \left( p^2 - \frac{1}{4} g_1 ^2 \phi _2 ^2 \right) - \left( 1 - \frac{1}{\xi _B} \right) p_\mu p_\nu \; .
\end{eqnarray}
 Taking the determinant of the matrix $\Sigma$, it is straightforward to derive the mass eigenvalues that enter the one-loop corrections to the tree-level potential:
\begin{eqnarray}
V_{\rm tr} (h,s,Q) &=& \frac{\mu ^2(Q)}{2}  h^2 + \frac{\lambda _h (Q)}{8}h^4 \nonumber \\ && +\frac{K_1(Q)}{2} h^2 s + \frac{K_2 (Q)}{4} h^2 s^2 \nonumber \\ && + \frac{M_S(Q)^2}{2} s^2 + \frac{\kappa (Q)}{3} s^3 + \frac{\lambda _s(Q)}{2} s^4 \;, \\
V_{\rm CW}(h,s,Q,\xi_W,\xi _B) &=& \sum _{i \in \rm scalars} n _i \frac{m_i ^4(h,s,Q,\xi_W,\xi _B)}{64 \pi ^2} \left[ \log \left( \frac{m_i ^2(h,s,Q,\xi_W,\xi _B)}{Q^2} \right)  - \frac{3}{2} \right] \nonumber \\ && -  \frac{3 m_t^4(h,Q)}{16 \pi ^2} \left(\frac{m_t^2(h,Q)}{Q ^2}  \right) \nonumber \\ && + \sum _{i \in \rm gauge} n _i \frac{m_i ^4(h,s,Q)}{64 \pi ^2} \left[ \log \left( \frac{m_i ^2(h,s,Q)}{Q^2} \right)  - \frac{5}{6} \right]\;, \\
V_T(h,s,Q,\xi_W,\xi _B) &=& \frac{T^4}{2\pi ^2} \sum _i n_i J_B \left(\frac{m_i(h,s,Q,\xi_W,\xi _B)}{T^2} \right) \nonumber\\ && 
+ 12 \frac{T^4}{2 \pi ^2} J_F \left( \frac{m_t(Q,h)}{T^2} \right) \;.
\end{eqnarray}

In the above, the field-dependent top and gauge boson masses take their usual form. The four Goldstone-like scalar masses\footnote{They reduce to the Goldstone modes that one derives in the Landau gauge when working in the $R_\xi$ gauges.} and two physical modes are, respectively,
\begin{eqnarray}
   m_{1,\pm } ^2 &=& \frac{1}{2}\left( \chi \pm  \sqrt{\chi ^2 - \Upsilon _W } \right) \;, \\
     m_{2,\pm } ^2 &=& \frac{1}{2}\left( \chi \pm  \sqrt{\chi ^2 - \Upsilon _Z } \right) \;, \\
     m_{h,\pm } &=& \frac{1}{2} \left(-b \pm \sqrt{b^2 - 4 c} \right) \;,
\end{eqnarray}
 where 
 \begin{eqnarray}
 \chi &=& \mu ^2(Q) + \frac{\lambda (Q)}{2} h^2 + \frac{K_2(Q)}{2}s^2 +K_1(Q) s \;, \\
 \Upsilon _W &=& \frac{1}{2} \left( 2 \mu ^2 (Q) + s (2 K_1(Q)+s K_2(Q)) + \lambda (Q) h^2 \right)g_2 ^2(Q) h^2 \xi _W \;, \\
  \Upsilon _Z &=& \frac{1}{2} \left( 2 \mu ^2 (Q) + s (2 K_1(Q)+s K_2(Q)) + \lambda (Q) h^2 \right)(g_2^2 (Q) \xi _W +g_1^2(Q) \xi _B)h^2 \;, \\
  b&=& -M_S^2(Q) - \frac{s}{2} \left( 2 K_1(Q) + s K_2 (Q) + 12 \lambda _s(Q) s + 4 \kappa (Q)  \right) - \mu ^2 (Q)\nonumber \\ && -\frac{h^2}{2}\left( K_2(Q)+ 3 \lambda (Q) \right)\;, \\
  c &=& \frac{1}{4} ( 2 (M_S^2(Q) + 2 s (3 \lambda _s (Q) s +\kappa (Q) ) ) ( 2 K_1(Q)s K_2(Q) s^2 2 \mu ^2(Q)) \nonumber \\ && + (-4 K_1^2(Q)-6K_1(Q)K_2(Q)s - 3 K_2^2(Q)s^2 \nonumber \\ && + 6( M_S^2(Q) + 2 s (3 \lambda _s(Q)s + \kappa (Q) )) \lambda (Q) + 2 K_2(Q) \mu ^2 (Q) ) h^2 + 3 K_2(Q) \lambda (Q) h^4) \;. \nonumber \\ 
 \end{eqnarray}
 Finally, the multiplicities for the scalar masses are $n_{1,+}=n_{1,-}=2$ and $n_{2,\pm} = n_{h,\pm} = 1$.
 The ``Arnold-Espinosa'' method involves resumming the bosonic masses which results in the addition of Daisy terms to the effective potential
\begin{equation}
    V_D = -\frac{T}{12 \pi } \sum _i n_i \left( \tilde{m_i} ^{3} - m_i^3  \right) \;,
\end{equation}
where we have suppressed gauge, field and scale dependence in the arguments. The scalar Debye masses are given by
\begin{eqnarray}
\Pi _h &=& \left( \frac{\lambda}{4} + \frac{g_1^2 +3g_2^2}{16}+\frac{y_t^2}{4}  \right) T^2 \;, \nonumber \\ \Pi _s &=&  \frac{\lambda _s }{2} T^2 \;.
\end{eqnarray}
The scalar masses that go into the Daisy resummation term of the potential are simply $\tilde{m}^2 = m_{i,\pm}^2+ \Pi _h $ for the four Goldstone-like mass terms. For the physical states, the tilded masses are the eigenvalues of the matrix
\begin{equation}
    \left( \begin{array}{cc}
        \frac{\partial ^2V_{\rm tr} }{dh^2} + \Pi _h &  \frac{\partial ^2V_{\rm tr} }{dh ds} \\
        \frac{\partial ^2V_{\rm tr} }{dh ds}  &  \frac{\partial ^2V_{\rm tr} }{ds^2} +\Pi _s 
    \end{array} \right) \ .
\end{equation}
To remarkable accuracy, one can write the two gauge boson Debye masses as 
\begin{eqnarray}
\Pi _W &=& \frac{11}{6} g_2^2 T^2 \;, \\
\Pi _Z &=& \frac{11}{6} \frac{g_1^4 + g_2^4}{g_1^2 + g_2^2} \;.
\end{eqnarray}
\section{Current constraints}\label{app:current}

\subsection{Current constraints through \texttt{HiggsBounds} and \texttt{HiggsSignals}}

To incorporate an array of current collider constraints into our analysis, we employ the \texttt{HiggsBounds} (v5.9.0)~\cite{Bechtle:2008jh,Bechtle:2011sb,Bechtle:2013gu,Bechtle:2015pma} and \texttt{HiggsSignals} (v2.5.1)~\cite{Bechtle:2013xfa, Stal:2013hwa, Bechtle:2014ewa} packages. \texttt{HiggsBounds} takes a selection of Higgs sector predictions for any model as input and then uses the experimental topological cross section limits from Higgs boson searches at LEP, the Tevatron and the LHC to determine if this parameter point has been excluded at 95\% C.L.. \texttt{HiggsSignals} performs a statistical test of the Higgs sector predictions of arbitrary models with the measurements of Higgs boson signal rates and masses from the Tevatron and the LHC. \texttt{HiggsBounds} returns a boolean corresponding to whether the Higgs sector passes the constraints at 95\% C.L. (true) or not (false). \texttt{HiggsSignals} returns a probability value ($p$-value) corresponding to the goodness-of-fit of the Higgs sector over several SM-like ``peak'' observables. At present these involve the LHC 13 TeV results of refs.~\cite{Aaboud:2018gay, Aaboud:2017rss, ATLAS:2019nvo, CMS:2018rbc, Sirunyan:2018hbu, Sirunyan:2017elk, Sirunyan:2017dgc, CMS:2019lcn, Sirunyan:2018shy, CMS:2018dmv}.

Effectively, \texttt{HiggsBounds} and \texttt{HiggsSignals} impose constraints through direct SM-like Higgs boson signals ($h_1$) and direct searches for heavy scalars ($h_2$). The SM-like Higgs boson signals impose constraints through the reduction of the rate via $\cos \theta$ and the $h_2$ searches impose constraints on $\sin \theta$ and the branching ratio to a particular final state, according to eq.~\ref{eq:rescaledcouplings}. In \texttt{HiggsBounds}, the most constraining analyses have been found to be the CMS combination of SM-like Higgs boson searches and measurements of its properties using 7 and 8 TeV proton-proton centre-of-mass data corresponding to an integrated luminosity of 5.1~fb$^{-1}$ and 12.2~fb$^{-1}$ respectively~\cite{CMS-PAS-HIG-12-045} ($pp\rightarrow h_1$), the ATLAS combination of searches for heavy resonances decaying into bosonic and leptonic final states using 36~fb$^{-1}$ of data at 13 TeV~\cite{Aaboud:2018bun} ($pp\rightarrow h_2 \rightarrow VV$), and CMS searches for new scalar resonances decaying to $Z$ boson pairs at 13 TeV with 35~fb$^{-1}$ of data~\cite{CMS-PAS-HIG-17-012} ($pp\rightarrow h_2 \rightarrow ZZ$).

\subsection{Current constraints through resonant $h_2 \rightarrow h_1 h_1$}\label{sec:hh}

To supplement the \texttt{HiggsBounds} and \texttt{HiggsSignal} constraints we consider in addition the searches of resonant SM-like Higgs boson pair production conducted by ATLAS~\cite{Aad:2019uzh} and CMS~\cite{Sirunyan:2018ayu}. These searches consider combinations of searches for resonant Higgs boson pair production at 13 TeV centre-of-mass energy, using 36.1~fb$^{-1}$ and 35.9~fb$^{-1}$ of data, respectively. Both analyses include the final states $h_1 h_1 \rightarrow (b\bar{b})(\gamma \gamma)$, $(b\bar{b})(\tau^+\tau^-)$, $(b\bar{b})(b\bar{b})$ and $(b\bar{b})(W^+W^-)$ and consider resonances with $m_2 > 250$~GeV. The CMS analysis contains in addition the $h_1 h_1 \rightarrow (b\bar{b})(ZZ)$ final state and the ATLAS analysis contains the $h_1 h_1\rightarrow (W^+W^-)(W^+W^-)$ and $(W^+W^-) (\gamma \gamma)$ final states. We consider the results of fig.~5(a) of ref.~\cite{Aad:2019uzh} and of fig.~3 of ref.~\cite{Sirunyan:2018ayu}, both constructed in the cases of narrow resonances.  To construct a constraint for a given parameter-space point of the real-singlet extended SM, we compare to the rescaled 13~TeV cross section calculated at next-to-next-to-to-leading order in QCD with next-to-next-to-leading logarithmic resummation  (NNLO+NNLL), which appear in the CERN ``Yellow Report 4''~\cite{deFlorian:2016spz}.

\subsection{Current Higgs boson signal strength constraints}

The on-peak SM-like Higgs boson measurements directly constrain the mixing angle. For the current measurements, these constraints are implemented in our analysis through the \texttt{HiggsSignals} package. The latest 13 TeV ATLAS and CMS SM-like Higgs boson global signal strengths, $\mu = \sigma^{\mathrm{measured}}/\sigma^{\mathrm{SM}}$ have been found to be $\mu_\mathrm{ATLAS} = 1.11 ^{+0.09}_{-0.08}$~\cite{Aad:2019mbh} and $\mu_{\mathrm{CMS}} = 1.02 ^{+0.07}_{-0.06}$~\cite{CMS:2020gsy}, with 80~fb$^{-1}$ and 137~fb$^{-1}$ of data, respectively. These measurements are currently not included in the \texttt{HiggsSignals} constraints. Since $\mu = \mu_{SM} \cos^2 \theta$ in the SM extended by a real scalar, the model cannot accommodate measurements of $\mu > 1$. Therefore, the fact that current experimental central values both lie in $\mu >1$ results in more stringent constraints than expected. In particular, at 95\% C.L. (2$\sigma$), the ATLAS measurement would imply that $\sin^2 \theta = -0.11^{+0.18}_{-0.16}$  and the CMS measurement, $\sin ^2 \theta = -0.02 ^{+0.14} _{-0.12}$. Since $\sin^2 \theta > 0$, taking only the positive values results in $\sin^2 \theta < 0.07$ and $\sin^2 \theta < 0.12$ for the ATLAS and CMS measurements respectively, at 95\% C.L.. Combining the two measurements by taking their mean and their errors in quadrature results in $\sin^2 \theta < 0.05$ for the combined current ATLAS and CMS constraint at 95\% C.L.. We impose this constraint in addition to the \texttt{HiggsSignals} constraints.

\section{High-Luminosity LHC}\label{app:highlumi}

\subsection{HL-LHC Higgs boson signal strength}
Both ATLAS and CMS have presented projections for the high-luminosity phase of the LHC, foreseen to collect 3000~fb$^{-1}$ of data at 14~TeV~\cite{TheATLAScollaboration:2014ewu, CMS:2018qgz}. In ref.~\cite{TheATLAScollaboration:2014ewu}, the ATLAS collaboration provides a projection on a global coupling modifier, which they call $\kappa$, of $\delta \kappa = 2.2\%$ for the ``halved systematics'' scenario (table 3 of ref.~\cite{TheATLAScollaboration:2014ewu}). Since the signal strength $\mu$ is proportional to $\kappa^2$, the expected uncertainty on $\mu$ would be $\delta \mu = 4.4\%$. The CMS analysis of ref.~\cite{CMS:2018qgz} does not provide a global signal strength or coupling modifier estimate, therefore we take their ``S2'' projection for the gluon-fusion Higgs boson production mode, with an expected uncertainty on $\mu$ of $\delta \mu = 3\%$ (fig.~3 of~\cite{CMS:2018qgz}). Combining the two in quadrature would result in a total ultimate uncertainty estimate at the HL-LHC of $\delta \mu \sim 2.7\%$. This would imply a 95\% C.L. constraint on the mixing angle of $\sin ^2 \theta < 0.054$. This is in fact almost identical to the current constraint. This is due to the fact that the current measured central value of the signal strength is $\mu > 1$, providing a more stringent constraint on the mixing angle than expected. Thus, in the absence of new phenomena, as the central value of the signal strength measurement $\mu \rightarrow 1$, it is likely that the current constraint will become weaker at the HL-LHC.

\subsection{HL-LHC Heavy Higgs boson searches}

\subsubsection{HL-LHC $\mathbold h_2 \rightarrow Z Z$}
Ref.~\cite{Cepeda:2019klc} presented an extrapolation of the results of ref.~\cite{Sirunyan:2018qlb} at 13 TeV, of searches of a heavy resonance decaying to a pair of $Z$ bosons. This was done in the narrow-resonance limit, matching the analysis of the present article. The extrapolation focuses on a single final state $ZZ \rightarrow (2\ell)(2q)$ (for $\ell$ representing electrons or muons and $q$ quarks). This final state is more sensitive in the high-mass region at LHC energies and therefore the extrapolation only provides constraints for $m_2 \geq 550$~GeV. If $m_2 < 550$~GeV, we will assume that no information is given by this analysis on $h_2 \rightarrow ZZ$. A parameter $f_\mathrm{VBF}$ is defined as the fraction of the electroweak production cross section with respect to the total cross section. The results are given in two scenarios: $f_\mathrm{VBF}$ floated, and $f_\mathrm{VBF}$ = 1. In the expected result, the two scenarios correspond to the gluon fusion (ggF) and vector-boson fusion production (VBF) modes, respectively. Here we consider the floating VBF scenario, corresponding to the ggF mode. We consider the ``YR18'' systematics scenario presented in ref.~\cite{Cepeda:2019klc}, where the systematic uncertainties are halved with respect to the values given in ref.~\cite{Sirunyan:2018qlb}. This corresponds to the lower-left plot of fig.~173 in ref.~\cite{Cepeda:2019klc}. Here and in the rest of this article, we have ``digitised'' the constraints of the figure using the package \texttt{EasyNData}~\cite{Uwer:2007rs}. As in section~\ref{sec:hh}, we compare to the rescaled 13 TeV cross section at N$^3$LO.

To estimate the expected HL-LHC limit for $m_2 < 550$~GeV through $h_2 \rightarrow ZZ$, we naively extrapolate the results of the 13 TeV analysis of ref.~\cite{CMS:2017vpy} that employed the $ZZ \rightarrow 4\ell$, $\rightarrow (2\ell)(2q)$ and $\rightarrow (2\ell)(2\nu)$ channels for 35.9~fb$^{-1}$ of data. This is done simply by rescaling the expected 95\%.C.L. cross section by the square root of the ratio of luminosities, $\sqrt{\mathcal{L}_\mathrm{curr.}/\mathcal{L}_\mathrm{HL-LHC}}$, where $\mathcal{L}_\mathrm{curr.}$ is the luminosity used to derive the constraints of the ATLAS and CMS analyses and $\mathcal{L}_\mathrm{HL-LHC} = 3000$~fb$^{-1}$. Further details of this analysis are outlined in section~\ref{sec:futzz}, where we re-purpose it for our future proton collider studies.

\subsubsection{HL-LHC $\mathbold h_2 \rightarrow W^+W^-$}
The ATLAS note~\cite{ATLAS:2018ocj} has presented the HL-LHC prospects for narrow-width diboson resonance searches in the $WW\rightarrow (\ell \nu) (qq)$ final state. The analysis considered a 14~TeV centre-of-mass energy and therefore we use the NNLO+NNLL ggF Higgs boson cross sections of ref.~\cite{deFlorian:2016spz} to deduce whether a parameter-space point is excluded or not. For this purpose, we have digitised the lower-left plot of fig.~5 in ref.~\cite{ATLAS:2018ocj}. As before, the $h_2 \rightarrow W^+W^-$ analysis considered relatively high-mass resonances, $m_2 \geq 500$~GeV, therefore we will assume that it provides no information on parameter-space points with $m_2 < 500$~GeV. To accommodate points with $m_2 < 500$, we extrapolate the ATLAS analysis of ref.~\cite{Aaboud:2017gsl}, where the final state $WW\rightarrow (e \nu) (\mu \nu)$ was investigated at 13 TeV with 36.1~fb$^{-1}$ of data. Further details of this analysis are provided in~\ref{sec:futww}.

\subsubsection{HL-LHC resonant Higgs boson pair production, $\mathbold
  h_2 \rightarrow h_1 h_1$}
To derive the approximate expected constraints at the HL-LHC originating from searches of resonant SM-like Higgs boson pair production, we perform a naive extrapolation of the ATLAS~\cite{Aad:2019uzh} and CMS~\cite{Sirunyan:2018ayu} analyses considered in section~\ref{sec:hh}. 

\section{Electron-positron colliders}\label{app:ee}

\subsection{Electroweak precision observables}

To incorporate the constraints of electroweak precision observables (EWPO), we follow the procedure outlined in refs.~\cite{Profumo:2014opa, Kotwal:2016tex, Huang:2017jws}. The existence of a real singlet scalar field will modify the Higgs field's contributions to the diagonal weak gauge boson vacuum polarisation diagrams and will induce additional contributions. The effect can be characterised via the $S$, $T$ and $U$ parameters~\cite{Hagiwara:1994pw} which can be calculated pertubatively in any model from the gauge boson propagator functions. A change in an EWPO $\mathcal{O}$ due to the presence of a real singlet field, $\Delta \mathcal{O}$, with respect to the SM value is then given by: $\Delta \mathcal{O} = \mathcal{O}_\mathrm{BSM} - \mathcal{O}_\mathrm{SM}$, where $\mathcal{O}_\mathrm{SM} = \mathcal{O}(m_1^2)$ is the contribution of the SM Higgs boson. At one loop, the change is then $\Delta \mathcal{O} = \mathcal{O}(m_2^2) \sin^2 \theta + \mathcal{O}(m_1^2) \cos ^2 \theta - \mathcal{O}(m_1)$. This leads to:
\begin{equation}
\Delta \mathcal{O} = ( \mathcal{O}(m_2^2) - \mathcal{O}(m_1^2)) \sin^2 \theta\;.
\end{equation}
The above expression implies weaker constraints for $m_2 \sim m_1$ and small mixing angles. The parameter $U$ has a negligible dependence on the mass of the scalar particle involved, and hence we only consider modifications of $S$ and $T$, i.e.\ $\Delta S$ and $\Delta T$, as in ref.~\cite{Huang:2017jws}. The relevant contributions come from the $S_B$ and $T_B$ functions found in the appendices C.1 and C.2 of ref.~\cite{Hagiwara:1994pw}.

To extract ``current'' constraints, i.e.\ those coming from LEP, we performed a correlated $\chi^2$ fit of the Gfitter results~\cite{Baak:2014ora}:
\begin{equation}
\Delta S  = 0.06\pm 0.09,\; \Delta T = 0.10 \pm 0.07,
\end{equation}
with correlation coefficient $\rho_{ST} = 0.91$. The $\chi^2$ is then given by:
\begin{equation}
\chi^2 (m_2, \sin^2 \theta) = \sum_{ij} \left[\Delta \mathcal{O}_i  (m_2, \sin^2 \theta) - \Delta \mathcal{O}_i^\mathrm{meas.} \right] (\boldsymbol{\sigma^2})^{-1}_{ij}\left[\Delta \mathcal{O}_j  (m_2, \sin^2 \theta) - \Delta \mathcal{O}_j^\mathrm{meas.} \right]\;,
\end{equation}
where $\Delta \mathcal{O}_i = {S,T}$ for $i = {1,2}$ respectively, $\Delta \mathcal{O}_i^\mathrm{meas.}$ is the measured difference of the corresponding observable from the SM expectation and the matrix $(\boldsymbol{\sigma^2})^{-1}$ is the inverse of the covariance matrix:
\begin{eqnarray}
\boldsymbol{\sigma^2} = 
\begin{pmatrix}
  \sigma_S^2 &  \rho_{ST} \sigma_S \sigma_T \\
\rho_{ST} \sigma_S \sigma_T  & \sigma_T^2 
\end{pmatrix}\;,
\end{eqnarray}
where $\sigma_S$ and $\sigma_T$ are the uncertainties on the measured differences coming from the global electroweak fit.

In addition, future lepton collider prospects of EWPO measurements are discussed in ref.~\cite{Baak:2014ora}. Assuming no deviation from the SM expectations, the International Linear Collider (ILC) ``GigaZ'' option at 90 to 200 GeV $e^+e^-$ centre-of-mass energy, is expected to yield:
\begin{equation}
\Delta S  = 0.000\pm 0.017 (\mathrm{exp.}) \pm 0.006 (\mathrm{th.}),\; \Delta T = 0.000 \pm 0.022 (\mathrm{exp.}) \pm 0.005 (\mathrm{th.}),
\end{equation}
where the first uncertainty is experimental and the second theoretical. We combine these in quadrature in what follows.

\begin{figure}[htp]
  \centering
  \includegraphics[width=0.48\columnwidth]{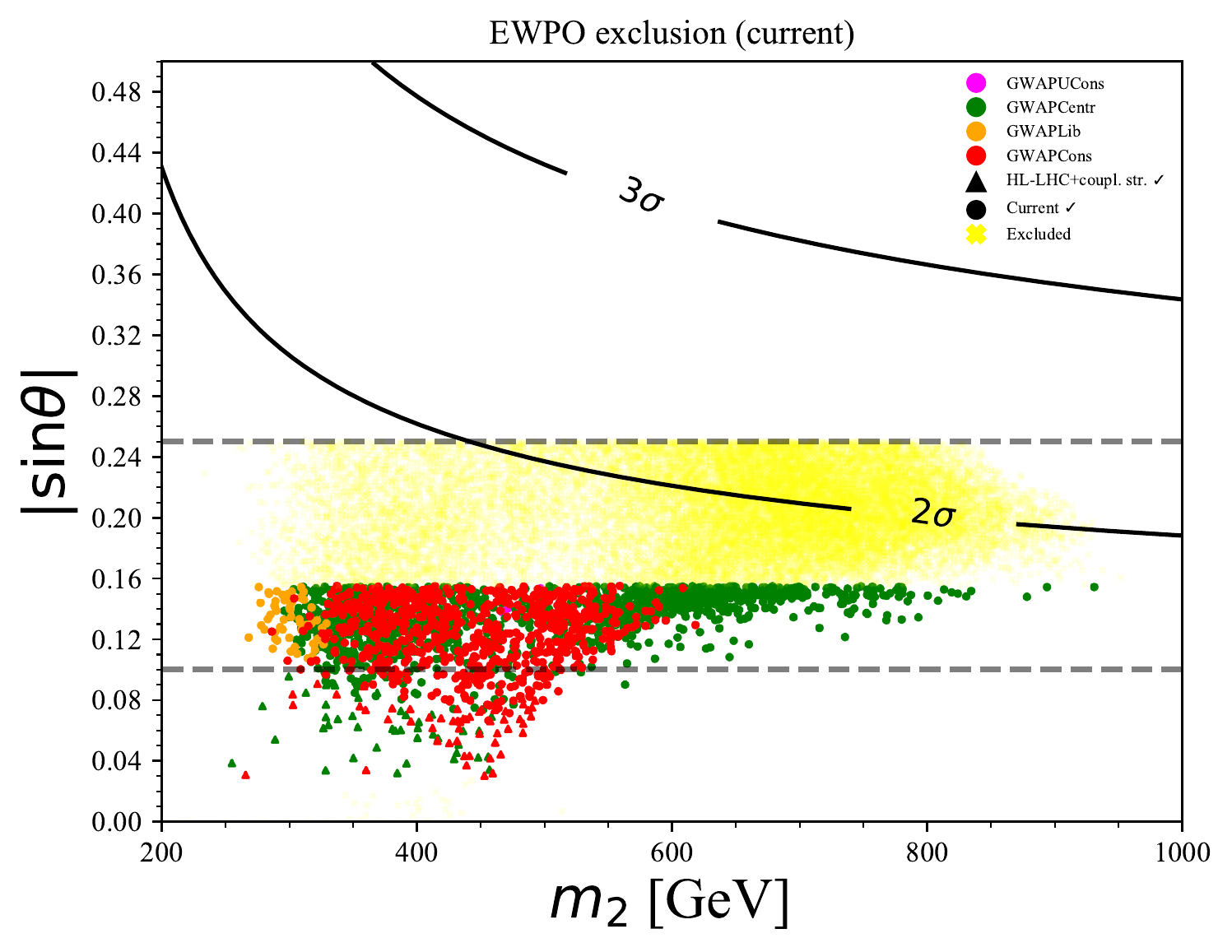}
  \includegraphics[width=0.48\columnwidth]{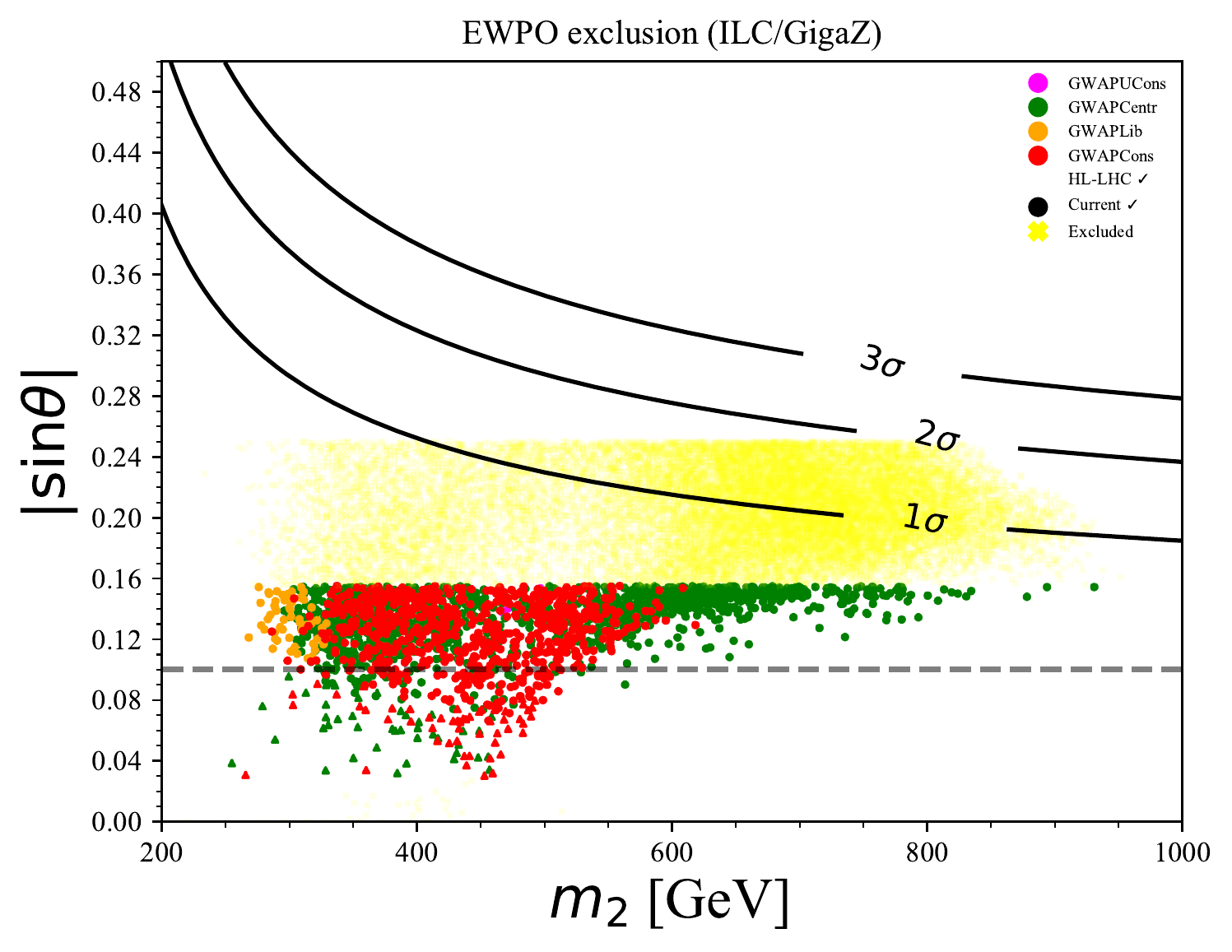}
\caption{The exclusion regions for a real singlet scalar particle with mixing angle $\theta$ and mass $m_2$ through EWPO precision observables obtained through LEP measurements (left) and future ILC/GigaZ projections (right). The curves correspond to constant $\chi^2 = 2.30, 6.18 , 11.83$, giving the $1, 2, 3\sigma$ exclusion regions, respectively. The grey dashed lines show the $|\sin \theta| = 0.25$ and $|\sin \theta| = 0.1$ boundaries, the former corresponding to the upper limit in our scans and the latter to the future collider limit from Higgs boson signal strength measurements. We note that points with $\sin \theta \gtrsim 0.16$ are excluded by current signal strength constraints. We show points that pass the current constraints ($\CIRCLE$) or the HL-LHC constraints plus future signal strength ($| \sin \theta | < 0.1$) constraints ($\blacktriangle$), as well as those that are currently excluded (faint $\color{yellow} \pmb{\pmb{\times}} \color{black}$).}
\label{fig:ewpo}
\end{figure}

Using the above considerations we have calculated the $\chi^2$ on the $(|\sin \theta|, m_2)$-plane, shown in fig.~\ref{fig:ewpo}, both using the current LEP results and the ILC/GigaZ projections. The curves correspond to constant $\chi^2 = 2.30, 6.18 , 11.83$, giving the $1, 2, 3\sigma$ exclusion regions, respectively. It is clear that by comparing the $2\sigma$ region excluded by current EWPO constraints to the $2\sigma$ region coming from future ones, that the current region appears to be more restricted than the future one. In fact, all points on the parameter space are currently incompatible at more than $1\sigma$ with LEP results through EWPO. That is due to the fact that the central values of the $S$ and $T$ parameters extracted LEP measurements are not centred about the SM expectations, thereby increasing the value of $\chi^2$. It is also clear that the constraints coming from EWPO observables will not be as stringent as the future signal strength measurements. Nevertheless, they could provide information on the overall picture, in case a signal is observed.  

\subsection{Direct searches for Heavy Higgs bosons at future lepton colliders}

A comprehensive study of prospects of constraining the real singlet extension of the SM at the Compact Linear Collider (CLIC) was performed in ref.~\cite{No:2018fev}. The article examined the decays $h_2 \rightarrow VV$ for $V=W$ and $V=Z$ gauge bosons, as well as $h_2 \rightarrow h_1 h_1$ at $e^+ e^-$ centre-of-mass energies of 1.4~TeV and 3~TeV and concluded that CLIC would allow an up to two orders of magnitude improvement with respect to the sensitivity achievable by HL-LHC in $m_2 \in [250, 1000]$~GeV. To incorporate constraints in our study, we have extracted the combined $h_2 \rightarrow VV$ constraints of fig.~7 of ~\cite{No:2018fev}, at 3~TeV, $\mathcal{L} = 2000$~fb$^{-1}$, (green curve) and the $h_2 \rightarrow h_1 h_1$ constraints of fig.~9 at 1.4~TeV ($\mathcal{L} = 1500$~fb$^{-1}$) and 3~TeV ($\mathcal{L} = 2000$~fb$^{-1}$), with the high $b$-tagging option (90\%), corresponding to the orange dashed and blue dashed curves respectively. The limits are given in terms of the $\kappa = \sigma(m_2)/\sigma(m_2)^\mathrm{SM}$ for the $h_2\rightarrow VV$ final states and in terms of $\kappa = \sigma(m_2)/\sigma(m_2)^\mathrm{SM} \times BR(h_2 \rightarrow h_1 h_1)$ for the $h_2 \rightarrow h_1 h_1$ final state. 
\begin{table}[h]
    \centering
    \begin{tabular}{|l|c|c|c|c|c|c|}
    \hline
    & \multicolumn{6}{c|}{Fraction excluded} \\ \hline
    Analysis & Ultra-Conservative & Conservative & Central & Liberal & Tight & Loose \\ \hline 
    $h_1 h_1$ (1.4~TeV) & 0.91 &  0.98  & 0.97 & 0.98 & 0.94 & 0.97 \\ \hline 
    $h_1 h_1$ (3~TeV) & $\sim$1.0 & $\sim$1.0 & $\sim$1.0 & $\sim$1.0 & $\sim$1.0 & $\sim$1.0 \\ \hline 
    $VV$ (3~TeV) & $\sim$0.0 & $\sim$0.0 & $\sim$0.0 & $\sim$ 0.0 & $\sim$0.0  & $\sim$0.0\\ \hline 
    \end{tabular}
    \caption{The fraction of currently-viable parameter-space points in $m_2 \in [200, 1000]$~GeV excluded by each of the CLIC analyses performed in ref.~\cite{No:2018fev}, for each of the categories defined in section~\ref{sec:paramspace}.}
    \label{tab:clic}
\end{table}

In table~\ref{tab:clic} we show the fraction of currently-viable points in $m_2 \in [200, 1000]$~GeV that is excluded by each of the CLIC analyses performed in ref.~\cite{No:2018fev}, for each of the categories defined in section~\ref{sec:paramspace}. Evidently, at CLIC, the $h_1 h_1$ analyses are much more powerful in excluding points than the $VV$ analysis. At a 3~TeV CLIC, the $h_1 h_1$ analysis on its own could exclude the whole of the viable parameter space. Therefore, CLIC should be able to provide strong constraints on the parameter space of the real singlet scalar model, but will most likely be unable to exclude the totality of the viable parameter space on its own, except at 3~TeV centre-of-mass energies. 

\section{Future proton colliders}\label{app:futurepp}

\subsection{Heavy Higgs boson searches at a 100 TeV proton collider}\label{sec:100tev}

\subsubsection{Monte Carlo analysis details}\label{sec:analysis}
To the best of our knowledge, the prospects of detecting heavy scalar resonances at a 100 TeV proton collider have not been previously examined in detail. To this end, we have performed detailed phenomenological  analyses at the Monte Carlo level, of the main decay channels of a heavy scalar resonance with SM-like couplings in the mass range $m_{h_2} \in [200, 1000]$~GeV, assuming a narrow width, i.e.\ $\Gamma_{h_2} \ll m_{h_2}$.\footnote{In practice, the width was set to $\Gamma_{h_2} = 1$~GeV in all Monte Carlo simulations.} We have considered both gluon-fusion (GF) production and vector boson-fusion (VBF) production of the $h_2$. 

All the parton-level events have been generated via the Monte Carlo (MC) event generator \texttt{MadGraph5\_aMC@NLO} (v2.7.2)~\cite{Alwall:2014hca, Hirschi:2015iia}, with a custom modification of the \texttt{loop\_sm} model to incorporate an additional scalar particle and its interactions with the SM particles. Both GF and VBF production channels have been simulated at leading order (LO). In the case of GF, we have calculated approximate next-to-next-to-next-to-leading (N$^3$LO) corrections via the \texttt{ihixs} program (v2.0)~\cite{Dulat:2018rbf}, with renormalisation/factorisation scales set to $\mu_R = \mu_F = m_2/2$ and the PDF set \texttt{NNPDF23\_nnlo\_as\_0119}~\cite{Ball:2012cx}. The corrections were calculated in the Higgs Effective Field Theory, including full NLO quark mass dependence and threshold resummation up to next-to-next-to-leading logarithmic accuracy.\footnote{For $m_2 < 340$~GeV we include the NNLO $1/m_\mathrm{top}$ terms.} In the case of VBF we calculated the next-to-leading (NLO) corrections via fixed-order runs in \texttt{MadGraph5\_aMC@NLO}. Next-to-leading order matching and multi-jet merging (where appropriate), QCD parton showering, hadronization and underlying event simulation were all performed within the general-purpose MC event generator \texttt{HERWIG} (v7.2.1)~\cite{Bahr:2008pv, Gieseke:2011na, Arnold:2012fq, Bellm:2013hwb, Bellm:2015jjp, Bellm:2017bvx, Bellm:2019zci}. We consider NLO matching through the \texttt{MC@NLO} method~\cite{Frixione:2002ik} and for a subset of processes we employ FxFx merging~\cite{Frederix:2012ps,Frederix:2015eii}. The Monte Carlo simulations were performed with the PDF \texttt{NNPDF23\_nlo\_as\_0119} for LO and \texttt{NNPDF23\_nlo\_as\_0118\_qed} for NLO. Events were analysed via the \texttt{HwSim} module~\cite{hwsim} for \texttt{HERWIG} which saves events in a \texttt{ROOT} compressed file format~\cite{Brun:1997pa}, with jets clustered using \texttt{FastJet} (v3.3.2)~\cite{Cacciari:2011ma}. The anti-$k_T$ algorithm~\cite{Cacciari:2008gp} with a radius parameter $R=0.4$ was chosen to be the default jet clustering algorithm, unless otherwise stated. 

To capture the detector effects, we restrict the pseudo-rapidity coverage of all objects to $|\eta| < 4$ and only consider particles with transverse momentum with $p_T > 400$~MeV as being detectable. We smear all the
final state reconstructed object momenta according to
$\Delta p_T = 1.0 \times \sqrt{p_T}$ for jets, $\Delta p_T = 0.2 \times \sqrt{p_T} + 0.017 \times p_T$ for photons~\cite{Kotwal:2016tex}, with $p_T$ in GeV. Muons and electron momenta are smeared according to~\cite{TheATLAScollaboration:performance1}. The lepton identification efficiency is assumed to be $\epsilon(p_T) = 0.85 - 0.191 \times \exp(1 - p_T/(20~\mathrm{GeV}))$ for electrons and $\epsilon = 0.54$ for muons in $|\eta| < 0.1$ and $\epsilon=0.97$ otherwise. The efficiency for measuring a jet was taken to be $\epsilon(p_T) = 0.75 + 0.2 \times p_T /(30~\mathrm{GeV})$~\cite{TheATLAScollaboration:performance1, TheATLAScollaboration:performance2}. The tagging of heavy-flavour jets is discussed, where applicable, in the individual analyses outlined below. 

Following the construction of a set of observables for each analysis, we employ \texttt{ROOT}'s ``Toolkit for Multivariate Data Analysis'' (TMVA) to obtain the optimal signal versus background discrimination. We use the ``Boosted Decision Tree'' (BDT) classifier method with AdaBoost.\footnote{See the TMVA manual~\cite{Speckmayer:2010zz} for further details on BDT methods with Gradient Boost.} To further reduce dependence on the size of our Monte Carlo samples, we have run the BDT classifier a number of times where appropriate, and obtained the mean signal cross section that yields the desired significance. We have requested that at least 100 signal events remain in all analyses.

The statistical significance, $\mathcal{S}$, of the analysis is calculated as:
\begin{equation}\label{eq:signif}
    \mathcal{S} = \frac{S}{\sqrt{B+(\alpha B)^2}} \;,
\end{equation}
where $\sigma_S$ is the signal cross section, $\sigma_B$ is the sum of the background cross sections, $S = \sigma_S \mathcal{L}$ and $B = \sigma_B \mathcal{L}$ are total signal and background events at an integrated luminosity $\mathcal{L}$ for a specific analysis channel, after the BDT classification has been applied and $\alpha$ is a fractional systematic uncertainty on the background processes, aiming to approximate the individual process systematic uncertainties. We do not include any systematic uncertainty on the signals, as it is very likely to be sub-dominant at the time of analysis of any future collider results. We consider $\alpha = 0.05$ as the systematic uncertainty estimate on all the backgrounds throughout our analysis. 

\subsubsection{\boldmath pp@100 TeV $h_2 \rightarrow Z Z$}\label{sec:futzz}
The final states arising through the decay of a heavy scalar to Z bosons, $h_2 \rightarrow ZZ$, provide a relatively clean avenue to discovery at hadron colliders, usually ranking as the second or third branching ratio. We have adapted the CMS analyses of~\cite{Sirunyan:2018qlb} to 100 TeV, where the final states $ZZ \rightarrow (2\ell)(2\nu)$, $ZZ \rightarrow 4\ell$ and $ZZ \rightarrow (2\ell)(2q)$ were examined.\\

\noindent \underline{$\mathbold{ h_2 \rightarrow Z Z \rightarrow (2\ell)(2\nu)}$}: The event selection for this final state consists of combining di-lepton $Z$ boson candidates with a relatively large missing transverse momentum ($\slashed{p}_T$). We require two oppositely-charged leptons of the same flavour, each with $p_T(\ell) > 50$~GeV. We further require their combined invariant mass within 30~GeV of the $Z$ boson mass and di-lepton transverse momentum, $p_T(\ell\ell) > 55$~GeV. In addition require $p_T^{\mathrm{miss}} > 125$~GeV. We veto events if $\Delta \phi(\vec{\slashed{p}}_T,~\mathrm{any~jet~with~}p_T > 30\mathrm{~GeV}) < 0.5$, where $\Delta \phi$ is the difference in angle between the $\vec{\slashed{p}}_T$ and any jet on the plane perpendicular to the beam axis. We also require the $Z$ boson candidate to satisfy $\Delta \phi (Z, \vec{\slashed{p}}_T) > 0.5$. We construct the transverse mass as:
\begin{equation}
m_T^2 = \left( \sqrt{ p_T(\ell\ell)^2 + m(\ell \ell)^2} + \sqrt{ \slashed{p}_T^2 + m_Z^2} \right)^2 - (\vec{p}_T(\ell \ell) + \vec{\slashed{p}}_T)^2 \;,
\end{equation}
where $m(\ell \ell)$ is the invariant mass of the di-lepton system. The final set of observables that are used in the discrimination of signal versus background consists of: $\bullet$ the transverse momenta of the leptons that form the $Z$ boson candidate, $p_T(\ell_1)$, $p_T(\ell_2)$, $\bullet$ the corresponding, di-lepton invariant mass, $m(\ell\ell)$, and transverse momentum, $p_T(\ell \ell)$, $\bullet$ their pseudo-rapidity distance $\Delta \eta = | \eta (\ell_1) - \eta (\ell_2) |$ and their distance $\Delta R = \sqrt{ \Delta \eta^2 + \Delta \phi^2 }$, $\bullet$ the transverse mass $m_T$ as defined above and $\bullet$ the magnitude of the missing transverse momentum, $\slashed{p}_T$. As backgrounds, we consider those that can yield the $2\ell$ final state with an associated missing transverse momentum, originating from the on-shell production of $ZZ$, $WZ$, $ZVV$ where $V=W,Z$, $t\bar{t}$ and $WW$ production, all matched via the \texttt{MC@NLO} method to the parton shower. We do not consider the mis-identification of jets or photons as leptons, and we do not include $\tau$ leptons in either signal or backgrounds, here and whenever we consider leptons in the following analyses. These would contribute additional sensitivity when considered in future analyses.\\

\noindent \underline{$\mathbold{ h_2 \rightarrow Z Z \rightarrow (2\ell)(2q)}$}: The leptons that form one of the $Z$ boson candidates are required to satisfy the same constraints as in the $(2\ell)(2\nu)$ analysis described above, with the only difference being $p_T(\ell\ell) > 100$~GeV instead. In addition to the jets with $R=0.4$, we cluster ``fat'' jets with the anti-$k_T$ algorithm with $R=0.8$. The hadronic $Z$ candidate can thus be formed either from the ``resolved'' $R=0.4$ jets or through a fat $R=0.8$ jet. The fat jets undergo ``pruning''~\cite{Ellis:2009me}, with parameters $\beta = 1.0$, $z_\mathrm{cut} = 0.1$ and $r_\mathrm{cut} =0.5$ and the ratio of the observables ``2-subjettiness'' to ``1-subjettiness'', $\tau_{21}$ is calculated for each jet~\cite{Thaler:2010tr}. For the fat jets that will form $Z$ boson candidates, $Z_\mathrm{had}^\mathrm{fat}$ we require that $\tau_{21} < 0.6$, $m(Z_\mathrm{had}^\mathrm{fat}) \in (70, 105)$~GeV, $p_T(Z_\mathrm{had}^\mathrm{fat}) > 170$~GeV, and the distance from any lepton to satisfy $\Delta R ( \ell, Z_\mathrm{had}^\mathrm{fat} ) > 0.8$. For the candidates formed through the resolved $R=0.4$ jets, $Z_\mathrm{had}^\mathrm{res}$, we require $m(Z_\mathrm{had}^\mathrm{res}) \in (40, 180)$~GeV and $p_T(Z_\mathrm{had}^\mathrm{res}) > 100$~GeV. We then choose the ``best'' candidate through the following procedure: 
\begin{itemize}
    \item If there exists a fat jet candidate with $p_T > 300$~GeV and $p_T(\ell\ell) > 200$~GeV then it takes precedence.
    \item Otherwise if there's no merged candidate with precedence, move to the resolved jets and pick the highest-$p_T$ one.
    \item If there are no resolved candidates, but there are fat ones, pick the highest-$p_T$ one.
\end{itemize}

Contrary to the analysis of~\cite{Sirunyan:2018qlb}, we do not impose any cut on the invariant mass of the $ZZ$ system. The final set of observables used in the BDT consists of: $\bullet$ the transverse momenta of the leptons that form the leptonic $Z$ boson candidate, $p_T(\ell_1)$, $p_T(\ell_2)$, $\bullet$ their invariant mass $m(\ell \ell)$ and transverse momentum, $p_T(\ell \ell)$, $\bullet$ their pseudo-rapidity distance $\Delta \eta = | \eta (\ell_1) - \eta (\ell_2) |$ and $\bullet$ their distance $\Delta R = \sqrt{ \Delta \eta^2 + \Delta \phi^2 }$, $\bullet$ the invariant mass of the hadronic $Z$ boson candidate, $m(Z_\mathrm{had}$), $\bullet$ its transverse momentum $p_T(Z_\mathrm{had})$, $\bullet$ the transverse momentum of the combined $ZZ$ leptonic and hadronic candidates, $p_T(Z_\mathrm{lep}Z_\mathrm{had})$, $\bullet$ the invariant mass of their combination $m(Z_\mathrm{lep}Z_\mathrm{had})$, and $\bullet$ the distance between them, $\Delta R(Z_\mathrm{lep}, Z_\mathrm{had})$. As backgrounds we consider those originating from the $Z$+jets events, $ZZ$, $WZ$, $WW$ and $t\bar{t}$. The $Z$+jets backgrounds have been matched/merged via the FxFx method, whereas the rest have been matched via \texttt{MC@NLO}.\\

\noindent \underline{$\mathbold{ h_2 \rightarrow Z Z \rightarrow (4\ell)}$}: Events are considered if they contain four leptons with transverse momenta satisfying, from hardest to softest, at least: $p_T(\ell_{1,2,3,4}) > 50, 50, 30, 20$~GeV. Further, events are only accepted if they contain two pairs of oppositely-charged same-flavour leptons and these are combined to form the $Z$ boson candidates, with the constraint $m(\ell \ell) \in [12, 120]$~GeV. If an event does not contain at least two $Z$ boson candidates, it is rejected. In the case of four same-flavour leptons, if there exist two viable lepton combinations, the combination $(\ell_i \ell_j)(\ell_k \ell_l)$ with the lowest value of $\chi^2 = (m(\ell_i \ell_j)-m_Z)^2 + (m(\ell_k \ell_l) - m_Z)^2$ is chosen, forming the candidates $Z_1$ and $Z_2$. We require that the combined invariant mass of the four leptons satisfies $m(\ell_i \ell_j\ell_k \ell_l) > 180$~GeV. 

The final set observables consists of: $\bullet$ the lepton transverse momenta, $p_T(\ell_{1,2,3,4})$, $\bullet$ the combined lepton invariant mass $m(\ell_i \ell_j\ell_k \ell_l)$, $\bullet$ the transverse momentum of the two $Z$ boson candidates, $p_T(Z_1)$, $p_T(Z_2)$, $\bullet$ their invariant masses, $m(Z_1)$, $m(Z_2)$, $\bullet$ their distance $\Delta R(Z_1, Z_2)$, $\bullet$ the distance between the leptons that form the two candidates, $\Delta R(\ell_i, \ell_j)$ and $\Delta R(\ell_k \ell_l)$, $\bullet$ the invariant mass of the combined $Z$ boson candidates $m(Z_1 Z_2)$ and $\bullet$ their combined transverse momentum, $p_T(Z_1 Z_2)$. We consider only the dominant backgrounds, originating from non-resonant and resonant SM four lepton production, matched at NLO via the \texttt{MC@NLO} method. In addition, we consider the LO gluon-fusion component of four lepton production that originates from the resonant loop-induced production of two $Z$ bosons, i.e.\ $gg \rightarrow ZZ$, deemed to be important at higher proton-proton centre-of-mass energies~\cite{Harlander:2018yns}. 

\begin{figure}[htp]
  \centering
  \includegraphics[width=0.45\columnwidth]{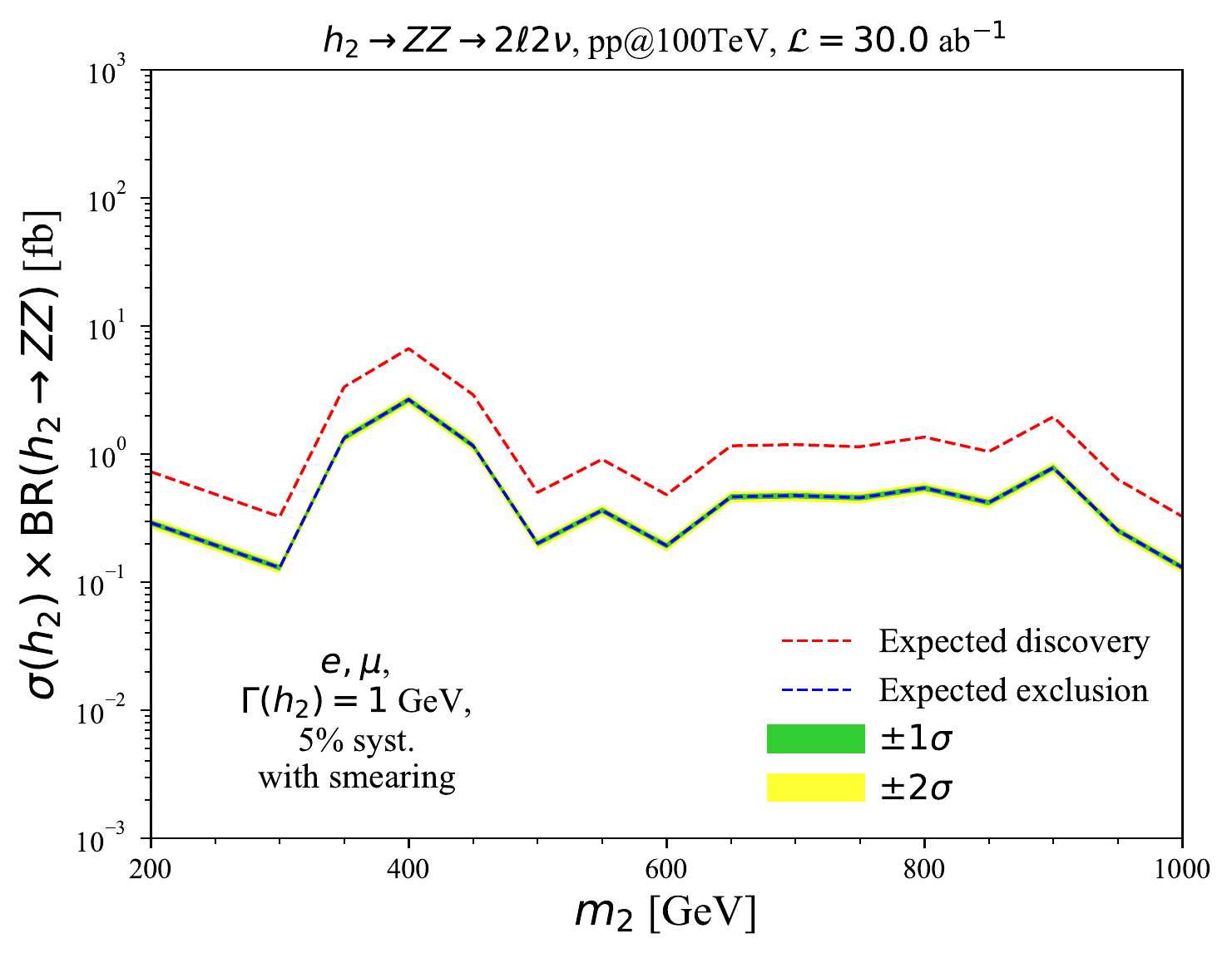}
  \includegraphics[width=0.45\columnwidth]{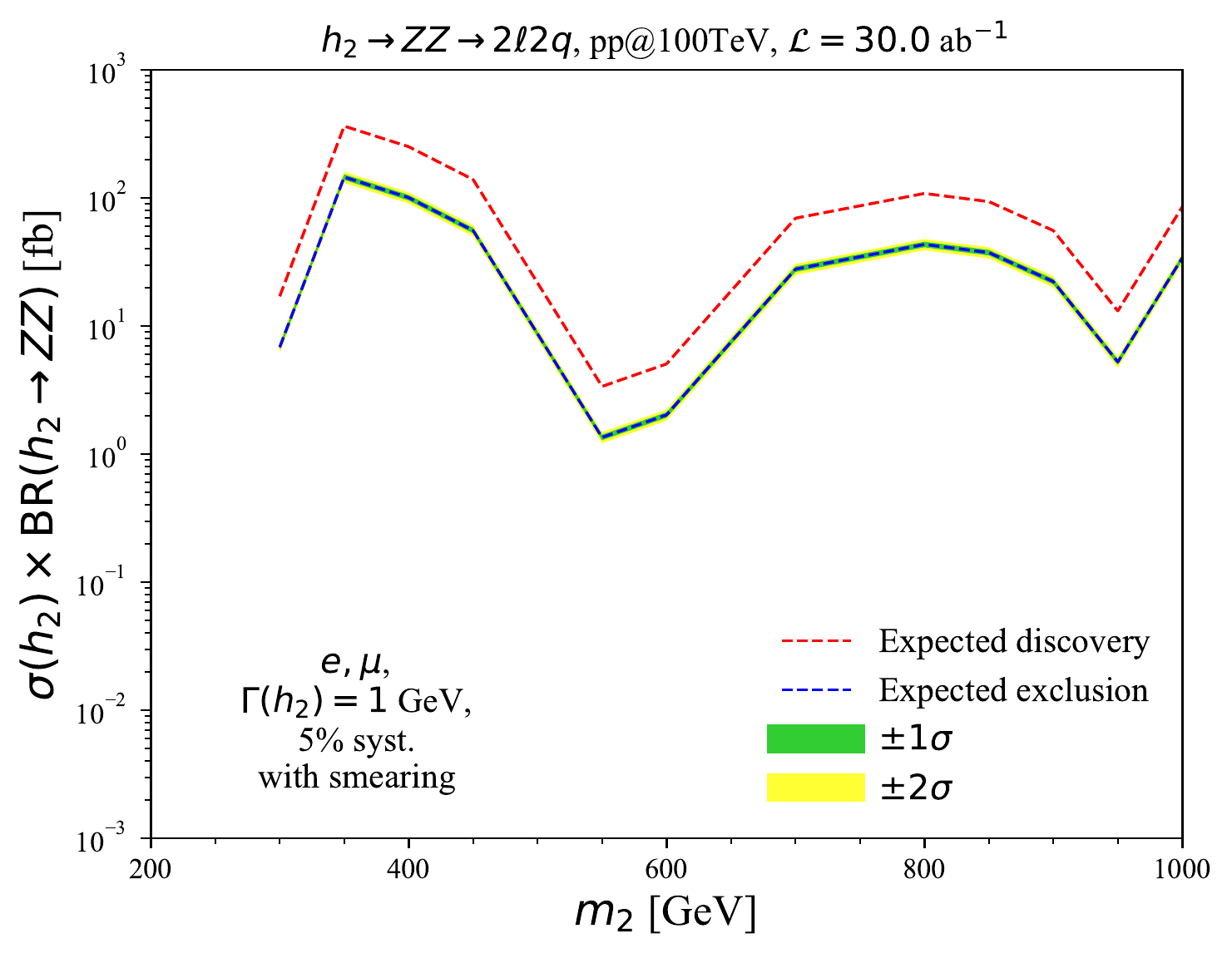}
  \includegraphics[width=0.45\columnwidth]{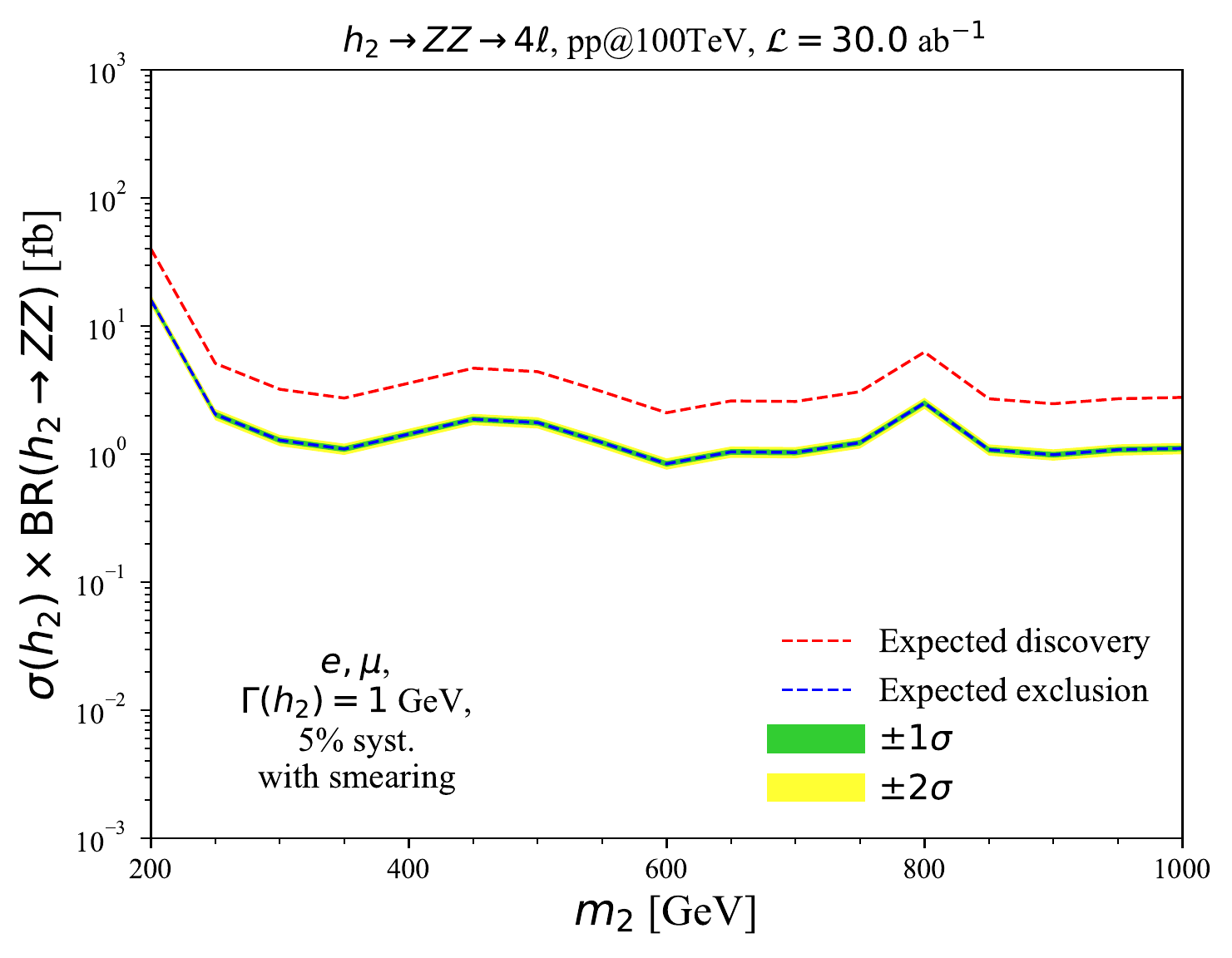}
\caption{The expected discovery ($5\sigma$, red dashes) and exclusion ($2\sigma$, blue dashes) cross sections multiplied by the branching ratio of $h_2 \rightarrow ZZ$, at a 100 TeV proton collider with an integrated luminosity of 30~ab$^{-1}$, coming from the three analyses $h_2 \rightarrow ZZ \rightarrow 2\ell 2\nu$ (top left), $h_2 \rightarrow ZZ \rightarrow 2\ell 2q$ (top right) and $h_2 \rightarrow ZZ \rightarrow 4\ell$ (bottom). The expected $1\sigma$ and $2\sigma$ variation bands are shown for the expected exclusion.}
\label{fig:analysisResultZZ}
\end{figure}

To obtain the best limit on $\sigma(h_2) \times \mathrm{BR}(h_2 \rightarrow ZZ)$, we pick the most constraining analysis at each value of $m_2$. This is in fact represented by the $h_2 \rightarrow ZZ \rightarrow (2\ell) (2\nu)$ analysis in the mass range we consider, as can be observed in fig.~\ref{fig:analysisResultZZ}. A combination of the channels should yield a more stringent limit, but we omit it in the absence of a full study of the correlations between channels. Furthermore, the $(2\ell)(2q)$ and $4\ell$ final states will allow for reconstruction of the $h_2$ and therefore a precise determination of its mass in case of discovery.

\subsubsection{\boldmath pp@100 TeV $h_2 \rightarrow W^+W^-$}\label{sec:futww}

For the analysis of the $W^+W^-$ final state we follow the ATLAS analysis of~\cite{Aaboud:2017gsl} focusing on the $WW\rightarrow (e \nu) (\mu \nu)$ final state. We require two oppositely-charged different-flavour leptons with transverse momenta $p_T(\ell_{1,2}) > 80, 60$~GeV with combined invariant mass $m(\ell_1 \ell_2) > 60$~GeV and maximum pseudo-rapidity difference $| \eta(\ell_1) - \eta(\ell_2) |$. We veto events that contain additional leptons with $p_T \geq 15$~GeV. To reject the backgrounds originating from $t\bar{t}$, we veto events containing $b$-jets with $p_T > 30$~GeV. The $b$-jet tagging probability was chosen to be $0.75$. We also allow for the mis-identification of jets as leptons, with flat probability $5\times 10^{-3}$. For the events that pass the above constraints, we construct the transverse masses of the two $W_i$'s ($i=1,2$) as:
\begin{equation}
    m_T^{W_i} = \sqrt{ 2 p_T(\ell_i) \slashed{p}_T ( 1 - \cos ( \phi^{\ell_i} - \slashed{\phi} )) }\;,
\end{equation}
where $p_T(\ell_i)$ are the lepton transverse momenta, $\slashed{p}_T$ is the missing transverse momentum, $\phi^{\ell_i}$ and $\slashed{\phi}$ are the azimuthal angles of the lepton $i$ and the missing transverse momentum vector, respectively. We require both of these observables $i=1,2$ to satisfy $m_T^{W_i} > 60$~GeV. We also construct an event transverse mass via:
\begin{equation}
    m_T = \sqrt{ (E_T(\ell \ell) + \slashed{p}_T)^2 - | \vec{p}_T(\ell\ell) + \vec{\slashed{p}}_T|^2 }\;, 
\end{equation}
where the combined transverse energy of the leptons was defined as $E_T(\ell \ell) = \sqrt{ |\vec{p}_T(\ell \ell)|^2 + m(\ell\ell)^2}$.

The final set of observables consists of: $\bullet$ the lepton transverse momenta, $p_T(\ell_{1,2})$, $\bullet$ the di-lepton transverse momentum, $p_T(\ell\ell)$, $\bullet$ the di-lepton invariant mass $m(\ell \ell)$, $\bullet$ the distance between the leptons $\Delta R (\ell_1, \ell_2)$, $\bullet$ their pseudo-rapidity distance $\Delta \eta = | \eta (\ell_1) - \eta (\ell_2) |$, $\bullet$ the $W$ transverse masses $m_T^{W_i}$, $\bullet$ the event transverse mass, $m_T$, and $\bullet$ the magnitude of the missing transverse momentum, $\slashed{p}_T$. As background processes we have considered the SM $WW$, $t\bar{t}$, $ZZ$, $WZ$ processes, simulated at NLO via \texttt{MC@NLO} and the $W$+jets process simulated via FxFx merging at NLO. 

\begin{figure}[htp]
  \centering
  \includegraphics[width=0.6\columnwidth]{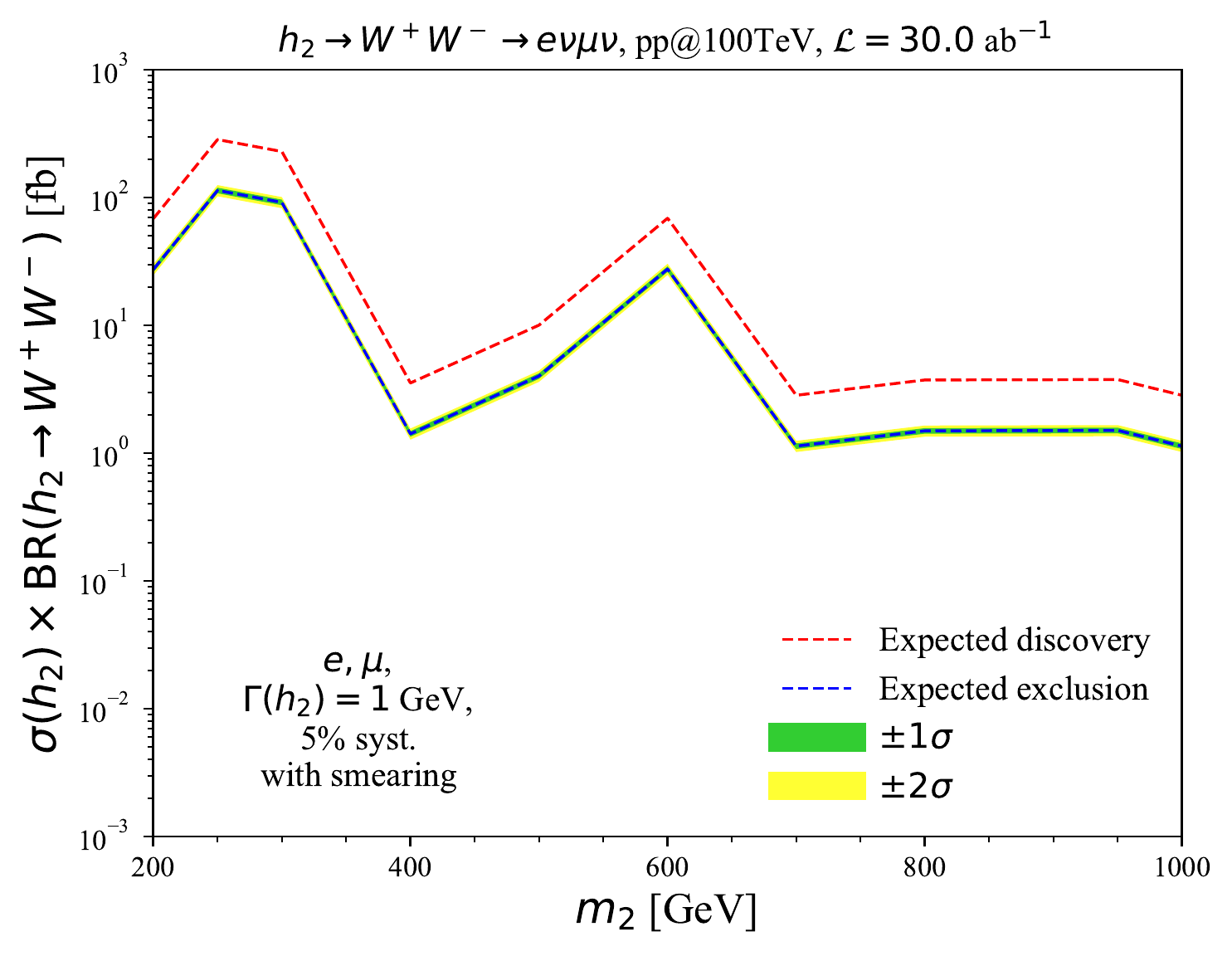}
\caption{The expected discovery ($5\sigma$, red dashes) and exclusion ($2\sigma$, blue dashes) cross sections multiplied by the branching ratio of $h_2 \rightarrow W^+ W^-$, coming from the $h_2 \rightarrow W^+ W^- \rightarrow e\nu \mu \nu$ analysis, at a 100 TeV proton collider with an integrated luminosity of 30~ab$^{-1}$. The expected $1\sigma$ and $2\sigma$ variation bands are shown for the expected exclusion.}
\label{fig:analysisResultWW}
\end{figure}

We show the resulting expected  ($5\sigma$, red dashes) and exclusion ($2\sigma$, blue dashes) cross sections times branching ratio, $\sigma(h_2) \times \mathrm{BR}(h_2 \rightarrow W^+ W^-)$, in fig.~\ref{fig:analysisResultWW}. The constraints obtained through the $h_2 \rightarrow W^+ W^- \rightarrow e\nu \mu \nu$ are comparable to those coming from the $h_2 \rightarrow ZZ$ analyses. These could be improved in future studies, e.g. by including hadronic decay modes of the $W$ bosons. 

\subsubsection{\boldmath pp@100 TeV resonant Higgs boson pair production, $h_2 \rightarrow h_1 h_1$}

The non-resonant $h_1 h_1$ final state has been investigated at a 100 TeV collider in~\cite{Mangano:2020sao}. Here we investigate resonant $p p \rightarrow h_2 \rightarrow h_1 h_1$ production at 100 TeV. We focus on by far the most sensitive final state, $h_1 h_1 \rightarrow (b\bar{b}) (\gamma \gamma)$. Future studies may include other final states to improve on this, see e.g.~\cite{Papaefstathiou:2012qe,deLima:2014dta,Papaefstathiou:2015iba}. We require all jets (including $b$-tagged) to have transverse momentum $p_T > 30$~GeV and to lie within $|\eta| < 3.0$. The $b$-jet tagging probability was set to 0.75, uniform over the transverse momentum, as for the previous analysis. The jet to photon mis-identification probability was set to $0.01 \times \exp{(p_T/30~\mathrm{GeV})}$, where $p_T$ is the jet transverse momentum~\cite{ATL-PHYS-PUB-2013-009}. We require that the invariant mass of the two $b$-jets lies in $m_{bb} \in [100, 150]$~GeV and that the invariant mass of the di-photon system within $m_{\gamma\gamma} \in [115, 135]$~GeV. 

The final set of observables constructed for the BDT consists of: $\bullet$ the invariant mass of the two $b$-jets, $m_{bb}$, $\bullet$ the invariant mass of the di-photon system, $m_{\gamma\gamma}$, $\bullet$ the invariant mass of the combined system of the two $b$-jets and the photons, $m_{bb\gamma\gamma}$, $\bullet$ the distance between the $b$-jets, $\Delta R (b,b)$ $\bullet$ the distance between the photons, $\Delta R (\gamma \gamma)$, $\bullet$ the distance between the two $b$-jet system and the di-photon system, $\Delta R (b b, \gamma \gamma)$, $\bullet$ the transverse momentum of each $b$-jet, $p_T(b_1,2)$, $\bullet$ the transverse momentum of each photon $p_T(\gamma_1)$, $p_T(\gamma_2)$, $\bullet$ the transverse momentum of the two $b$-jet system, $p_T(bb)$, $\bullet$ the transverse momentum of the di-photon system $p_T(\gamma\gamma)$, the transverse momentum of the combined $b$-jet and photon systems, $p_T(bb\gamma\gamma)$ and $\bullet$ the distances between any photon and any $b$-jet, $\Delta R (b_i, \gamma_j)$ with $i, j=1,2$. As backgrounds we consider $\gamma\gamma$+jets, $\gamma$+jets, by producing, respectively, $\gamma\gamma j$ and $\gamma j j$ via MC@NLO, $t\bar{t} \gamma \gamma$ via MC@NLO, $b\bar{b}\gamma\gamma$ and $bj\gamma\gamma$ at LO. We also consider backgrounds originating from single Higgs boson production: $b\bar{b} h_1$, $Z h_1$, $t\bar{t} h_1$, where we assume that the branching ratios possess their SM values. We also consider the non-resonant part of $h_1 h_1$ as a background, assuming that the self-coupling maintains a value close to the SM value. In the case of discovery of a new scalar particle, the full interference pattern should be considered in a more detailed analysis. We leave this endeavour to future work. 

\begin{figure}[htp]
  \centering
  \includegraphics[width=0.6\columnwidth]{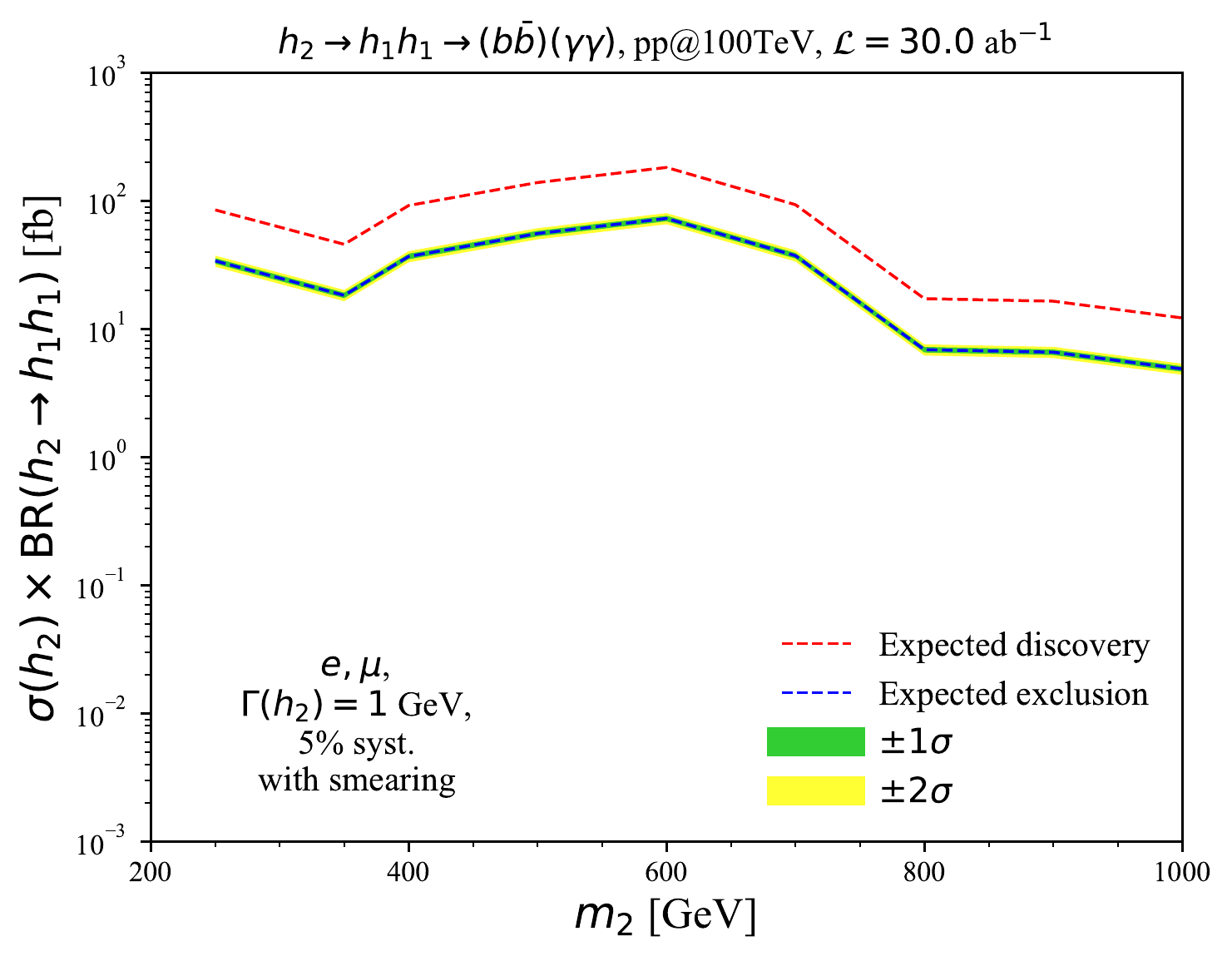}
\caption{The expected discovery ($5\sigma$, red dashes) and exclusion ($2\sigma$, blue dashes) cross sections multiplied by the branching ratio of $h_2 \rightarrow h_1 h_1$, coming from the $h_2 \rightarrow h_1 h_1 \rightarrow e\nu \mu \nu$ analysis, at a 100 TeV proton collider with an integrated luminosity of 30~ab$^{-1}$. The expected $1\sigma$ and $2\sigma$ variation bands are shown for the expected exclusion.}
\label{fig:analysisResultHH}
\end{figure}

We show the expected discovery ($5\sigma$, red dashes) and exclusion ($2\sigma$, blue dashes) cross sections multiplied by the branching ratio of $h_2 \rightarrow h_1 h_1$ in fig.~\ref{fig:analysisResultHH}. The magnitude of the constraints is comparable to that of the gauge boson analyses. We note here that the same process was also considered in~\cite{Kotwal:2016tex} in the context of the real singlet extension of the SM. Our results appear to be comparable but somewhat different. This can be attributed to differences in the analysis approach: we have calculated the $K$-factors for both signal and backgrounds, we have considered the vector-boson fusion process as part of the signal process as well as a different set of observables for which we ran the BDT.  We also use the \texttt{HERWIG} Monte Carlo event generator, hence a completely different ``tune'' for the parton showering and the description of non-perturbative effects. We expect the aforementioned facts, in conjunction with different parton density functions for the incoming proton beams, that contain rather large uncertainties when extrapolated to higher energies, will introduce large differences in the significances, at the level observed between our results and the results of~\cite{Kotwal:2016tex}. We expect all of these issues to be evaluated in detail at the onset of operation of any future collider through more precise phenomenological analyses. 

\subsection{Extrapolation of Heavy Higgs boson searches to 27 TeV}\label{app:27tev}

To obtain an estimate of the performance of a potential upgrade of the LHC to 27 TeV centre-of-mass energy, we consider an extrapolation of the detailed analyses obtained in the previous sections. We write the collider energy ($E$) dependence of the cross sections in the significance, given in Eq.~\ref{eq:signif}, as
\begin{equation}
\mathcal{S}(E, \mathcal{L}) 
 =  \frac{\sigma_S(E) \mathcal{L}} {\sqrt{\sigma_B(E) \mathcal{L} + \alpha^2 \sigma_B^2(E) \mathcal{L}^2}} \;.
\end{equation}\label{eq:signifE}

To extrapolate the obtained limits through our detailed analyses to a different energy $E'$ and luminosity $\mathcal{L}'$, we solve Eq.~\ref{eq:signif} for $\sigma_S(E')$:
\begin{equation}
    \sigma_S({E'}) = \frac{\mathcal{S} (E', \mathcal{L}')}{\mathcal{L}'} \sqrt{\sigma_B(E') \mathcal{L} + \alpha^2 \sigma_B^2(E')\mathcal{L}'^2}\;.
\end{equation}
Under the approximation that the impact of the analysis cuts remains the same between 100~TeV and 27~TeV, to obtain a limit on the signal cross section at the new energy $E'$ we only need to determine the change in the total background cross section $\sigma_B(E) \rightarrow \sigma_B(E')$ with energy. We achieve this by assuming that the background cross sections scale in the same way as either the $gg \rightarrow h_2$ process or a (hypothetical) $q \bar{q} \rightarrow h_2$ process. Since the analysis cuts will emphasise regions of phase space in which the backgrounds behave similarly to the signal, this should be a reasonable approximation. We assume that the $gg$ and $q\bar{q}$ scalings provide the error estimate on this part of our extrapolation. In the derivation of the actual limits, we use the one that provides the weakest signal constraints, so as to provide a conservative performance of a 27~TeV machine under this procedure. Another noteworthy assumption is that there will remain a sufficient number of signal events at 27~TeV after the analysis cuts are applied. This assumption ignores the fact that the extrapolation is performed from higher centre-of-mass energy and luminosity. This leads to potentially optimistic results and these should be assessed in a fully-fledged phenomenological analysis at 27~TeV. 

The results for the 27~TeV extrapolation of the individual channels are shown in fig.~\ref{fig:signif27channels}. The $ZZ$ final state represents the highest significance all over parameter space as for the case of the 100 TeV collider.

\section{Results for the alternative categorisation of parameter-space points}\label{app:altresults}

\begin{figure}[!htb]
  \centering
  \includegraphics[width=0.48\columnwidth]{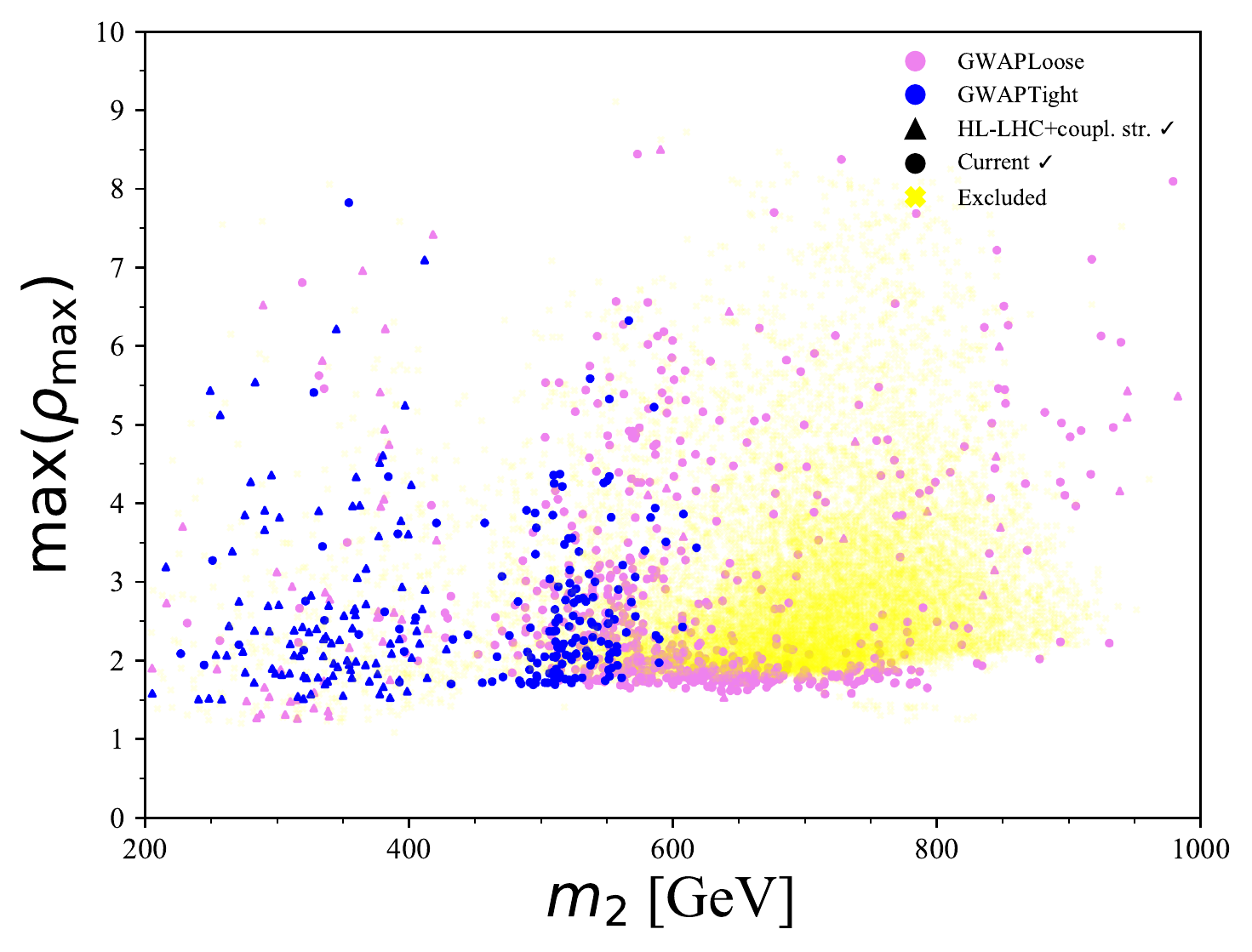}
 \includegraphics[width=0.48\columnwidth]{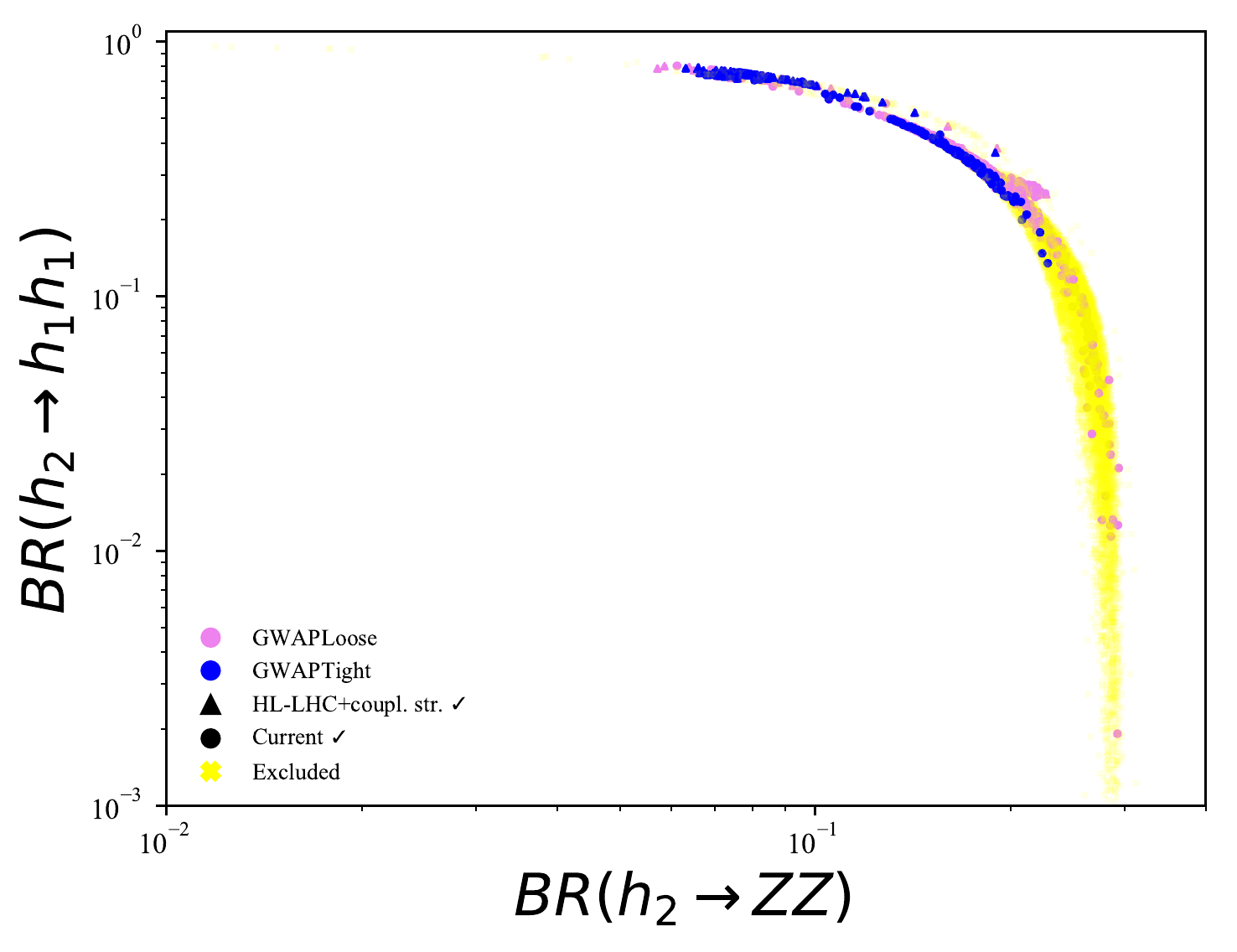}
\caption{A selection of results for the ``Tight'' and ``Loose'' categorisations section~\ref{sec:paramspace}. Left: The maximum value of $\rho_\mathrm{max} = \phi_C/T_C$ over the set of eight variations for each of the parameter-space points against the mass of $h_2$, $m_2$, right: The branching ratio of BR$(h_2 \rightarrow h_1 h_1)$ plotted against BR$(h_2 \rightarrow ZZ)$. We show points that pass the current constraints ($\CIRCLE$) or the HL-LHC constraints plus future signal strength ($| \sin \theta | < 0.1$) constraints ($\blacktriangle$), as well as those that are currently excluded (faint $\color{yellow}\pmb{\times} \color{black}$).}
\label{fig:rhomaxbr_cat2}
\end{figure}

\begin{figure}[!htb]
  \centering
  \includegraphics[width=0.48\columnwidth]{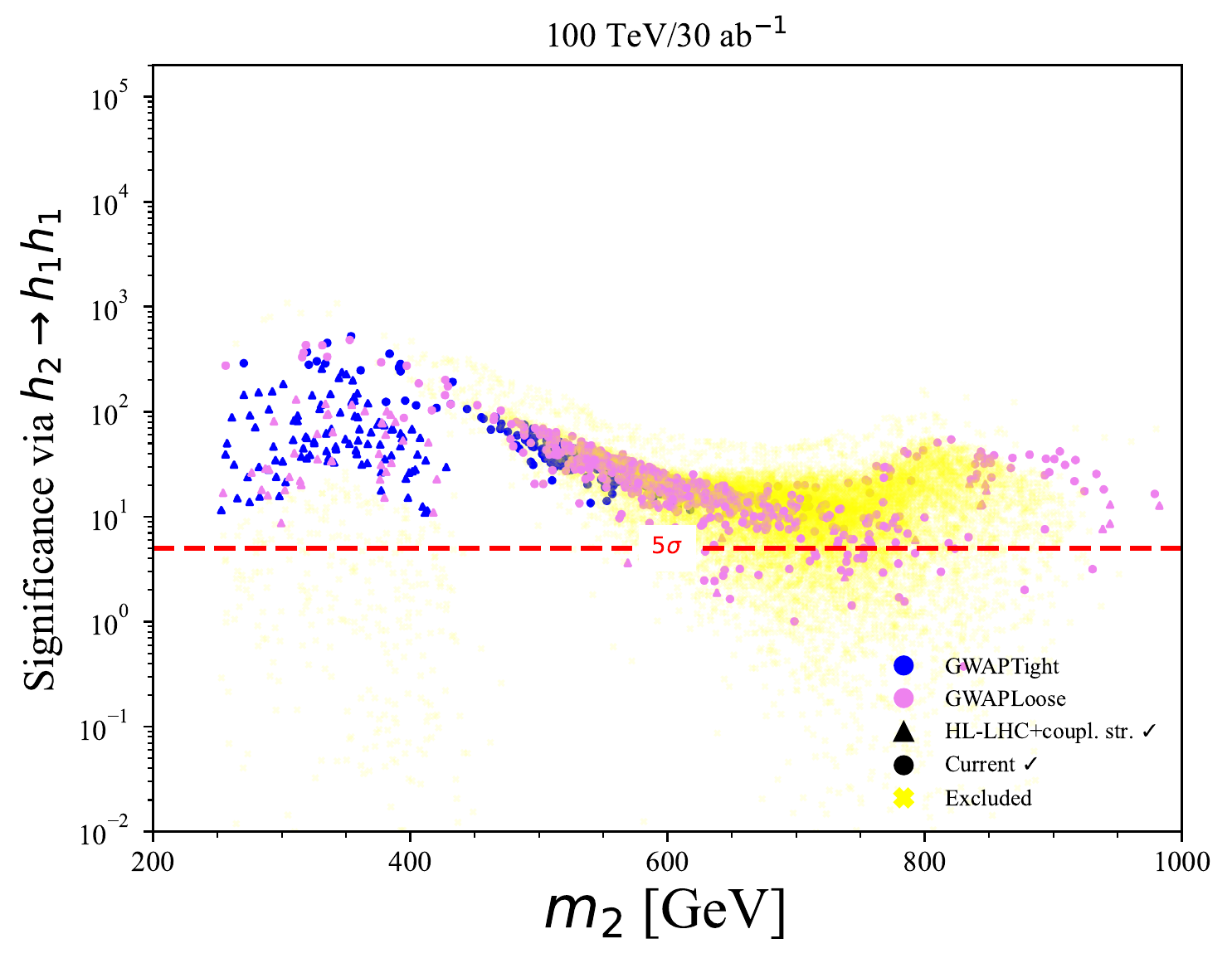}
    \includegraphics[width=0.48\columnwidth]{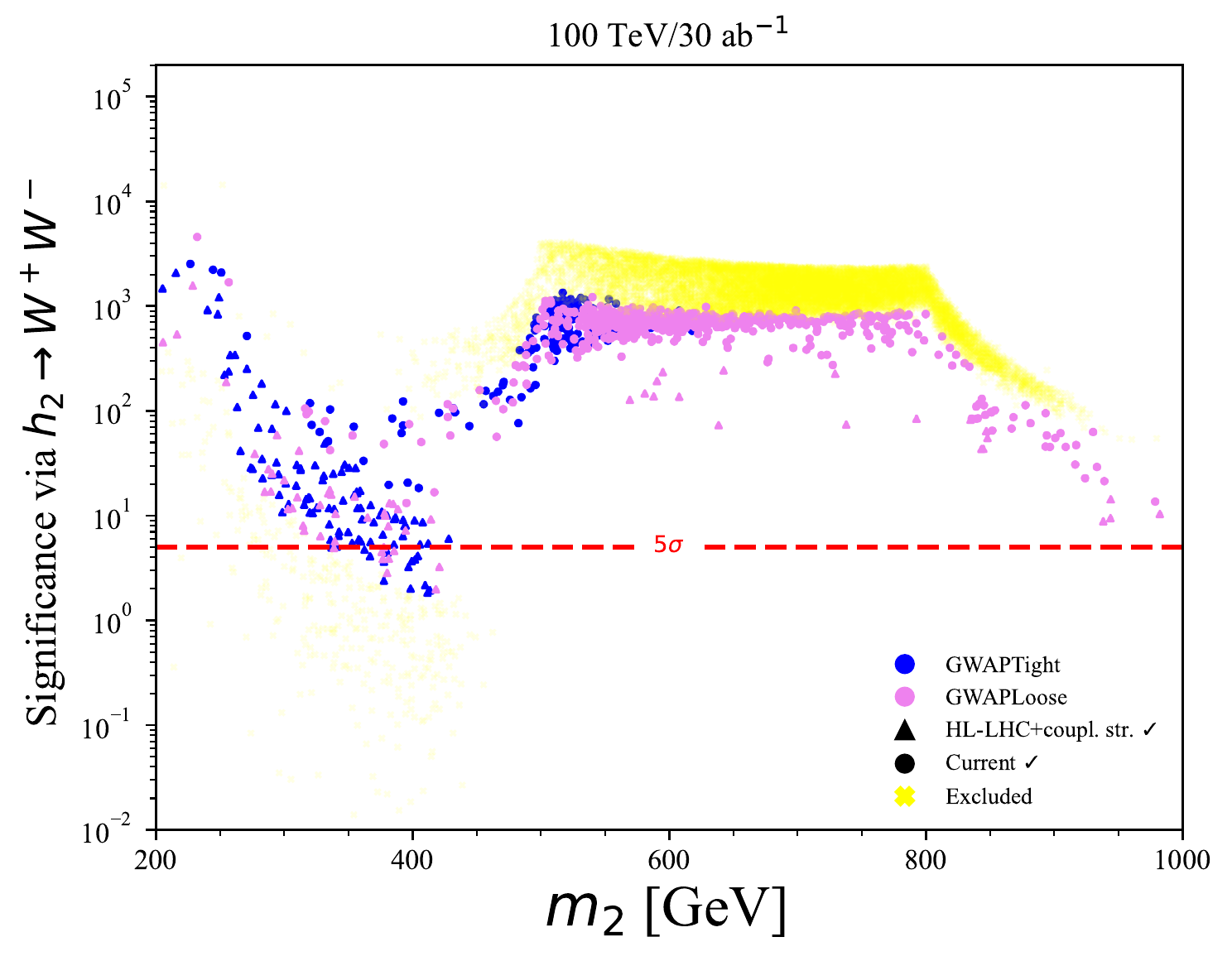}
      \includegraphics[width=0.48\columnwidth]{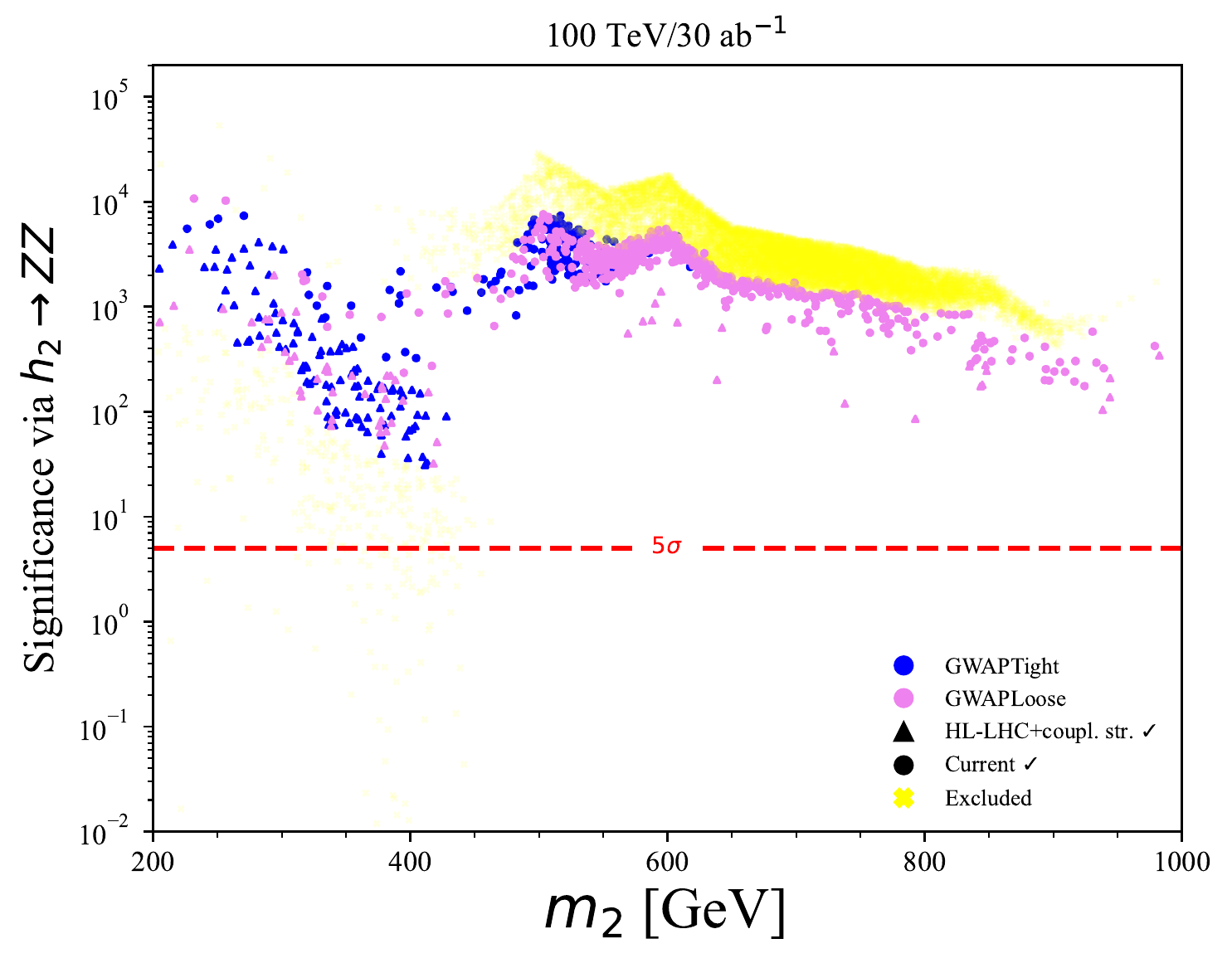}
\caption{A selection of results for the ``Tight'' and ``Loose'' categorisations section~\ref{sec:paramspace}. Top left: the statistical significance for a heavy Higgs boson ($h_2$) signal at a 100 TeV collider with an integrated luminosity corresponding to $30$~ab$^{-1}$ through $pp \rightarrow h_2 \rightarrow h_1 h_1$, top right: through $pp \rightarrow h_2 \rightarrow W^+W^-$ and bottom: through $pp \rightarrow h_2 \rightarrow ZZ$. We indicate the $5\sigma$ (discovery) boundaries by the red dashed lines. We show points that pass the current constraints ($\CIRCLE$) or the HL-LHC constraints plus future signal strength ($| \sin \theta | < 0.1$) constraints ($\blacktriangle$), as well as those that are currently excluded (faint $\color{yellow}\pmb{\times} \color{black}$).}
\label{fig:signif_cat2}
\end{figure}

Figures~\ref{fig:rhomaxbr_cat2} and~\ref{fig:signif_cat2} present a selection of results for the alternative categorisation of ``Tight'' and ``Loose'' points described in section~\ref{sec:paramspace}, found during the parameter-space scans via \texttt{PhaseTracer}. The results show that the ``Centrist'' and ``Loose'' categories behave in a similar fashion, whereas the ``Liberal'' category of the first classification has no correspondence in the second and allows for larger theoretical uncertainties.  

\section{Fine-tuning studies}\label{app:finetuning}

\begin{figure}[!htb]
  \centering
  \includegraphics[width=0.8\columnwidth]{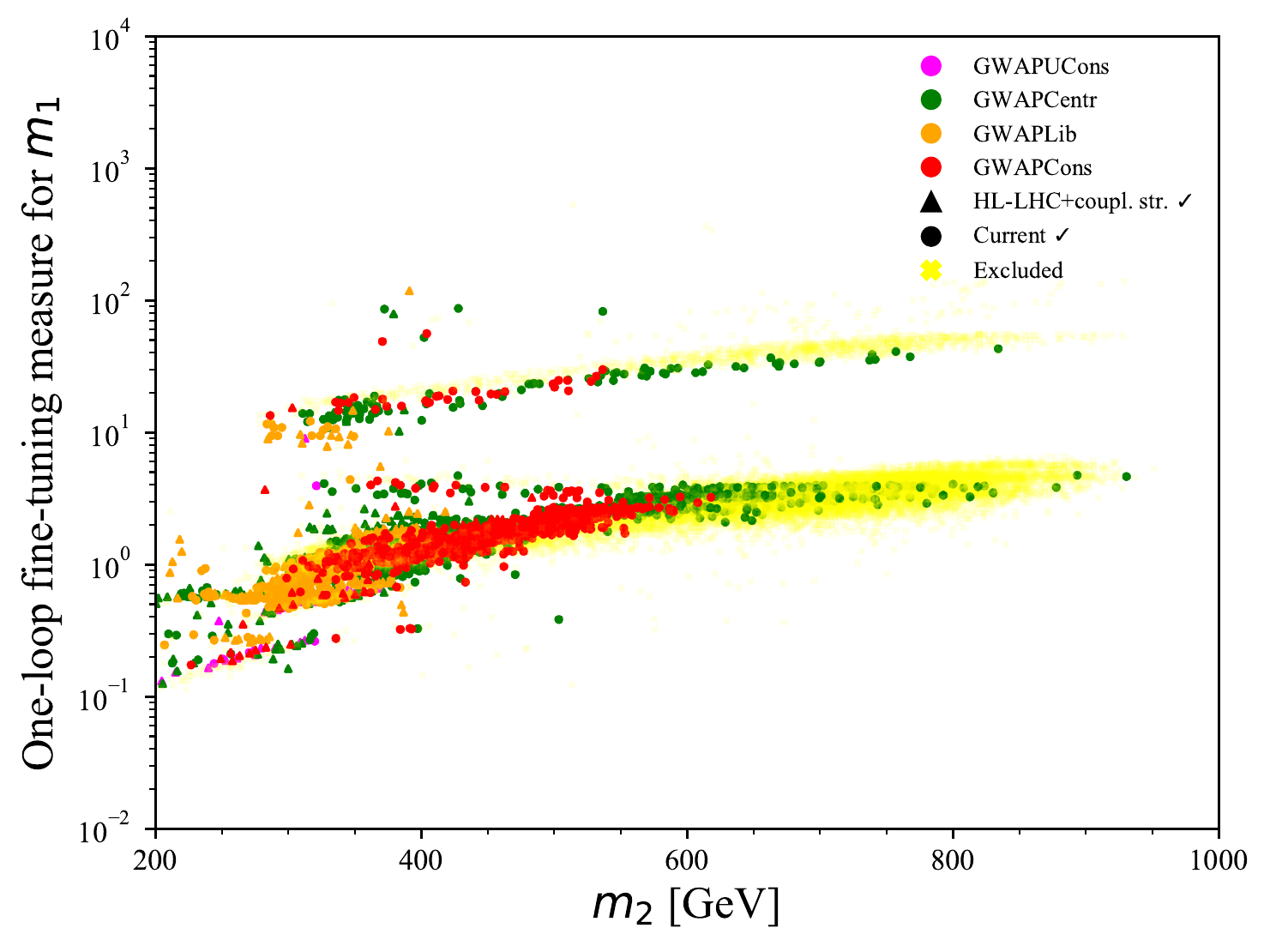}
\caption{The value of the one-loop fine-tuning measure plotted against the one-loop mass of $h_2$, $m_2$. We show points that pass the current constraints ($\CIRCLE$) or the HL-LHC constraints plus future signal strength ($| \sin \theta | < 0.1$) constraints ($\blacktriangle$), as well as those that are currently excluded (faint $\color{yellow}\pmb{\times} \color{black}$).}
\label{fig:finetuning}
\end{figure}

Any parameter-space point has to contain a light Higgs boson, $h_1$, close enough to $125.1$~GeV, if it is to be phenomenologically relevant. It is therefore reasonable to investigate whether the introduction of additional interactions requires fine tuning of the parameters to achieve the observed Higgs boson mass. To do this, we employ the Barbieri-Giudice measure~\cite{Barbieri:1987fn} (see also~\cite{Boer:2019ipg}):
\begin{equation}
    \Delta = \max_{i,j} \left| \frac{\partial \log \mathcal{O}_i}{\partial \log p_j} \right| = \max_{i,j} \left| \frac{\mathcal{O}_i}{p_j} \frac{\partial  \mathcal{O}_i}{\partial p_j} \right| \;,
\end{equation}
for a set of observables $\mathcal{O}_i$ and a set of model parameters $p_j$. One can roughly interpret the logarithm of $\Delta$ as the number of significant digits that need to be tuned in at least one of the parameters of the theory. 

We are interested in fine tuning in the Higgs boson mass and hence we consider derivatives of the one-loop expression for $\mathcal{O} = m_1$ with respect to the relevant free parameters in this case. 
\begin{equation}
    \Delta = \max_{j} \left| \frac{p_j}{m_1} \frac{\partial m_1} {\partial p_j} \right| \;.
\end{equation}
We consider this quantity for each of the parameter-space points and plot against the one-loop value of $m_2$ at renormalisation scale $\mu = m_Z$. The derivatives have been calculated numerically.

The results are shown in fig.~\ref{fig:finetuning}. It is evident that the parameter-space points that we consider do not possess substantial fine tuning. This is in line with the tree-level studies of ref.~\cite{Boer:2019ipg}, where it was pointed out that it is possible to have a large hierarchy of scales in theories with extended Higgs sectors, without having fine tuning.

\bibliographystyle{JHEP}
\clearpage
\bibliography{references.bib}
\end{document}